\documentclass[11pt]{article}
\usepackage[utf8]{inputenc}
\usepackage{amsmath,amssymb,amsthm}
\usepackage{bbm}
\usepackage{bm}
\usepackage{graphicx}
\usepackage{algorithm}
\usepackage{algpseudocode}
\usepackage{hyperref}
\usepackage{xcolor}
\usepackage{geometry}
\usepackage{booktabs}
\usepackage{natbib}
\title{A scalable Bayesian functional factor model for high-dimensional longitudinal molecular data}
\author{Salima Jaoua$^1$, Daniel Temko$^2$, Hélène Ruffieux$^{2,*}$ \\
  $^1$University of Zurich, Switzerland\\
  $^2$MRC Biostatistics Unit, University of Cambridge, United Kingdom\\
  $^*$Corresponding author: H. Ruffieux, \href{mailto:helene.ruffieux@mrc-bsu.cam.ac.uk}{helene.ruffieux@mrc-bsu.cam.ac.uk}.}
\date{}

\begin{document}
\maketitle

\begin{abstract}
Large-scale longitudinal molecular profiling 
is now firmly established in biomedical research, 
prompted by the need to  
 uncover coordinated biomarker trajectories reflecting the dynamics of underlying biological mechanisms  
 and characterise patient heterogeneity in disease progression. 
 While a range of statistical tools exist for either longitudinal modelling or high-dimensional analysis, there is no unified framework tailored to address these questions jointly.  Motivated by a longitudinal COVID-19 study conducted in Cambridge hospitals, we propose a Bayesian functional factor model to address this gap. 
 The framework combines latent factor modelling with functional principal component analysis to represent shared temporal programmes across subsets of variables while capturing individual variation through low-dimensional functional scores. We specify sparsity-inducing priors that yield interpretable factor structure 
and allow the effective number of factors to be inferred via overspecification. 
An annealed variational algorithm ensures 
efficient joint posterior inference at scale. The approach achieves 
accurate recovery of temporal structure in simulations 
with up to $20\,000$ variables. 
 Applied to a panel of 57 circulating biomarkers and to whole-blood expression of 13\,177 genes from the COVID-19 cohort, the model recovers coordinated inflammatory, interferon and immune-cell signatures, with subject-level scores that follow disease severity.  
 The methodology is implemented in the \texttt{R} package  \texttt{bayesSYNC}. %
\end{abstract}


\section{Introduction}

High-throughput longitudinal profiling is increasingly central to modern biomedical research. 
Large clinical studies now routinely collect repeated measurements of molecular markers such as genes, proteins, metabolites and immune features over time for each individual. 
These data provide a dynamic view of biological systems and offer insights into the temporal reprogramming of molecular mechanisms throughout 
disease progression, treatment response and recovery. 
A primary scientific objective in such settings is to identify a small number of underlying 
processes (biological programs or pathways) 
that drive coordinated 
changes across many variables (biomarkers), while characterising subject variability 
in the temporal expression of these processes. 
Addressing these questions jointly requires representations that are parsimonious across variables and over time, yet remain interpretable in terms of underlying biological mechanisms.

Two strands of methodology are particularly relevant in this context. 
Latent factor models provide effective dimension reduction for high-dimensional multivariate data by representing observed variables through a small number of latent components that induce structured dependence \citep[e.g.,][]{anderson1956statistical}. 
Functional principal component analysis (FPCA) offers a principled decomposition of subject-specific temporal variability into orthogonal modes of variation \citep[e.g., ][]{ramsay05, yao05, wang16}. 
Bayesian formulations of both frameworks enable coherent uncertainty quantification and borrowing of information across subjects and time points  \citep[e.g., ][]{bernardo2003bayesian, carvalho2008high, bhattacharya2011sparse, suarez17, nolan23}. 

Despite their complementary nature, these approaches have largely developed in parallel. 
Factor models typically treat observations as multivariate but not explicitly functional. 
FPCA  
captures temporal dynamics but generally analyses each variable separately or produces globally dense multivariate components \citep{nolan2025efficient} that are difficult to interpret when the number of variables exceeds a handful. In many biomedical applications, neither perspective alone is sufficient.
 This calls for approaches that jointly model complex cross-variable dependence and within-process temporal heterogeneity, while remaining scalable in practical settings 
where the number of variables can be in the 
tens of thousands and trajectories are sparsely and irregularly sampled. 
Existing approaches based on Gaussian-process latent factor models provide flexible modelling of smooth temporal variation. In particular, MEFISTO \citep{velten2022identifying} has become a popular tool in systems biology and related applications \citep[e.g.,][]{arutyunyan2023spatial, inamo2025comparative}. While MEFISTO can be configured to accommodate subject-specific latent trajectories, it does not handle shared temporal modes of variability that vary across arbitrary groups of samples. 
In addition, Gaussian-process covariance operations become computationally demanding as the number of subjects grows, which can restrict scalability in high-dimensional settings. 

Motivated by these considerations and, more specifically, by a longitudinal COVID-19 study conducted at Cambridge University Hospitals, we develop a scalable Bayesian functional factor model that combines sparse latent factor modelling with functional principal component representations of the factor processes. 
The model induces sparsity in the loading matrix through spike-and-slab priors, 
while FPCA expansions provide a parsimonious and orthogonal characterisation of temporal variability within each factor. The resulting high-dimensional posterior poses substantial computational and inferential challenges; to address these, we develop an annealed variational inference scheme with closed-form updates and provide a 
parallel implementation in the \texttt{R} package \texttt{bayesSYNC}.

\section{Motivating clinical study: a longitudinal COVID-19 application}
\label{sec_motivation}

The methodology presented in this paper was directly motivated by the Cambridge Institute of Therapeutic Immunology and Infectious Disease-National Institute for Health and Care Research (CITIID-NIHR) Bioresource COVID-19 
initiative conducted 
at Cambridge University Hospitals during the first wave of the pandemic. The study involved dense longitudinal molecular profiling of individuals with PCR-confirmed SARS-CoV-2 infection and uninfected controls \citep{bergamaschi2021longitudinal, ruffieux2023patient}. Five clinical severity strata were defined according to hospitalisation status and level of respiratory support: A, asymptomatic ($n=18$); B, mild symptomatic ($n=40$); C, hospitalised without supplemental oxygen ($n=50$); D, hospitalised with supplemental oxygen ($n=38$); and E, hospitalised with assisted ventilation ($n=69$).
 For the purposes of the present analysis, we examined 
 patient trajectories 
 over seven weeks post-symptom onset, with repeated measurements obtained at 
 patient-specific time points. Asymptomatic patients (class A) are not included in our
analyses, because their time origin is the first positive swab rather than symptom onset, and
combining the two origins would misalign the trajectories and blur the interpretation of
recovery.
 The resulting data take the form of high-dimensional multivariate functional observations 
on inflammatory proteins, cytokines, complement components, immune cell populations and metabolites, as well as whole-blood gene expression, thereby capturing multiple facets of the immune response along the disease trajectory.

Although 
early COVID-19 cohort studies revealed marked inter-individual variability in clinical trajectories, immune responses and recovery dynamics \citep{lucas2020longitudinal, brodin2021immune, huang20216}, 
the mechanisms driving this heterogeneity remain,  to date, only partially explained. Progress requires 
identifying coordinated temporal patterns across markers that reflect distinct underlying biological processes and determining how patients differ in their expression of these processes over time. In practical terms, this gives rise to 
three interrelated questions: (i) can a small number of latent functional processes explain the joint
evolution of these markers?
(ii) which subsets of markers contribute to each process?
and (iii) how do subject-specific deviations along these processes relate to
clinical severity and recovery? 

\section{Bayesian functional factor model} \label{sec_model}

\subsection{Background} 

For $i=1,\ldots,N$, let
\[
y_i = \big(y_i^{(1)},\ldots,y_i^{(p)}\big)
\]
be independent realisations of a $p$-dimensional random function
\[
y = \big(y^{(1)},\ldots,y^{(p)}\big),
\]
where each coordinate function $y^{(j)}$, $j = 1, \ldots, p$, is a square-integrable stochastic process on a compact interval 
$\mathcal{T}\subset\mathbb{R}$, equipped with Lebesgue measure. We view $y$ as a random element of the Hilbert space
$\mathcal{H}_p = L^2(\mathcal{T})^p$,
equipped with inner product
\[
\langle f,g\rangle_{\mathcal H_p}
=
\sum_{j=1}^p \int_{\mathcal T} f^{(j)}(t) g^{(j)}(t)\,dt,
\qquad
f,g\in\mathcal H_p.
\]
Each 
$y_i^{(j)}$ represents the longitudinal trajectory of variable 
$j$ for subject $i$. Throughout, $p$ is assumed to be large. In the context of longitudinal genomic studies, the realisations $\{y_i\}_{i=1}^N$ correspond to repeated trajectories of individual biomarkers (genes or downstream gene products such as proteins or metabolites), possibly with thousands to tens of thousands of features per subject. 
From a biological perspective, such high-dimensional functional observations are often thought to be driven by a few underlying biological pathways – e.g., immune activation, cell differentiation or stress response – that evolve dynamically and jointly shape the observed trajectories. 
This motivates a parsimonious representation that captures shared temporal structure while allowing
individual biomarkers to load selectively to these latent processes. 

\subsection{Latent factor representation for high-dimensional curves}\label{sec_latent}
We model the joint variation of the $p$ functional variables through a low-dimensional latent representation 
that induces structured dependence across variables. 
Specifically, we posit the decomposition
\begin{equation}
y_i = \mu + \bm B h_i \,+\, e_i,
\qquad i=1,\ldots,N,
\label{eq:latent_factor}
\end{equation}
where 
$\mu=(\mu^{(1)},\ldots,\mu^{(p)})^\top \in \mathcal{H}_p$ is the mean function,
$
h_i=(h_i^{(1)},\ldots,h_i^{(Q)})^\top \in \mathcal{H}_Q$ collects the latent factor curves for subject $i$ 
and $\bm{B}=(b_{jq})\in\mathbb{R}^{p\times Q}$ is a loading matrix with $Q\ll p$, acting linearly on $\mathcal H_Q$. The term $e_i=(e^{(1)}_i,\ldots,e^{(p)}_i)^\top \in \mathcal{H}_p$ collects variable-specific deviations, taken to be mean zero, uncorrelated across variables and across subjects, and uncorrelated with the latent processes. Decomposition~\eqref{eq:latent_factor} thus separates variation that is shared across variables, and is therefore carried by the $Q$ latent processes, from variation specific to a single variable.
The latent processes $\{h^{(q)}\}_{q=1}^Q$ are assumed mutually uncorrelated, that is, 
$$\mathbb{C}\text{ov}\left( h^{(q)}(s), h^{(q')}(t)\right) = 0, \qquad q' \neq q, \; s,t \in \mathcal T.$$ 
This assumption separates the sources of temporal variation across factors and ensures that each factor captures a distinct mode of joint variability.
 Together, these assumptions determine the covariance structure of the observed processes:
\[
\mathbb{C}\mathrm{ov}\!\left(y^{(j)}_i(s), y^{(j')}_i(t)\right)
=
\sum_{q=1}^Q b_{jq}b_{j'q}\,
\mathbb{C}\mathrm{ov}\!\left(h^{(q)}(s), h^{(q)}(t)\right)
+
\delta_{jj'}\,
\mathbb{C}\mathrm{ov}\!\left(e^{(j)}(s), e^{(j)}(t)\right),
\qquad s,t \in \mathcal T,
\]
where $\delta_{jj'}$ is the Kronecker delta. The covariance between two different variables ($j \neq j'$) thus involves the loadings and the latent processes alone, and does not depend on the temporal structure of $e_i$, whereas the covariance of a variable with itself ($j = j'$) combines them with $e_i$.

Decomposition \eqref{eq:latent_factor} extends sparse latent factor models to the functional setting by allowing
both the observed variables and the latent factors to evolve over time.
The entry $b_{jq}$ quantifies the contribution of latent factor $q$ to variable $j$.
Biologically, this representation reflects the assumption that  biomarkers participate in a limited
number of coordinated, time-varying biological pathways, 
with heterogeneous contributions across  biomarkers.
Sparsity in $\bm B$, if assumed, encodes the additional assumption that each such process influences only a subset
of  biomarkers 
(see Section~\ref{sec_spa}).

\subsection{FPCA representation of latent factor processes}\label{sec_fpca}

While the latent factor model captures cross-variable dependence through the loading matrix,
additional structure is required to characterise within-factor temporal variability across subjects.
We therefore model each latent factor trajectory using functional principal component analysis
(FPCA). FPCA represents 
subject-specific variation within each factor process 
as orthogonal modes ordered by explained variance, 
which  separates dominant temporal patterns from higher-order
fluctuations.

For $q=1,\ldots,Q$, assume that $h^{(q)}$ is a mean-zero square-integrable stochastic process
in $L^2(\mathcal T)$ with continuous covariance function
\[
\sigma^{(q)}(s,t)=\mathbb E\{h^{(q)}(s)h^{(q)}(t)\}, \qquad s,t \in \mathcal T.
\]
Given independent realisations $h^{(q)}_1,\ldots,h^{(q)}_N$ of $h^{(q)}$, define the associated
covariance operator $\Gamma^{(q)}:L^2(\mathcal T)\to L^2(\mathcal T)$ by
\[
\Gamma^{(q)} f(t)
=
\int_{\mathcal T} \sigma^{(q)}(s,t) f(s)\,ds, \qquad t \in \mathcal T, \; f\in L^2(\mathcal T).
\]
The operator $\Gamma^{(q)}$ is self-adjoint, positive semidefinite and compact.
By the spectral theorem, it admits a countable sequence of nonnegative eigenvalues
$\lambda^{(q)}_1 \ge \lambda^{(q)}_2 \ge \cdots  \ge 0$
and an associated complete orthonormal basis
$\{\psi^{(q)}_\ell\}_{\ell\ge1}$ of $L^2(\mathcal T)$ \citep{ramsay05}. 
By the Karhunen--Lo\`eve theorem, each realisation $h^{(q)}_i$ can be expanded as
\begin{equation}
h^{(q)}_i(t)
=
\sum_{\ell=1}^\infty \zeta^{(q)}_{i\ell}\,\psi^{(q)}_\ell(t),
\qquad i=1,\ldots,N,
\label{eq:kl_factor}
\end{equation}
where the scores
$\zeta^{(q)}_{i\ell}=\langle h^{(q)}_i,\psi^{(q)}_\ell\rangle_{L^2}$ 
satisfy
$\mathbb E(\zeta^{(q)}_{i\ell})=0$ and 
$\mathrm{Var}(\zeta^{(q)}_{i\ell})=\lambda^{(q)}_\ell$, and 
are uncorrelated across $\ell$ and independent across $i$ when the processes $\{h_i^{(q)}\}_{i=1}^N$ are independent realisations. 
The decomposition in \eqref{eq:kl_factor} is unique up to a change of sign of the scores and
eigenfunctions, provided the eigenvalues are distinct. 
Intuitively, the scores measure how strongly each realisation expresses the different orthogonal modes of temporal variation.
Because the eigenvalues decrease in magnitude, the leading components capture most of the temporal
variability. In practice, the expansion is truncated at a finite number $L$ of components
(see Section~\ref{sec_orth}).

\subsection{Observation model and spline representation}\label{sec_obs}

We now relate the FPCA-based latent functional model to the observed data.
For notational convenience, we assume that all curves are observed on a common grid
$\bm{t}=(t_1,\ldots,t_n)^\top$, that is,
$\bm{t}_i=\bm{t}$ and $n_i=n$ for all
$i=1,\ldots,N$;  our implementation, however, supports irregular time grids across subjects, and this flexibility is exploited in our numerical experiments (Sections~\ref{sec_sim} and \ref{sec_application}).

Throughout, functional quantities are defined as elements of
$L^2(\mathcal T)$ and are denoted using functional notation, i.e., as previously introduced, 
$y_i^{(j)}$ denotes the observed trajectory of variable $j$ for subject $i$, $\mu^{(j)}$ the mean function of variable $j$, 
$\psi^{(q)}_\ell$ the $\ell$th eigenfunction associated with latent factor $q$ 
and $h_i^{(q)}$ the trajectory of factor $q$ for subject $i$. 
When evaluated on the observation grid
$\bm t$, we write
$\bm y^{(j)}_i$, $\bm\mu^{(j)}$,  $\bm\psi^{(q)}_\ell$ and $\bm h_i^{(q)}$ for their finite-dimensional representations. 
We use a common spline basis to represent the functions 
$\mu^{(j)}$ 
and $\psi^{(q)}_\ell.$ 
Let $\bm{c}(t)~\in~\mathbb{R}^{1 \times K}$ denote the row vector of spline basis functions at time $t$. Evaluating this basis on the grid $\bm t$ yields the design matrix 

$$ \bm{C}(\bm t) =\begin{bmatrix}
\bm{c}(t_1)\\
\vdots \\
\bm{c}(t_n)
\end{bmatrix} \in  \mathbb{R}^{n \times K}.
$$
We then write
\[
\mu^{(j)}(t) =
\bm{c}(t)\boldsymbol{\nu}^{(j)}_\mu,
\qquad
\psi^{(q)}_\ell(t) =
\bm{c}(t)\boldsymbol{\nu}^{(q)}_{\psi,\ell},
\]
where $\boldsymbol{\nu}^{(j)}_\mu \in \mathbb{R}^K$ and $\boldsymbol{\nu}^{(q)}_{\psi,\ell} \in \mathbb{R}^K$ are the corresponding spline coefficients.

For the trajectories evaluated  at times $\bm{t}$, 
we assume the Gaussian observation model  
\begin{equation}
\bm{y}_i^{(j)} \mid \boldsymbol{\mu}^{(j)}, \bm{b}_j,
\{\bm{h}_i^{(q)}\}
\sim
\mathcal{N}\!\left(
\boldsymbol{\mu}^{(j)} +
\sum_{q=1}^Q b_{jq} \bm h_i^{(q)},
\;
\sigma^2_{\varepsilon,j}\bm{I}_n
\right),
\label{eq:likelihood}
\end{equation}
where
\[
\bm{h}_i^{(q)}
=
\sum_{\ell=1}^L
\zeta^{(q)}_{i\ell}\boldsymbol{\psi}^{(q)}_\ell,
\qquad
\boldsymbol{\psi}^{(q)}_\ell
:= \bm{C}(\bm t)\boldsymbol{\nu}^{(q)}_{\psi,\ell}
\in \mathbb R^n,
\qquad
\boldsymbol{\mu}^{(j)} := \bm{C}(\bm t)\boldsymbol{\nu}^{(j)}_\mu \in \mathbb R^n,
\]
 and where $\sigma^2_{\varepsilon,j}$ is a variable-specific residual
variance, gathering the measurement error and the variable-specific deviations
$e^{(j)}_i$ of~\eqref{eq:latent_factor}. Writing
$\sigma^{(j)}_e(s,t) = \mathbb E\{e^{(j)}(s)e^{(j)}(t)\}$ for the covariance
function of the latter, model~\eqref{eq:likelihood} specifies
$\sigma^{(j)}_e(s,t) = \sigma^2_{\varepsilon,j}\,\delta(s-t)$, so that the
variable-specific deviations are taken to be uncorrelated along $t$ while their
variance is allowed to differ across variables.

 Under this model, the marginal covariance between observed trajectories satisfies
\[
\mathbb{C}\mathrm{ov}\!\left(
y_i^{(j)}(s),y_i^{(j')}(t)
\right)
=
\sum_{q=1}^Q
b_{jq} b_{j'q}\,
\sigma^{(q)}(s,t)
+
\sigma^2_{\varepsilon,j}\,
\delta_{jj'}\,
\delta(s-t),
\]
where $\sigma^{(q)}(s,t)$ denotes the covariance function of latent process
$h^{(q)}$ introduced in Section~\ref{sec_fpca}, and $\delta(s-t)$ equals one if
$s = t$ and zero otherwise. The expression is understood at the observation times
$\bm t$ of model~\eqref{eq:likelihood}, at which the variable-specific deviations are
independent with variance $\sigma^2_{\varepsilon,j}$, and it is the covariance of
Section~\ref{sec_latent} with these deviations uncorrelated along $t$. Specifying the
deviations only at the observation times avoids a white-noise process in continuous
time, which would not be square integrable.

Representing
each latent process by a finite number $L$ of components on a finite spline basis
leaves a residual that varies smoothly in $t$, so a more faithful specification
would give $\sigma^{(j)}_e$ a smooth form, for instance a Mat\'ern covariance. We
retain the simpler specification for two reasons. First, $e_i$ does not enter the blocks
with $j \neq j'$ of the covariance of Section~\ref{sec_latent}, so misspecifying its
temporal structure leaves unaffected the covariance between different variables,
which carries the shared structure. Under the usual conditions for factor models
(enough variables loading on each factor), these blocks determine $\bm B$ and the
eigenfunctions up to the indeterminacies discussed in Section~\ref{sec_orth}. The
scores of individual subjects are not determined in this way, and their estimates
may be affected by the misspecification, as may the efficiency of estimation and
the calibration of the posterior uncertainty. Second, a smooth $\sigma^{(j)}_e$ would
require $p$ additional covariance functions to be estimated and would make the
variational updates for the spline coefficients considerably more costly, since the
observation-level precision would no longer be diagonal.

Section~\ref{sec_misspec} assesses the consequences of this
simplification by simulation, under a variable-specific error correlated along $t$, which
perturbs the blocks with $j = j'$, and under an unmodelled smooth component of the latent
trajectories, representing the discarded tail of the expansion, which perturbs the blocks with
$j \neq j'$. It also describes a diagnostic that can be computed on real data, the correlation
between residuals at pairs of observation times, binned by the lag between them, which is zero
at every non-zero lag under~\eqref{eq:likelihood} and positive and decaying with the lag when
the errors are correlated along $t$.

\subsection{Bayesian specification and sparsity}\label{sec_spa}

We work in a Bayesian framework. Let $\boldsymbol{\zeta}^{(q)}_i = (\zeta^{(q)}_{i1},\ldots,\zeta^{(q)}_{iL})^\top$. 
We place independent Gaussian priors on the scores and spline coefficients:
\[
\boldsymbol{\zeta}^{(q)}_i \sim \mathcal{N}(\bm{0},\bm{I}_L),
\]
\[
\boldsymbol{\nu}^{(j)}_\mu \mid \sigma^2_{\mu,j} \sim \mathcal{N}(\bm{0},\sigma^2_{\mu,j}\bm{I}),
\qquad
\boldsymbol{\nu}^{(q)}_{\psi,\ell} \mid \sigma^2_{\psi,q\ell} \sim \mathcal{N}(\bm{0},\sigma^2_{\psi,q\ell}\bm{I}),
\]
with half-Cauchy priors on all variance parameters. The eigenvalue scale is absorbed into $\sigma^2_{\psi,q\ell}$, with unit score variance imposed to avoid non-identifiability between scores and basis coefficients. Component-specific variance contributions are then recovered in post-processing after orthonormalisation (see Section~\ref{sec_orth}).

To induce sparsity in the loading matrix, we adopt a spike-and-slab prior
\[
b_{jq}\mid\gamma_{jq} \sim \gamma_{jq}\mathcal N(0,1)+(1-\gamma_{jq})\delta_0,
\qquad
\gamma_{jq}\mid\omega_q \sim \mathrm{Bernoulli}(\omega_q),
\]
where $\delta_0$ denotes the Dirac probability measure at zero. This  specification conveniently 
yields marginal posterior inclusion probabilities  of the form $\mathrm{pr}(\gamma_{jq} = 1 \mid \bm y) = \mathbb{E}(\gamma_{jq}\mid \bm y)$, which provide a direct measure of evidence for factor--variable associations. Here and throughout, $\mathrm{pr}(\cdot)$ denotes the probability of an event and $\pi(\cdot)$ a density or probability mass function, the symbol $p$ being reserved for the number of variables. Continuous shrinkage priors such as the horseshoe
\citep{carvalho10} would be a reasonable alternative for estimating the loadings, but they
set no loading exactly to zero, so that assigning variables to factors would require a
further thresholding rule. The spike-and-slab instead gives each loading a posterior
inclusion probability, on which both the assignment of variables to factors and the pruning
of factors (Section~\ref{sec_Q}) are based.

The inclusion probability $\omega_q$ is factor-specific and controls the
expected proportion of nonzero loadings within factor $q$. We complete the hierarchical specification by placing a hyperprior on $\omega_q$, which allows the sparsity levels of the different latent factors be 
learnt from the data
through the posterior distribution of $\omega_q$, 
\begin{equation}\label{eq_beta}\omega_q \sim \mathrm{Beta}(1,p).\end{equation}
Since $\mathbb{E}(\omega_q)=1/(1+p)$, the expected number of nonzero loadings per factor 
remains of order one as $p$ increases. This specification is common in high-dimensional Bayesian variable selection
\citep[see, e.g., ][]{scott2010bayes} and serves as a default in applications
where each latent factor is expected to involve only a small subset of variables.
The Beta hyperparameters 
can however  be modified to encode alternative prior beliefs
about the overall degree of sparsity. 

\begin{figure}[t!]
\centering
\includegraphics{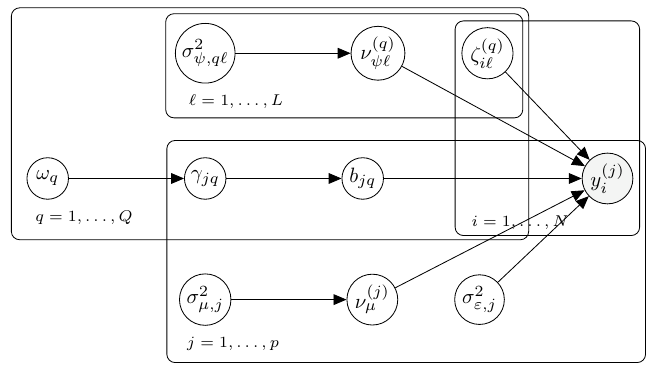}
  \caption{\footnotesize Graphical representation of model (\ref{eq_full_model}). The shaded nodes are observed, the others are inferred.}\label{fig_gm}
\footnotesize Alt text: Directed acyclic graph of the model. The shaded observed node $y_i^{(j)}$ receives arrows from the mean-function coefficients $\nu^{(j)}_{\mu}$, the error variance $\sigma^2_{\varepsilon,j}$, the loading $b_{jq}$, the eigenfunction coefficients $\nu^{(q)}_{\psi\ell}$ and the scores $\zeta^{(q)}_{i\ell}$. Further arrows run from $\sigma^2_{\mu,j}$ to $\nu^{(j)}_{\mu}$, from $\omega_q$ to $\gamma_{jq}$ to $b_{jq}$, and from $\sigma^2_{\psi,q\ell}$ to $\nu^{(q)}_{\psi\ell}$. Plates indicate replication over variables $j$, factors $q$, components $\ell$ and subjects $i$.
\end{figure}

For clarity, we now gather the full hierarchical specification of the model introduced above,
with the latent functional representation, observation model and prior distributions.
For 
observations $\bm y_i^{(j)}\in \mathbb R^n$ at times $\bm t$ of curves
$y_i^{(j)}$, $i=1,\ldots,N$, $j=1,\ldots,p$,
the hierarchical model is
\begin{align}
\hspace{0.3cm}\bm y_i^{(j)}
\mid 
\bm\nu^{(j)}_\mu, &\bm b_j,
\{\bm\nu^{(q)}_{\psi,\ell}\},
\{\zeta^{(q)}_{i\ell}\},
\sigma^2_{\varepsilon,j}
\sim
\mathcal N\!\left(
\bm C(\bm t)\bm\nu^{(j)}_\mu
+
\sum_{q=1}^Q b_{jq}
\sum_{\ell=1}^L
\zeta^{(q)}_{i\ell}
\bm C(\bm t)\bm\nu^{(q)}_{\psi,\ell},
\;
\sigma^2_{\varepsilon,j}\bm I_n
\right),\nonumber
\\[0.2em]
\zeta^{(q)}_{i\ell}
&\sim
\mathcal N(0,1),
\hspace{4.07cm}\sigma_{\varepsilon,j}
\sim
\mathrm{Half\text{-}Cauchy}(A), \quad q = 1, \ldots, Q,\; \ell = 1, \ldots, L,\, 
\nonumber\\
\bm\nu^{(j)}_\mu
\mid \sigma^2_{\mu,j}
&\sim
\mathcal N(\bm 0,\sigma^2_{\mu,j}\bm I_K),
\hspace{3.1cm}
\sigma_{\mu,j}
\sim \mathrm{Half\text{-}Cauchy}(A),
\label{eq_full_model}\\
\bm\nu^{(q)}_{\psi,\ell}
\mid \sigma^2_{\psi,q\ell}
&\sim
\mathcal N(\bm 0,\sigma^2_{\psi,q\ell}\bm I_K),
\hspace{2.82cm}
\sigma_{\psi,q\ell}
\sim \mathrm{Half\text{-}Cauchy}(A), 
\nonumber\\
b_{jq}\mid\gamma_{jq}
&\sim
\gamma_{jq}\mathcal N(0,1)
+
(1-\gamma_{jq})\delta_0,
\qquad
\gamma_{jq}\mid\omega_q
\sim
\mathrm{Bernoulli}(\omega_q),
\qquad
\omega_q
\sim
\mathrm{Beta}(1,p),\nonumber
\end{align}
where $A>0$, and $\bm C(\bm t)\in\mathbb R^{n\times K}$ denotes the spline
design matrix evaluated on the observation grid $\bm t$. A graphical representation of the model is provided in Figure~\ref{fig_gm}. 

In practice, we use a slightly modified implementation for the latent functions' representation using a semiparametric penalised spline construction. Specifically, the design matrix is partitioned into an
unpenalised low-order polynomial component and a penalised
component controlling smoothness, namely, $\bm C(\bm t) \equiv [ \textbf{1}_{n} , \bm t ,\bm  z_1(\bm t) , \ldots, \bm z_{K'}( \bm t)] $ with $K' = K-2$, and the corresponding spline coefficients have priors of 
the form $$\boldsymbol{\nu}^{(j)}_\mu
\sim \mathcal N\!\left( \bm 0, \bm \Sigma_{\mu}^{(j)}\right), 
\qquad \bm \Sigma_\mu^{(j)} = \text{blockdiag}(\sigma_\beta^2 \bm I_2, \sigma_{\mu,j}^{2}\bm I_{K'}),$$
with user-defined hyperparameter $\sigma^2_\beta$; we use an analogous specification for $\bm\nu^{(q)}_{\psi,\ell}$. This formulation is equivalent to the O’Sullivan spline
penalty \citep{osullivan86, wand08}, with smoothness learnt through the variance
parameters $\sigma_{\mu,j}^{2}$ and $\sigma^2_{\psi,q\ell}$. The number of basis functions is chosen according
to the rule
$K' = \max\left\{\min\left(\left\lfloor n / 4 \right\rfloor, 40\right), 7\right\},$ adapted from 
\citet{ruppert2002selecting}. 
Finally, for computational stability, 
we adopt an equivalent auxiliary variable scale-mixture formulation for the variance parameters. 
For instance, for the mean function variance we write
$$ \sigma^2_{\mu,j} \mid a_{\mu,j} \sim \text{IG}\left(\frac{1}{2}, \frac{1}{a_{\mu,j}}\right),\qquad a_{\mu,j} \sim\text{IG}\left(\frac{1}{2}, \frac{1}{A^2}\right), $$
which marginally induces a half-Cauchy prior on $\sigma_{\mu,j}$ with scale $A$.
This auxiliary-variable formulation restores conditional conjugacy. 
An identical construction is used for $\sigma^2_{\varepsilon,j}$ and $\sigma^2_{\psi,q\ell}$. In practice, we use large default values for both the half-Cauchy scale parameter and the standard deviation controlling the unpenalised polynomial components ($A = 10^5$ and $\sigma_\beta= 10^5$), which yields diffuse prior specifications. Under such choices, inference is largely driven by the likelihood, while the priors provide sufficient regularisation to ensure numerical stability.

\subsection{Identifiability and post-processing}\label{sec_orth}

The factor representation in \eqref{eq:latent_factor} is identifiable only up to orthogonal
transformations, since rotations of the latent factors and corresponding score transformations leave the likelihood unchanged. 
Moreover, orthonormality of the eigenfunctions and uncorrelatedness of the scores are not enforced
during Bayesian estimation.
Following inference, we therefore apply a post-processing step based on the posterior mean estimates. Specifically, we evaluate the factor trajectories on a dense time grid, perform an eigendecomposition of their empirical covariance to obtain an orthonormal basis and rotate the score accordingly,   
reordering components by explained variance. This fixes the rotational indeterminacy within each factor and restores the FPCA interpretation; see \citet{nolan23} for theoretical justification.

The post-processing does not resolve the indeterminacy
between factors, the loadings $\bm B \bm R$ and factors $\bm R^\top h_i$ giving the same fit
for any orthogonal $Q \times Q$ matrix $\bm R$. This indeterminacy is resolved by the
spike-and-slab prior, since a rotation generally makes the columns of a sparse loading matrix
dense, so that rotated solutions, although they fit the data equally well, have much lower
prior probability than the sparse one. Conditions under which a sparsity pattern
determines the loadings up to the order and signs of the columns are given by \citet{rohe23}
and \citet{hosszejni26}. We do not verify them here, and fix the order and signs by
convention, ordering the components by explained variance and choosing the sign of each
factor for ease of interpretation, since the data do not determine it. The scale is shared in the same way between the loadings and the trajectory of each
factor, $b_{jq} c$ and $h^{(q)}/c$ giving the same fit for any $c > 0$, and this
indeterminacy is resolved by the prior, through the unit variance of the slab.

This post-processing yields an ordered decomposition that allows  inference to be carried out with a conservatively large upper bound $L_{\max}$ on the number of 
components used to represent each factor. The proportion of variance explained (PVE) by component $\ell = 1, \ldots, L_\text{max},$ is estimated as
\[
\widehat{\mathrm{PVE}}^{(q)}_\ell
=
\frac{\widehat{\lambda}^{(q)}_\ell}
{\sum_{\ell'=1}^{L_{\max}}\widehat{\lambda}^{(q)}_{\ell'}}\,, 
\]
for each factor $q = 1, \ldots Q$, where $\widehat{\lambda}^{(q)}_\ell$ denotes the estimated $\ell$th  eigenvalues of the covariance operator. These estimates are obtained from the empirical variance of the post-processed scores, exploiting the FPCA identity $\mathbb{V}\text{ar}(\zeta^{(q)}_{i\ell})=\lambda^{(q)}_\ell$ (see Section~\ref{sec_fpca}). We then retain components until the cumulative PVE plateaus or exceeds a high threshold
(e.g., 99\%).
This procedure, analogous to standard
practice in classical principal component analysis, allows learning factor-specific numbers of components 
$L^{(q)}$.

\section{Inference algorithm}\label{sec_inference}

Posterior inference for model~\eqref{eq_full_model} is challenging due to the combination of
high-dimensional functional parameters, latent factor scores and a sparse loading
matrix governed by spike-and-slab priors.
The resulting posterior distribution is high-dimensional and strongly multimodal,
which makes generic Markov chain Monte Carlo (MCMC) methods computationally prohibitive in the
settings considered here.
We therefore adopt a variational inference approach, which provides a scalable
deterministic alternative while retaining posterior uncertainty quantification.

\subsection{Variational approximation}\label{sec_vb}

Let $\bm \Theta$ denote the collection of all 
model parameters.
Variational inference approximates the posterior distribution
$\pi(\bm \Theta \mid \bm y)$ with a tractable distribution $q(\bm \Theta)$ by minimising the
reverse Kullback--Leibler divergence 
 \begin{equation}\label{eq_kl}\mathrm{KL}\left(q \,\| \,\pi\right) = - \int q(\bm \Theta) \log\left\{ \frac{\pi(\bm \Theta\mid \bm y)}{q(\bm \Theta)}\right\} \mathrm{d}\bm \Theta\,. \end{equation}

Equivalently, this corresponds to maximising the evidence lower bound (ELBO),
\[
\mathcal{L}(q)
=
\mathbb{E}_q\{\log \pi(\bm y,\bm \Theta)\}
-
\mathbb{E}_q\{\log q(\bm \Theta)\},
\]
where $\mathbb{E}_q(\cdot)$ denotes expectation with respect to the variational
distribution $q(\bm \Theta)$.

Here we use  \emph{mean-field} variational inference \citep{opper2001advanced}, in which the variational distribution  $q(\bm \Theta)$ is assumed to factorise over some partition of the parameters $\{\bm \Theta_m\}_{m=1,\ldots, M}$, namely,
\begin{equation}\label{eq_mean_field} q(\bm \Theta) = \prod_{m=1}^M q_m(\bm \Theta_m)\,, \end{equation}
while leaving the functional form of each $q_m(\bm \Theta_m)$ unspecified. We employ a mean-field factorisation that is tailored to the structure of the functional factor model and that follows closely that developed in
\citet{nolan23} for the parameters involved in the FPCA expansion of each factor.
In particular, we assume posterior independence of the mean and latent functions
(the global parameters) from the factor scores (the subject-specific parameters).
Posterior independence between variance parameters and their associated
hyperparameters arises naturally from the auxiliary-variable representation of
half-Cauchy priors and is justified by asymptotic independence arguments
\citep{menictas13} and induced factorisations based on graph-theoretic results
\citep[Section~10.2.5]{bishop06}.
Further factorisations for the spline coefficient vectors follow from
Proposition~4.1 in \citet{nolan23}. Crucially, we model the spike-and-slab pairs $(b_{jq},\gamma_{jq})$ jointly.
This structured factorisation preserves the point-mass mixture structure of the
spike-and-slab prior and avoids collapsing the multimodal marginal posterior of the
loadings into a single Gaussian component, which would otherwise lead to poor
approximations of posterior inclusion probabilities.
With a single variable and a single factor, $p = Q = 1$
and $b_{11} = 1$, the model corresponds to the FPCA model of \citet{nolan23}. The updates for
the mean function coefficients $\bm\nu^{(j)}_\mu$ and for the variance parameters, with their
auxiliary-variable representation, keep the form given there, the contributions of the
factors being subtracted from the data before the mean function is updated. The other
updates have no counterpart in that setting, because they couple the $Q$ processes to one
another and to the loading matrix. The updates for the eigenfunction coefficients
$\bm\nu^{(q)}_{\psi,\ell}$ and for the scores $\bm\zeta^{(q)}_i$ pool information across all
$p$ variables through the second moments of the loadings, and use residuals from which the
contributions of the other factors have been subtracted; the pairs $(b_{jq},\gamma_{jq})$ are
updated jointly, and the factor-level inclusion probabilities $\omega_q$ govern the pruning of
Section~\ref{sec_Q}. These are derived in Supplementary Material~S1, with the corresponding
contributions to the evidence lower bound in Supplementary Material~S2. Likewise, the
post-processing of Section~\ref{sec_orth} applies the orthonormalisation of \citet{nolan23}
within each factor, whereas the separation of the factors from one another has no counterpart
in the single-factor setting.

Under this factorisation, coordinate ascent updates follow from the standard
variational identity
\[
\log q_m(\bm \Theta_m)
=
\mathbb{E}_{q_{-m}}\{\log \pi(\bm y,\bm \Theta)\}
+
\text{const.},
\]
where $\mathbb{E}_{q_{-m}}(\cdot)$ denotes expectation with respect to all
variational factors except $q_m(\bm \Theta_m)$ and \text{const.} is a constant with respect to $\bm \Theta_m$.
Thanks to conditional conjugacy throughout the model hierarchy as described in 
Section~\ref{sec_spa}, all variational updates admit closed-form expressions (see Supplementary Material S1). This fully analytical structure is crucial in the high-dimensional setting that concerns us, as it ensures scalable computation by enabling efficient deterministic updates without resorting to numerical optimisation or stochastic approximation at each iteration. In contrast to Gaussian-process-based latent factor formulations or fully MCMC-based implementations, the proposed variational algorithm scales to settings with $p > 10^4$ without requiring covariance inversion at each iteration.

\subsection{Annealed variational inference}

Although variational inference is computationally efficient, the ELBO is generally
non-concave and spike-and-slab priors induce multiple local optima corresponding to
different sparsity configurations of the loading matrix.
To improve exploration of this multimodal objective, we couple the variational
algorithm with a simulated annealing scheme
\citep{rose1990deterministic, ueda1998deterministic}, specifically the deterministic
variant developed for variational Bayes (VB) by \citet{katahira08}.
Related annealed variational approaches have been successfully implemented in
high-dimensional Bayesian models, including applications in genomics and
computational biology
\citep[e.g.,][]{ruffieux2020global, ruffieux2021epispot}.

Annealing introduces a temperature parameter $T \ge 1$ into the variational objective,
\[
\mathcal{L}_T(q)
=
\mathbb{E}_q\{\log \pi(\bm y,\bm\Theta)\}
-
T\,\mathbb{E}_q\{\log q(\bm \Theta)\}.
\]
For $T>1$, the entropy term is inflated, flattening the objective surface and reducing
the influence of local modes.
As $T$ decreases gradually to $1$, the algorithm transitions smoothly to the
original variational objective.

We adopt a 
temperature schedule starting from a user-specified maximum
temperature $T_{\max}$ and decreasing progressively to $T=1$.
At each temperature level, we update the variational parameters  before
moving to the next temperature until $T = 1$, at which point the algorithm is run until convergence (which is monitored by evaluating the ELBO at each iteration). 
Annealing introduces no additional per-iteration computational cost, as the same
closed-form updates apply at all temperatures. We provide full derivations and update equations in the Supplementary Material~S1~\&~S2.

Our annealed variational inference algorithm is implemented as an \texttt{R} package called \texttt{bayesSYNC}. The implementation supports parallel execution, which can substantially accelerate computation in practice. 
The annealing schedule is specified by a three-dimensional control vector, defining
(i) the spacing of the temperature grid (geometric by default; harmonic or linear alternatives are also available);
(ii) the initial temperature $T_{\max}\in(0,2)$ ($1.9$ by default); and
(iii) the number of temperature levels ($100$ by default). Unless otherwise stated, all results reported in the numerical experiments of Sections~\ref{sec_sim} and \ref{sec_application} are obtained using annealed variational inference with the default annealing schedule. We find this schedule to provide reliable posterior exploration; in practice, initial temperatures between $1$ and $2$ and temperature grids of size $10$ to $100$ yield stable inferences, with results showing little sensitivity to the precise choice of these parameters within these ranges. A dedicated empirical assessment of the benefits of this annealing scheme relative to standard variational inference is provided in Section~\ref{sec_sim_anneal}.

\subsection{Learning the number of factors}\label{sec_Q}

Thus far, we have treated the number of latent factors $Q$ as fixed. 
In practice, the true dimensionality of the latent factor structure 
is unknown and poor
specification may lead to overly simplistic or overly complex representations.
The Bayesian literature has proposed several approaches for adaptive rank selection,
notably through priors enforcing increasing shrinkage on higher-index factors, such
as the multiplicative gamma process shrinkage (MGPS) prior
\citep{bhattacharya2011sparse, montagna2012bayesian} and the cumulative shrinkage process
(CUSP) prior \citep{legramanti2020bayesian, kowal2023semiparametric, fruhwirth2023generalized}.

Rather than introducing additional hyperpriors for ordered shrinkage, we directly leverage
the specific structure of our modelling framework.
Regularisation is already enforced through multiple mechanisms: component-specific
variance parameters on spline coefficients shrink unnecessary FPCA components,
while spike-and-slab priors on the loading matrix encourage sparsity across variables.
Together, these features tend to prune away irrelevant 
factors when the model is
overspecified.

We take advantage of this and deliberately fit the model with a conservatively large number of factors, $Q_{\mathrm{max}}$, subsequently discarding those
unsupported by the data.
Specifically, exploiting the mean-field factorisation, we define \emph{posterior factor inclusion probabilities} as the posterior
probabilities that each factor $q \in \{1, \ldots, Q_{\mathrm{max}}\}$ is ``active'', i.e., 
\[
\mathrm{PPI}_q^{\mathrm{factor}}
=
\mathrm{pr}\!\left(\sum_{j=1}^p \gamma_{jq} \ge 1 \mid \bm y \right)
\;\approx\;
1 - \prod_{j=1}^p \bigl(1 - \mathrm{PPI}_{jq}\bigr),
\]
where $\mathrm{PPI}_{jq} = \mathbb E_q(\gamma_{jq})$ is the variational posterior mean of $\gamma_{jq}$ which approximates 
the posterior
probability of inclusion of loading $b_{jq}$ for variable $j$ and factor $q$.

We retain only those factors for which $\mathrm{PPI}_q^{\mathrm{factor}}$ exceeds a
pre-specified threshold \citep[e.g., $0.5$ corresponding to a \emph{median probability model}, ][]{barbieri2004optimal}, discarding the remainder.
 The numerical experiments show that this strategy recovers the relevant factors, and Section~\ref{sec_dis} discusses how it behaves when thousands of variables load on each factor.  We adopt it 
throughout the analyses presented in this paper.

\section{Simulation studies}
\label{sec_sim}

We present simulation studies that assess the ability of our approach to recover
the latent time-varying structure, the 
sparse loading patterns and the effective
number of active factors, under regimes motivated by longitudinal molecular data.
The experiments are designed to reflect key challenges encountered in practice, including high dimensionality, and
irregular and sparse observation grids. 

We consider four complementary assessments: (i) qualitative recovery on a single large-scale
dataset, (ii) systematic assessment across multiple replicates, with particular emphasis
on the benefits of annealed variational inference, and (iii) a comparison with the functional latent factor approach based on Gaussian processes, MEFISTO \citep{velten2022identifying}, in moderate-dimensional regimes where such a comparison is computationally feasible, and (iv) an assessment of robustness to departures from the model.

\subsection{Data-generation framework and inference settings}
\label{sec_sim_data}

Throughout, we simulate data for $N$ subjects and $p$ functional variables observed on $\mathcal T=[0,1]$. For subject $i$, observations are collected at
$n_i$ time points, where $n_i$ is drawn uniformly from a prespecified range, and the
observation times are sampled independently and uniformly over $\mathcal T$. This yields
heterogeneous and sparse temporal designs across subjects.

The data are generated from a latent factor model of the form 
\[
y_i(t) = \mu(t) + \bm B \, h_i(t) + \varepsilon_i(t),
\qquad t \in \mathcal T, \; i=1,\ldots,N,
\]
where $\mu \in \mathcal H_p$ is a smooth mean function, $h_i \in \mathcal H_{Q}$  corresponds to  
$Q$ latent factor processes and $\bm B$ is a
$p \times Q$ loading matrix. The noise processes
$ \varepsilon_i$ are independent across subjects and variables and follow Gaussian
white noise with unit variance. 

We vary the sparsity of the loading matrix $\bm B$ across simulation settings; the precise regimes are described in the corresponding sections. Depending on the scenario, latent factors may influence only a small subset of variables or act
more broadly across variables, with sparsity levels sampled in a factor-specific manner. We ensure that each latent factor is associated with at
least one variable to avoid degenerate components.

Each factor process $h^{(q)}_i$ is represented using a finite functional principal
component expansion,
\[
h^{(q)}_i(t) = \sum_{\ell=1}^{L} \zeta^{(q)}_{i\ell} \, \psi^{(q)}_\ell(t),
\]
where the eigenfunctions $\psi^{(q)}_\ell$ are constructed from orthonormalised B-spline bases.
The spline bases are chosen to be sufficiently smooth, with degree 2 or 3 varying across eigenfunctions and with equally spaced
knots over $\mathcal T$. 
The scores $\zeta^{(q)}_{i\ell}$ are independent Gaussian random variables with
mean zero and variances decreasing with $\ell$, namely,  $\sigma^2_{\zeta,\ell} = 1/\ell^2$, ensuring that higher-order components explain
progressively less variability. Details specific to the mean functions and the choice of basis functions are described in the
corresponding simulation sections below. 
The number of replicate datasets is given for each scenario in the corresponding section.

For inference, we deliberately use an over-specified model in all simulation experiments.
Specifically, we fit the model with $Q_{\mathrm{max}} = 5  > Q$ latent factors and
$L_{\mathrm{max}} = 5 > L$ FPCA components for each factor, and rely on shrinkage and factor-selection
mechanisms to recover the effective dimensionality as described in Sections~\ref{sec_orth} and~\ref{sec_Q}. In practice, some applications may require exploring larger candidate values of $Q_{\mathrm{max}}$ and $L_{\mathrm{max}}$; this remains computationally efficient, as the algorithm scales approximately linearly in $L$ and near-linearly in $Q$ per iteration.  Unless otherwise stated, inference is
performed using the default annealing scheme of the variational algorithm implemented in \texttt{bayesSYNC}. 
Since factor models are identifiable only up to permutation and sign changes, we apply permutation and sign matching to the orthonormalised estimated quantities (eigenfunctions, scores and factors), exclusively for aligning them with the simulated quantities 
and allow meaningful computation of performance metrics. Estimated factors are assigned to simulated factors by
searching over all assignments for the one whose posterior inclusion probabilities are
closest, in Frobenius norm, to the simulated sparsity pattern, and the sign of each factor is
the one that brings its loadings closer to the simulated loadings. Each simulated component is
then matched to the estimated component whose scores are most correlated, in absolute value,
with its simulated scores, the sign of that correlation setting the sign of the component.

We assess estimation accuracy for all functional quantities using the integrated squared error,
$\mbox{ISE}(f,\hat f)=\int_{\mathcal{T}} |f(x)-\hat f(x)|^2 \mathrm{d}x$, where $f(\cdot)$ denotes the data-generating function and $\hat f(\cdot)$ its corresponding posterior estimate (after alignment, for the latent factor processes and eigenfunctions).

\subsection{Qualitative recovery in a large-scale setting}
\label{sec_sim_single}

To illustrate the model outputs, we begin by presenting results from a single replicate of a large-scale simulated dataset. 
We consider $N=100$ subjects and $p=20\, 000$ variables, a dimensionality chosen to be
representative of genome-wide gene expression measurements in humans. The data are generated using
$Q=3$ latent factor processes, each represented using $L=2$
FPCA components. We simulate smooth and periodic variable-specific mean functions, with
phase shifts varying across variables. 
The latent factor eigenfunctions and scores are simulated as described in Section~\ref{sec_sim_data}. In this experiment, the number of observation times per subject is drawn independently as
$n_i \sim \mathrm{Unif}\{5, 10\}$.
We consider a semi-sparse regime in which each latent factor influences,
on average, approximately one variable out of eleven, i.e., with factor-specific probability of association drawn from a $\text{Beta}(1, 10)$ distribution. 
The resulting data can be
interpreted as longitudinal gene expression trajectories where the latent factors represent
shared biological pathways evolving over time, with only a small subset of genes responding to a given biological process.

\begin{figure}[t!]
\centering
\includegraphics[scale=0.53]{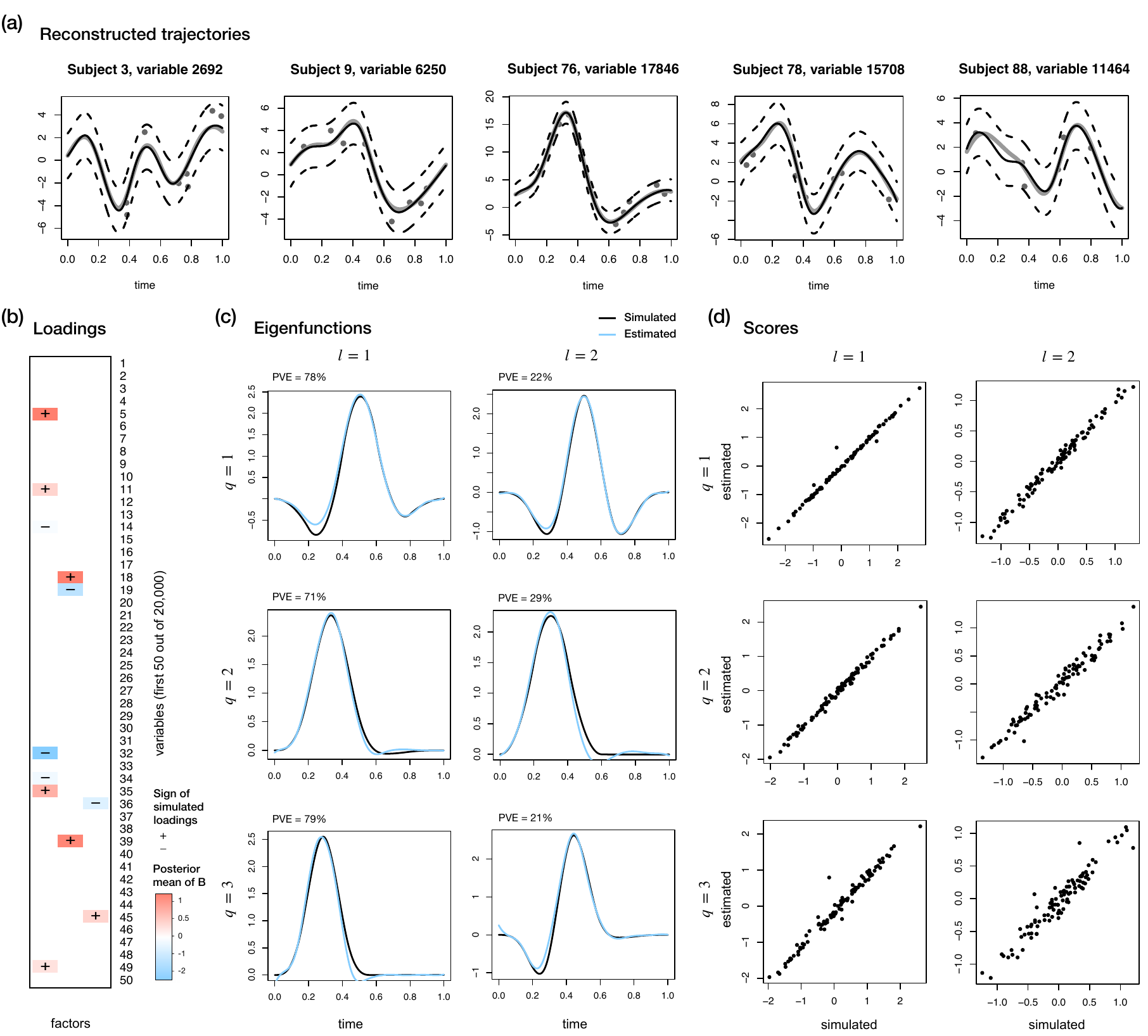}
\caption{\footnotesize
Qualitative recovery in a large-scale simulated setting ($N=100$ subjects, $p=20\,000$ variables, $Q=3$ latent factors, $L=2$ FPCA components). Results are shown for a single simulated dataset. 
(a)~Observed trajectories for a randomly selected subset of subjects and variables together with fitted trajectories and 95\% pointwise prediction bands obtained from annealed variational inference, with the true simulated trajectories as thick grey lines. The model is fitted to all subjects and variables, and the subset is shown for legibility only. 
(b)~Estimated loading matrix for the retained factors, displayed for the first 50 variables as an illustrative subset; rows correspond to variables and columns to latent factors, and $+$ and $-$ mark the nonzero simulated loadings and their signs. 
(c)~Comparison of true and estimated eigenfunctions for each factor after post-processing and alignment. 
(d)~Comparison of true and estimated subject-level FPCA scores.}\label{fig2}
\footnotesize Alt text: Four panels for one simulated data set. (a) Observed points, fitted trajectories with shaded prediction bands and thick grey simulated trajectories for a few subjects and variables. (b) Heat map of the estimated loadings of 50 variables on three factors, blue for negative and red for positive values, with plus and minus signs marking the nonzero simulated loadings. (c) Simulated and estimated eigenfunctions of each factor. (d) Estimated against simulated scores for each factor and component, close to the identity line.
\end{figure}

Figure~\ref{fig2}(a) displays the simulated trajectories for a random subset of
subjects and variables, together with the fitted trajectories obtained using variational
inference with 95\% pointwise prediction bands. The fitted trajectories provide a smooth reconstruction that appears to conform with the observed values and with the true simulated trajectories, even under sparse sampling. 

We next examine parameter recovery. Applying a median probability model threshold of $0.5$ to the posterior factor inclusion probabilities correctly retains three factors in this scenario.  Figure~\ref{fig2}(b) shows the estimated
loading matrix for these factors, restricted to the first $50$ variables, as an illustrative subset of the $20\,000$ variables analysed. 
The estimated sparsity patterns closely match the true ones, including
both the locations and signs of the nonzero loadings (the nonzero simulated loadings are marked $+$ or $-$ in the figure according to their sign, and their magnitudes are shown alongside the estimates in Supplementary Material~S3).

Figures~\ref{fig2}(c)\&(d) compare the true and estimated eigenfunctions and 
FPCA scores for each factor. The dominant modes of temporal variation are accurately recovered, with the model correctly identifying that the first two components capture all of the temporal variability within each latent function. 
Estimation of the second component scores tends to be slightly less precise,
which is expected as it reflects their smaller contribution to the overall variability.

\subsection{Benefits of annealed variational inference}
\label{sec_sim_anneal}

We now assess performance across 50 replicates of the large-scale setting described
in the previous section, with $N=100$, $p=20\,000$ and
$Q=3$, focusing on the role of annealed variational inference.

\begin{figure}[t!]
\centering
\includegraphics[scale=0.415]{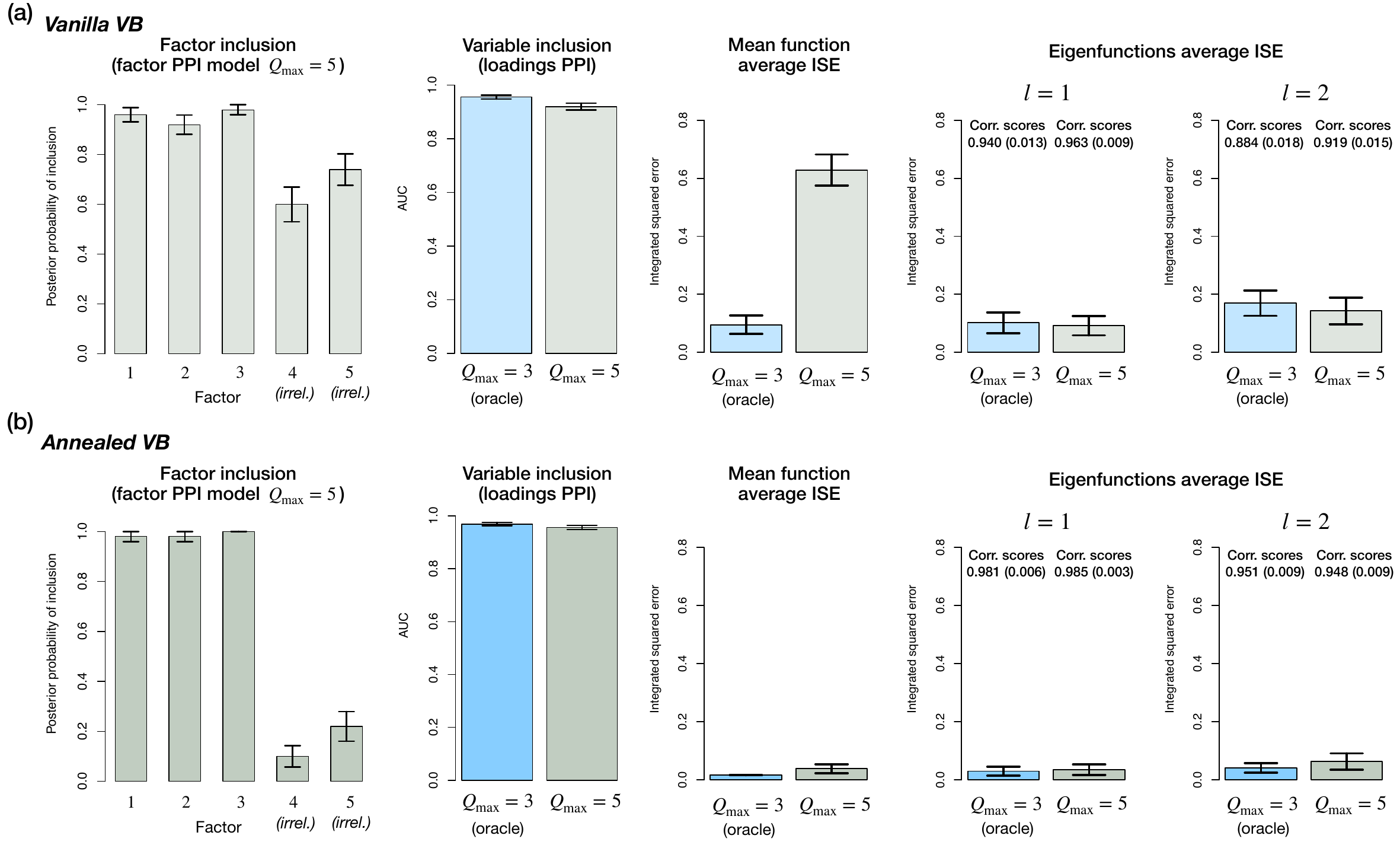}
\caption{\footnotesize
Effect of annealing on estimation accuracy and factor selection in the large-scale simulation setting ($N=100$, $p=20\,000$,  $Q=3$, $L=2$). Results are averaged over 50 independently generated datasets; error bars denote $\pm$ one standard error across replicates.
(a)~Vanilla variational Bayes (no annealing). From left to right: posterior factor inclusion probabilities (PPI$^{\text{factor}}_q$) under an overspecified model with $Q_{\max}=5$; variable inclusion performance for loadings (area under the receiver operating characteristic (ROC) curve, AUC, based on variable-factor loading PPI$_{jq}$); mean function integrated squared error (ISE); and eigenfunction ISE for $\ell=1,2$, under $Q_{\mathrm{max}}=3$ (oracle) and $Q_{\mathrm{max}}=5$. Reported score correlations are means (standard errors) across replicates.
(b)~Annealed variational Bayes under the same specification, with identical performance metrics.}
\label{fig3}
\footnotesize Alt text: Two rows of bar charts with standard-error bars, for vanilla (a) and annealed (b) variational inference. Each row shows the posterior inclusion probabilities of the five candidate factors, the AUC for the recovery of the loadings, and the integrated squared errors of the mean functions and of the first two eigenfunctions, for $Q_{\max}=3$ and $Q_{\max}=5$. Without annealing the two superfluous factors have non-negligible inclusion probabilities; with annealing these are close to zero.
\end{figure}

We first consider standard variational inference without annealing and compare its application using a correctly specified number of factors $Q=3$ (``oracle model'') with the default overspecified setting including $Q_\mathrm{max}=5$ candidate factors. As shown in Figure~\ref{fig3}(a), although the three true factors are typically identified with high posterior
inclusion probabilities, the two additional factors are often retained with non-negligible
probability. While the reconstructed loadings and eigenfunctions are comparable to those
obtained under the oracle model, the estimation error for the mean function is
substantially larger. This behaviour suggests a lack of identifiability, whereby
irrelevant factors partially compensate for the mean structure. 

To assess whether this behaviour results from entrapment in suboptimal local modes and whether annealing can mitigate this, we repeat the analysis using annealed variational inference. As illustrated in Figure~\ref{fig3}(b), the compensation
effect is essentially eliminated: the three true factors are still selected with high confidence,
while the remaining factors are now correctly inferred to be irrelevant. Moreover, estimation
accuracy improves across all components of the model, including the mean function, even under
the oracle specification.

\begin{figure}[t!]
\centering
\includegraphics[scale=0.44]{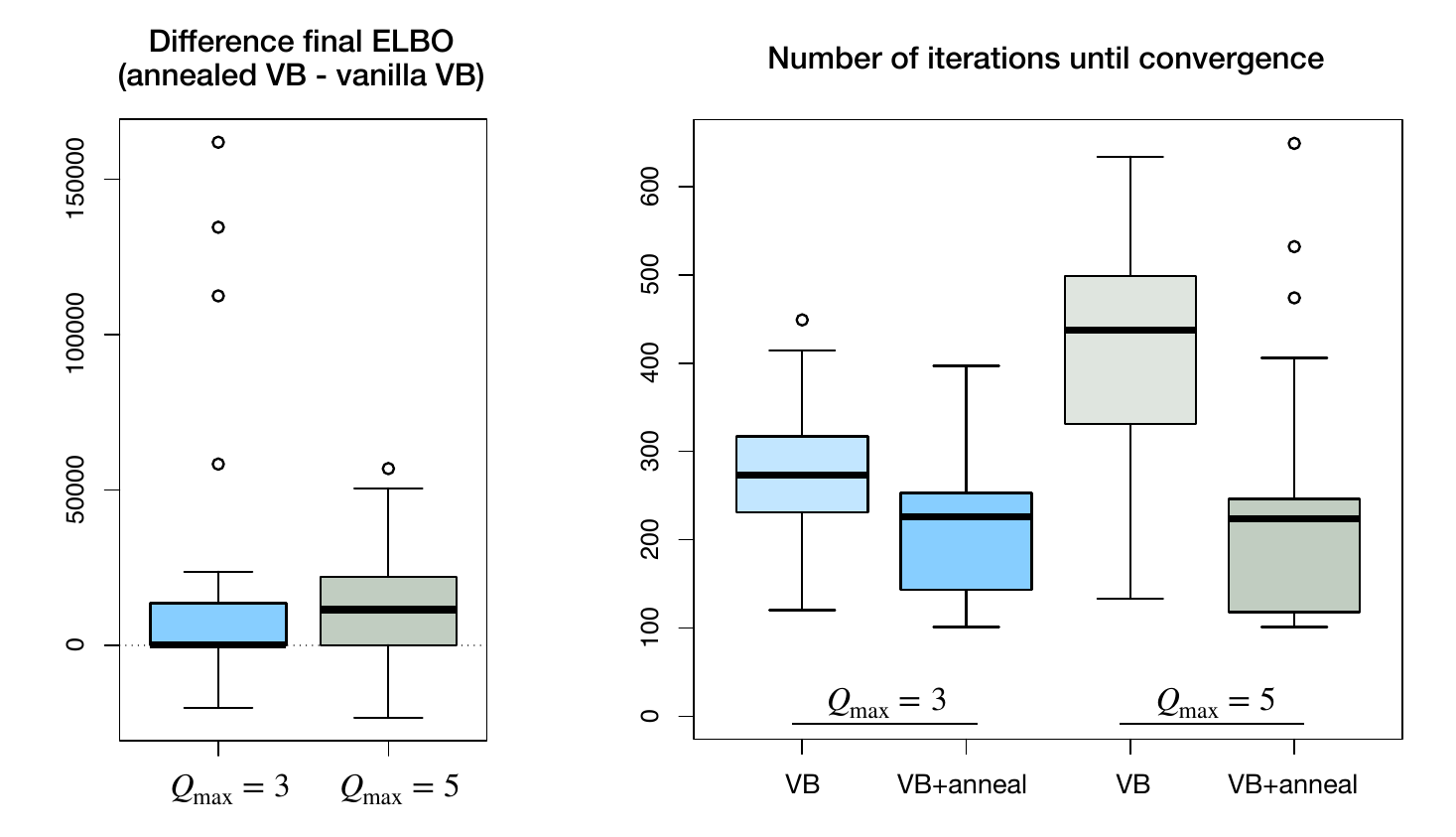}
\caption{\footnotesize
Optimisation behaviour with and without annealing, summarised over 50 simulation replicates. 
Left: Difference in final ELBO (annealed VB minus vanilla VB) under overspecified models with $Q_{\mathrm{max}}=3$ and $Q_{\mathrm{max}}=5$. Positive values indicate improved optima under annealing.
Right: Number of iterations until convergence for vanilla VB and annealed VB under $Q_{\mathrm{max}}=3$ and $Q_{\mathrm{max}}=5$. Boxplots summarise the empirical distribution across replicates.}
\label{fig4}

\footnotesize Alt text: Two boxplot panels over the replicates. Left, the difference in final ELBO between annealed and vanilla inference for $Q_{\max}=3$ and $Q_{\max}=5$. Right, the number of iterations to convergence for the four combinations of inference scheme and $Q_{\max}$.
\end{figure}

These improvements are reflected in both lower reconstruction error and higher evidence lower
bounds as shown in Figure~\ref{fig4}, consistent with a tighter bound on the marginal likelihood and a better variational fit. Interestingly, annealing also reduces the number of iterations required for convergence,
suggesting that it facilitates exploration of the variational landscape leading to 
faster 
convergence
to better modes.

\subsection{Comparison with a Gaussian-process based latent factor approach}
\label{sec_sim_mefisto}

In this section, we present numerical comparisons with MEFISTO \citep{velten2022identifying}, a Gaussian process–based factor analysis approach that is routinely applied in molecular analyses \citep[e.g.,][]{arutyunyan2023spatial, inamo2025comparative}. MEFISTO models latent factors as smooth
Gaussian processes but does not explicitly decompose within-factor temporal variability into
orthogonal modes, as our method does through FPCA. As a result,
the outputs of the two methods are not directly equivalent, with our approach providing a
richer representation of subject-specific temporal heterogeneity.
Additionally, whereas our model uses a spike-and-slab prior to induce element-wise sparsity in the loadings, MEFISTO by default assumes Gaussian loadings with factor-specific precision parameters under an automatic relevance determination (ARD) framework, in which all loadings of a factor share a precision parameter with a gamma prior, so that a factor with little support in the data is shrunk towards zero as a whole, without any individual loading being set to zero.

\citet{velten2022identifying} propose two settings for their method, that differ in how latent trajectories are modelled. The default setting estimates a single latent trajectory per factor that is shared across all subjects. Temporal evolution is therefore global and subjects differ only through observation noise and loading structure, which does not provide a representation of subject-level variation in temporal dynamics. We refer to this specification as ``MEFISTO-base''. The second setting can estimate subject-specific latent trajectories by defining each subject as a separate ``group''. This setting, which lets the latent trajectories vary across subjects, is the closer to our model. We refer to this setting simply as ``MEFISTO''.

While MEFISTO-base involves estimating only one Gaussian process per factor, the subject-specific setting requires modelling a substantially larger joint Gaussian process across all observed time points and subjects. This increases both computational and memory demands. 
To ensure that the experiments can be completed within a reasonable time frame ($<36$ hours) on our institutional high-performance computing cluster, we restrict this comparison to moderate-dimensional settings with $p=100$ variables and $N \in \{20,25,30,35,40\}$ subjects. We generate data with
$Q=2$ latent factors and $L=3$ FPCA components for each factor. As MEFISTO and MEFISTO-base do not explicitly model variable-specific time-varying mean functions, we simulate zero-mean variables. Compared to Sections~\ref{sec_sim_single} and \ref{sec_sim_anneal}, we consider a slightly sparser observation grid for each subject, i.e., with $n_i \sim \mathrm{Unif}\{2,10\}$, but a denser regime to simulate the factor loadings, i.e., in which each factor
influences all variables with equal probability, corresponding to a uniform (non-sparse)
loading structure. Nevertheless, to remain agnostic to the true loading structure, we apply our model using the default sparsity-inducing prior on $\omega_q$ given in~\eqref{eq_beta}, thereby deliberately assessing robustness under model misspecification. All methods are run using an over-specified model with $Q_\mathrm{max}=5$ latent factors. 


\begin{figure}[t!]
\centering
\includegraphics[scale=0.53]{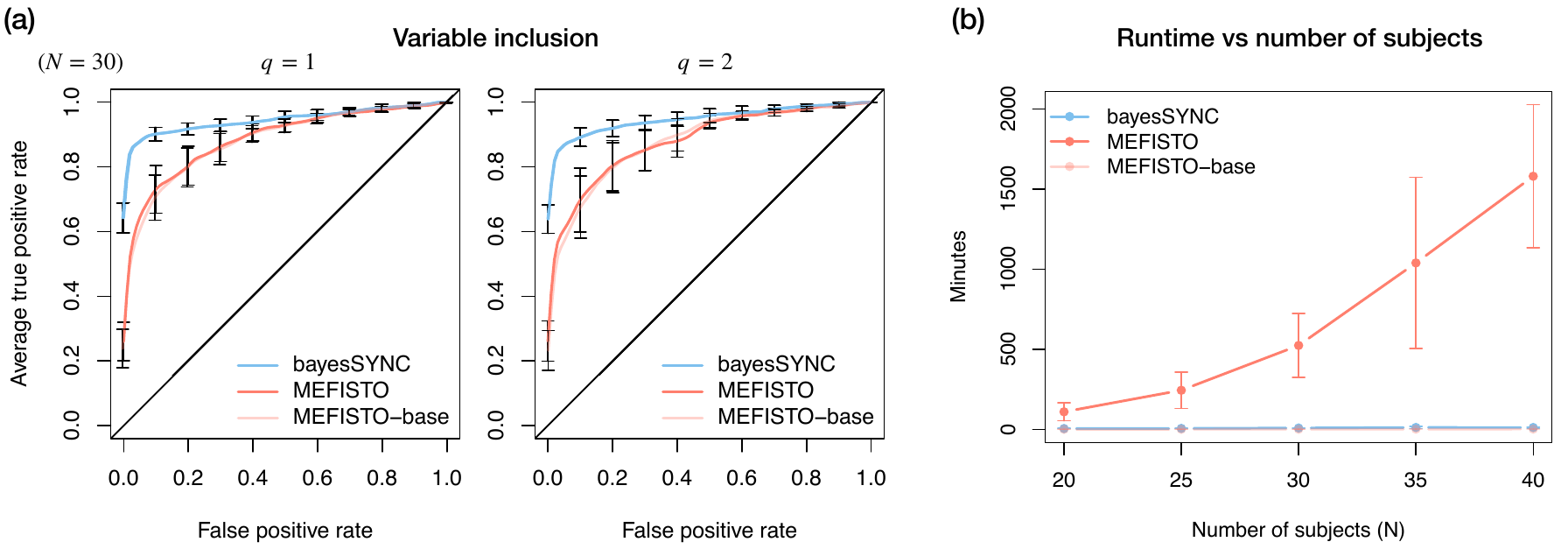}
\caption{\footnotesize Comparison of our method (bayesSYNC) and the Gaussian process-based method with or without subject-specific latent factor trajectories (MEFISTO and MEFISTO-base). 
(a)~Average ROC curves for recovery of loadings in the scenario with $N=30$ subjects, computed over 25 simulation replicates. The curves for MEFISTO and MEFISTO-base overlap. Error bars represent $\pm 1.96 \times$ standard error across replicates.
(b)~Runtime (minutes) as a function of sample size $N$, averaged over replicates on an Intel Xeon CPU, 2.60 GHz machine. Error bars denote $\pm$ one standard error, based on 25 replicates ($14$ replicates only for MEFISTO in the $N=40$-scenario). 
The runtimes displayed for bayesSYNC and MEFISTO-base overlap. 
}\label{fig5}
\footnotesize Alt text: Two panels. (a) Average ROC curves for the recovery of the loadings of the two factors at $N=30$, for bayesSYNC, MEFISTO and MEFISTO-base, the two MEFISTO curves overlapping. (b) Mean runtime in minutes against $N$ from 20 to 40, with standard-error bars; the runtime of MEFISTO increases steeply with $N$, whereas those of bayesSYNC and MEFISTO-base stay at a few minutes and overlap.
\end{figure}


\begin{table}[t!]
\fontsize{7}{9}\selectfont
\centering
\caption{\footnotesize
Comparison of our method (bayesSYNC) and the Gaussian process-based method with or without subject-specific latent factor trajectories (MEFISTO and MEFISTO-base) across sample sizes ($N=20,25,30,35,40$) under an overspecified model with $Q_{\max}=5$. Results are averaged over 25 simulation replicates (for MEFISTO, 14 only for $N = 40$ due to runs failing to complete within 36 hours); values are mean with standard error in parentheses. AUC refers to recovery of non-zero loadings based on posterior means of the loading entries. ISE denotes integrated squared error for reconstructed trajectories. Coverage and width refer to 95\% pointwise prediction bands. The number of factors corresponds to the average number of retained factors; the true number of factors is $Q = 2$. Best results for AUC, ISE, coverage and width are highlighted in bold.}
\label{tab:mefisto}
\begin{tabular}[t]{llccccc}
\toprule
\textbf{Metric} & \textbf{Method} & $\bm{N = 20}$ & $\bm{N = 25}$ & $\bm{N = 30}$ & $\bm{N = 35}$ & $\bm{N = 40}$\\
\midrule
\textbf{AUC $b_{jq}$} & bayesSYNC & \textbf{0.912 (0.010)} & \textbf{0.924 (0.009)} & \textbf{0.949 (0.006)} & \textbf{0.939 (0.008)} & \textbf{0.933 (0.011)}\\
 & MEFISTO & 0.881 (0.011) & 0.899 (0.011) & 0.886 (0.012) & 0.909 (0.009) & 0.925 (0.008)\\
& MEFISTO-base & 0.860 (0.013) & 0.889 (0.013) & 0.882 (0.011) & 0.908 (0.007) & 0.893 (0.010)\\
\addlinespace
\textbf{ISE $y_i^{(j)}$} & bayesSYNC & \textbf{0.398 (0.007)} & \textbf{0.361 (0.006)} & \textbf{0.275 (0.004)} & \textbf{0.293 (0.004)} & \textbf{0.300 (0.004)}\\
 & MEFISTO & 0.642 (0.012) & 0.642 (0.011) & 0.675 (0.011) & 0.772 (0.012) & 0.617 (0.012)\\
 &  MEFISTO-base & 1.264 (0.015) & 1.448 (0.015) & 1.409 (0.013) & 1.550 (0.014) & 1.505 (0.011)\\
\addlinespace
\textbf{Coverage (\%) $y_i^{(j)}$} & bayesSYNC & 95.9 (0.0) &  \textbf{95.8 (0.0)} & \textbf{95.7 (0.0)} & \textbf{95.6 (0.0)} & \textbf{95.6 (0.0)}\\
 & MEFISTO & 96.3 (0.0) & 96.4 (0.0) & 96.4 (0.0) & 96.3 (0.0) & 96.2 (0.0)\\
 & MEFISTO-base & \textbf{94.5 (0.1)} & \textbf{94.2 (0.0)} &  \textbf{94.3 (0.0)} & 94.1 (0.0) & 94.1 (0.0)\\
\addlinespace
\textbf{Width $y_i^{(j)}$} & bayesSYNC & \textbf{4.203 (0.004)} & \textbf{4.223 (0.003)} & \textbf{4.138 (0.002)} & \textbf{4.147 (0.002)} & \textbf{4.158 (0.002)}\\
 & MEFISTO & 4.391 (0.004) & 4.455 (0.004) & 4.460 (0.004) & 4.402 (0.003) & 4.352 (0.004)\\
 & MEFISTO-base & 5.293 (0.010) & 5.325 (0.008) & 5.439 (0.008) & 5.382 (0.007) & 5.397 (0.006)\\
\addlinespace
\textbf{Number factors} & bayesSYNC & 1.520 (0.154) & 1.680 (0.125) & 1.960 (0.091) & 1.880 (0.088) & 1.880 (0.088)\\
 & MEFISTO & 2.760 (0.119) & 2.320 (0.111) & 2.280 (0.108) & 2.360 (0.140) & 2.286 (0.125)\\
 & MEFISTO-base & 2.783 (0.108) & 2.333 (0.115) & 2.280 (0.108) & 2.348 (0.149) & 2.320 (0.095)\\
\bottomrule
\end{tabular}

\end{table}

Figure~\ref{fig5}(a) shows average ROC curves for loading recovery of the two factors for the scenario with $N=30$ subjects. The three methods achieve excellent recovery, yet with our approach achieving superior performance 
for both factors. This improvement likely reflects the use of a spike-and-slab prior, which induces element-wise sparsity in the loadings, whereas the ARD prior of the MEFISTO methods primarily shrinks entire factors and does not explicitly enforce selection at the level of individual loadings. Notably, the ROC curves for MEFISTO and MEFISTO-base essentially overlap, indicating that the accuracy of loading recovery is largely unaffected by whether latent trajectories are shared across subjects or estimated individually. 
Estimated subject-specific latent factor trajectories also suggest qualitative differences in the recovered temporal structure and subject-specific heterogeneity (Supplementary Material~S4). 

Table~\ref{tab:mefisto} reports averaged results across replicates across the grid of $N$. Besides the higher accuracy in recovering loadings which is confirmed for all values
of $N$, our approach also achieves a lower error in trajectory estimation (ISE) compared to MEFISTO, likely driven by the superior loading reconstruction. MEFISTO-base exhibits markedly larger reconstruction errors. As for MEFISTO, part of this difference stems from poorer loading recovery, but it is further amplified by the assumption of a single latent trajectory shared across all subjects. 

Coverage of the 95\% pointwise prediction bands is consistently close to
the nominal level for our method. Although variational approximations are often associated with underestimation of posterior uncertainty, these results indicate that the structured mean-field approximation, combined with the hierarchical separation of factor-level and FPCA-level variation, yields well-calibrated prediction bands in this setting.  MEFISTO achieves slightly conservative coverage. 
In contrast, MEFISTO-base tends to slightly underestimate coverage in some settings, although it remains competitive with our method for moderate sample sizes ($N=20,25,30$). However, this apparent competitiveness should be interpreted together with the width of the prediction bands. The prediction intervals produced by MEFISTO-base are substantially wider than those of both MEFISTO and our method. This inflation reflects the fact that a shared latent trajectory must absorb variability arising from heterogeneous subject-level dynamics, resulting in larger posterior uncertainty around the common trajectory. 

Regarding model selection, our method recovers the true number of latent
factors ($Q=2$) in most settings, although a slight tendency to underestimate
$Q$ is observed for smaller sample sizes.
In contrast, both MEFISTO and MEFISTO-base more frequently overestimate the number of factors
in this regime. This behaviour is consistent with ARD-based shrinkage, which can attenuate factor magnitudes without fully eliminating redundant components, whereas the combination of spike-and-slab regularisation and annealed inference in our approach more decisively suppresses weak or redundant factors.

Runtime comparisons  are shown in Figure~\ref{fig5}(b). Although our \texttt{R} package implements our method with full support for parallel execution, we report serial runtimes here to allow a fair comparison with the MEFISTO implementations, which do not leverage parallelisation.  Our approach
scales approximately linearly with $N$, with some additional overhead from nested updates over latent factors and FPCA components. MEFISTO-base is substantially faster than MEFISTO because it estimates a single Gaussian process trajectory per factor rather than one trajectory per subject. Its runtime is therefore of the same order of magnitude as our method (a few minutes) and in fact marginally shorter in these experiments (see also Supplementary Material~S5). Again, this should also be weighed against the difference in model scope: MEFISTO-base provides no representation of subject-level temporal variation in the latent space (Supplementary Material~S4), whereas our method estimates it and further decomposes its variability through FPCA components.

In contrast, for MEFISTO, the use of Gaussian process priors over each subject-specific latent factor trajectories requires modelling all observed time points across all subjects jointly within a single covariance matrix per factor. As the number of subjects increases, these covariance matrices become larger and the associated matrix operations (such as decompositions and hyperparameter optimisation) become increasingly expensive in terms of both computation and memory. Because observation grids are irregular and differ across subjects, the covariance matrix cannot be decomposed into smaller independent blocks and the full joint structure must be handled during inference.  Notably, for $N=40$, 11 out of 25 MEFISTO runs did not complete within the 36-hour time limit, and this cost is also why the comparison is limited to 25 replicates per setting. This computational burden substantially restricts its applicability at sample sizes typical of contemporary molecular studies, where cohorts commonly exceed this range.

\subsection{Robustness under model misspecification}\label{sec_misspec}

The scenarios of Section~\ref{sec_sim_data} generate data from
the model that is subsequently fitted. We now depart from it in four ways, keeping every
other aspect of the design fixed and pairing the replicates across settings, so that each
comparison is made on the same simulated loadings, scores, eigenfunctions, mean functions and
observation times, and only the departure differs. Three departures concern the structure of
the model. The variable-specific error is given an exponential correlation along $t$ with
range $\rho$, its variance being unchanged; a smooth component that the truncation at $L$
cannot represent is added to the latent trajectories, its variance being $\kappa$ times the
variance of the trajectories; and the scores of matching components of different factors are given correlation
$\varrho$, each factor keeping its own eigenfunctions. The fourth departure concerns the
observation process. Under informative drop-out, the probability that a subject leaves the
study at a given visit increases with the current value of the first latent process, and a
non-informative arm losing the same expected proportion of observations serves as a
control. All are fitted with the model as specified, that is with independent
errors, uncorrelated latent processes and no drop-out mechanism. We use $N = 100$ subjects,
$p = 1\,000$ variables and 50 replicates per setting, with $Q_{\max} = 5$.

We report the number of factors retained separately from the accuracy of the retained factors.
The well-specified baseline misses one of the three true factors in 12\% of replicates, and
each missed factor lowers its average accuracy. A setting that retains more factors than there
are true ones rarely misses a factor, so averaging over all replicates would make it look more
accurate than the baseline. The accuracy
columns of Table~\ref{tab_misspec} are therefore restricted to the replicates in which every
true factor was recovered.

Recovery of the loading matrix is insensitive to all four departures, the area under the
curve remaining between 0.962 and 0.981 against 0.976 in the baseline, so that the true
factors are found with the right variables under every departure.

\begin{table}[t!]
\centering
\caption{\footnotesize
Robustness under model misspecification, over 50 paired replicates with $N = 100$,
$p = 1000$, three true factors and $Q_{\max} = 5$. Every setting is generated from the same
simulated quantities as the baseline within a replicate, and fitted with the model as
specified. ``Retained'' is the mean number of factors retained and ``All found'' the
percentage of replicates in which each true factor was matched to a retained estimated factor. The accuracy columns
are restricted to those replicates, so that settings retaining different numbers of factors
are compared like for like; means are given with Monte Carlo standard errors over replicates.
AUC refers to recovery of the nonzero loadings, ISE to the integrated squared error of the
eigenfunctions, and ``Score corr.'' to the correlation between estimated and simulated subject
scores.}
\label{tab_misspec}
\scriptsize
\begin{tabular}{llrrrrr}
\toprule
 & & \multicolumn{2}{c}{Factors} & \multicolumn{3}{c}{Accuracy, runs recovering all factors} \\
\cmidrule(lr){3-4} \cmidrule(lr){5-7}
Mechanism & Setting & Retained & All found & AUC & Eigenfn.\ ISE & Score corr. \\
\midrule
Baseline & - & 2.90 & 88\% & 0.976 (0.002) & 0.007 (0.001) & 0.976 (0.001) \\
\addlinespace
Correlated& $\rho = 0.15$ & 4.90 & 100\% & 0.969 (0.002) & 0.008 (0.001) & 0.972 (0.002) \\
errors  & $\rho = 0.40$ & 5.00 & 100\% & 0.962 (0.004) & 0.065 (0.029) & 0.952 (0.011) \\
\addlinespace
Unmodelled & $\kappa = 0.1$ & 2.84 & 90\% & 0.980 (0.001) & 0.017 (0.001) & 0.909 (0.003) \\
  latent variation& $\kappa = 0.3$ & 2.98 & 94\% & 0.981 (0.001) & 0.050 (0.002) & 0.812 (0.005) \\
\addlinespace
Correlated  & $\varrho = 0.3$ & 2.92 & 88\% & 0.976 (0.002) & 0.008 (0.001) & 0.973 (0.003) \\
 factors& $\varrho = 0.6$ & 2.92 & 86\% & 0.972 (0.003) & 0.015 (0.008) & 0.973 (0.003) \\
\addlinespace
Drop-out & Non-informative & 3.04 & 82\% & 0.964 (0.004) & 0.033 (0.003) & 0.814 (0.006) \\
 & Informative & 2.96 & 80\% & 0.963 (0.003) & 0.044 (0.006) & 0.793 (0.006) \\
\bottomrule
\end{tabular}

\end{table}

The departures differ in their consequences. A variable-specific error correlated along $t$
leaves the estimates themselves largely intact but inflates the number of retained factors,
from 2.90 in the baseline to 4.90 and 5.00, the latter meaning that every replicate reached
$Q_{\max}$. The estimated residual variance does not absorb the departure, falling slightly
from 1.009 to 0.974 against a true value of one, so the unmodelled temporal structure is
taken up by additional factors rather than by the error term. Unmodelled latent variation mainly
affects the scores, the correlation of the estimated scores with the true ones falling from
0.976 to 0.909 and 0.812 as $\kappa$ increases. Correlation between the latent processes has almost no effect, the
retained number, the loadings and the scores all remaining at their
baseline values, so that in this design the assumption of uncorrelated latent processes
costs little, a point we return to in Section~\ref{sec_dis}.

Under drop-out, the mean number of observations per subject falls from 7.5 to 4.5 in both
arms. The two arms keep the same proportion of
observations on average, about 60\%, and differ only in whether who leaves depends on the
latent trajectory, so the difference between them measures the effect of informativeness
alone. Informativeness leaves the loadings and the number of retained factors unchanged, the
paired differences over all 50 replicates, $-0.003$ and $-0.08$, being smaller than their Monte Carlo standard errors of
0.003 and 0.09. It does attenuate the subject-level scores, the slope of the regression of the
estimated on the true scores falling by 0.043 (standard error 0.010) and their correlation with
the truth by 0.029 (0.009). Subjects with more severe trajectories leave the
study earlier, so that their scores rest on fewer observations and are shrunk further towards
the mean.

Finally, the correlation between residuals at pairs of observation times, binned by the lag
between them, separates the settings well. It stays within 0.015 of zero in the baseline and
under every departure other than the correlated error, whereas under correlated errors it is
positive and decays with the lag, from 0.81 to 0.42 at $\rho = 0.15$ and from 0.90 to 0.71 at
$\rho = 0.40$. The
assumption, and with it the main source of additional factors, can therefore be checked on
real data, and we report this diagnostic for the
application in Supplementary Material~S6.

\section{Application to longitudinal COVID-19 molecular data}
\label{sec_application}

We now return to the longitudinal COVID-19 dataset
described in Section~\ref{sec_motivation},  with the aim of identifying latent molecular processes shaping inter-individual heterogeneity in disease severity and recovery. 

We report two analyses of this cohort. The first uses the
panel of 57 circulating biomarkers and illustrates the model at a scale where every factor and
every fitted trajectory can be examined individually, and it is the panel on which we compare our
results with MEFISTO. The second uses whole-blood transcriptomic data on more than thirteen
thousand genes, and shows the method operating at the scale for which it was designed.

\subsection{Biomarker panel analysis}\label{sec_metabo}
We apply our approach to data from symptomatic PCR-positive patients (clinical severity classes B to E) during the first
seven weeks after symptom onset.
The panel comprises 57 circulating biomarkers, including C-reactive protein (CRP), complement
components, cytokines, immune cell populations, glycoproteins and
metabolic markers, measured at irregular and patient-specific
time points.
The analysis includes the 88 patients for whom all markers
were measured within this window, giving 136 samples, with one to four samples per patient
(median one).

We estimate model parameters using annealed variational inference as
described in Section~\ref{sec_inference}, with $Q_{\mathrm{max}}=5$
candidate latent factors and up to $L_{\mathrm{max}}=5$ candidate functional principal
components for each factor process. 
We retain factors whose posterior probability of inclusion exceeds $0.5$,
as described in Section~\ref{sec_Q}.
Under this criterion, two latent factors are selected.
Markers are likewise assigned to a factor when their
posterior inclusion probability exceeds 0.5. The choice of threshold has little effect
here, since most of these probabilities are below 0.1 or above 0.9, and the sets of markers
selected at thresholds between 0.25 and 0.95 overlap with those at 0.5 with a Jaccard index of
at least 0.91.

\begin{figure}[t!]
\centering
\includegraphics[scale=0.54]{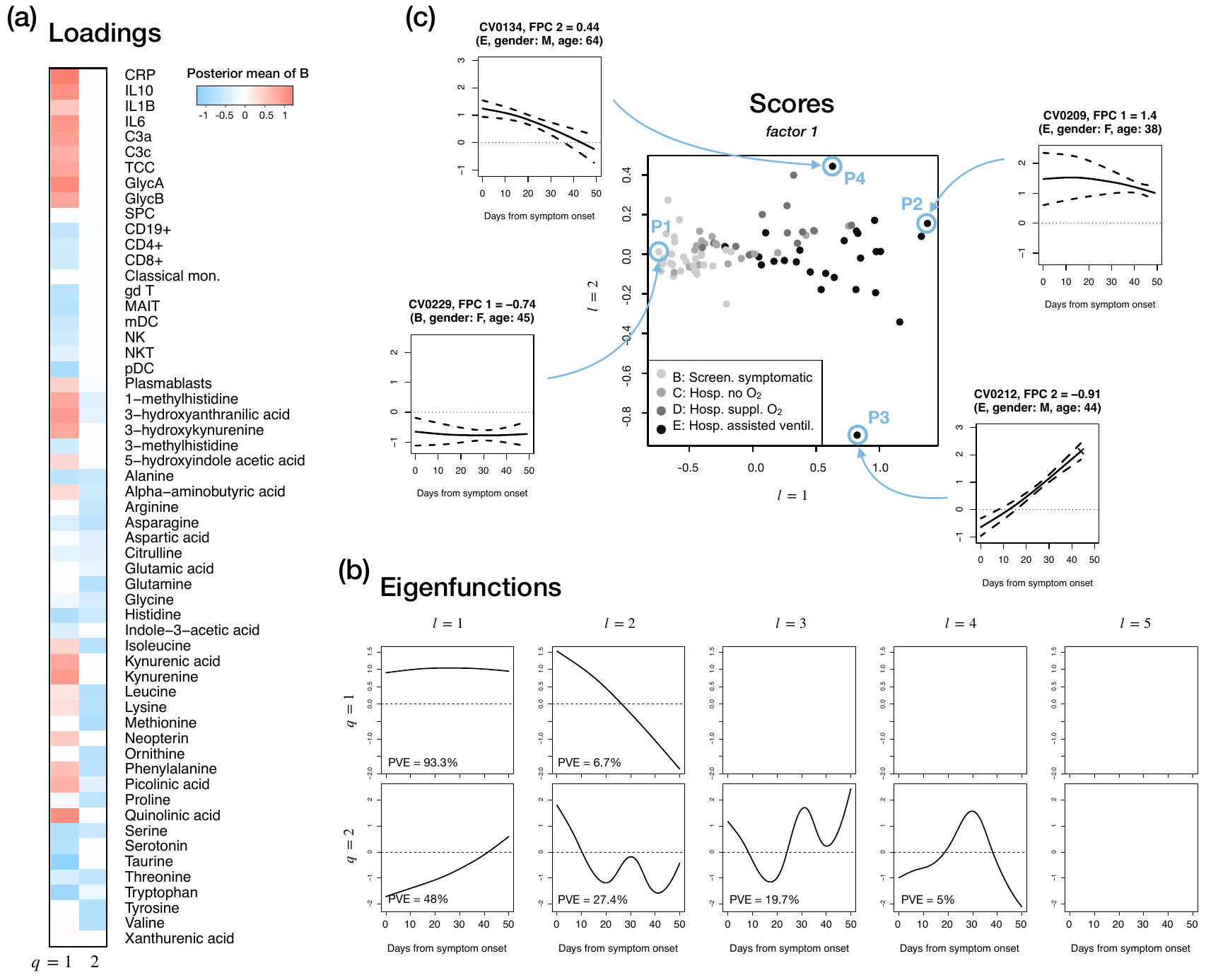}
\caption{\footnotesize
Application to the longitudinal COVID-19 cohort.
(a)~Estimated loading matrix for the two retained latent factors ($Q_{\max}=5$; factor inclusion threshold $>0.5$). Rows correspond to biomarkers and columns to factors.
(b)~Estimated eigenfunctions for each factor process, with the proportion of variance explained indicated for successive components.
(c)~Subject-level FPCA scores for the first factor, coloured by clinical severity group, along with reconstructed factor trajectories, along with $95\%$ pointwise credible bands, for the four patients (P1--P4) with most extreme scores.}
\label{fig6}

\footnotesize Alt text: Three panels for the biomarker panel analysis. (a) Heat map of the estimated loadings of the 57 markers on the two retained factors, blue for negative and red for positive values. (b) Estimated eigenfunctions of each factor, labelled with the proportion of variance each explains. (c) Scatter plot of the first two FPCA scores of the first factor, shaded by severity class, with the reconstructed factor trajectories and 95\% credible bands of four patients with extreme scores.
\end{figure}

Figure~\ref{fig6}(a) displays the estimated loading matrix for the two
retained factors. The inferred structure is moderately sparse and reveals two
distinct biological signatures. The first factor is characterised by strong positive loadings for classical
inflammatory markers, including CRP, several cytokines,
complement components and the glycoproteins GlycA and GlycB. In addition,
multiple metabolites from the kynurenine pathway, such as kynurenine,
quinolinic acid and 3-hydroxykynurenine, also contribute positively, whereas
tryptophan and serotonin load with opposite sign. This configuration is
biologically coherent, as pro-inflammatory cytokines activate indoleamine
2,3-dioxygenase (IDO), which promotes tryptophan degradation through the
kynurenine pathway and reduces serotonin synthesis. We then applied a gene set enrichment analysis (GSEA) to the ranked  estimated loadings using Kyoto Encyclopedia of Genes and Genomes (KEGG) pathways \citep{korotkevich2016fast}, which confirmed enrichment of \textit{Tryptophan metabolism} at false discovery rate (FDR) $<20\%$ 
($p=0.013$; FDR $=0.17$).  These findings support a hypothesis we advanced in previous applied work, namely that a coordinated upregulation of inflammatory signalling and kynurenine pathway activation represents a shared mechanistic axis. This interpretation is biologically plausible given evidence that persistent cytokine-driven IDO activation and downstream neuroactive kynurenine metabolites are implicated in the neurological sequelae observed in long COVID \citep{ruffieux2023patient}.  

\begin{figure}[t!]
\centering
\includegraphics[scale=0.56]{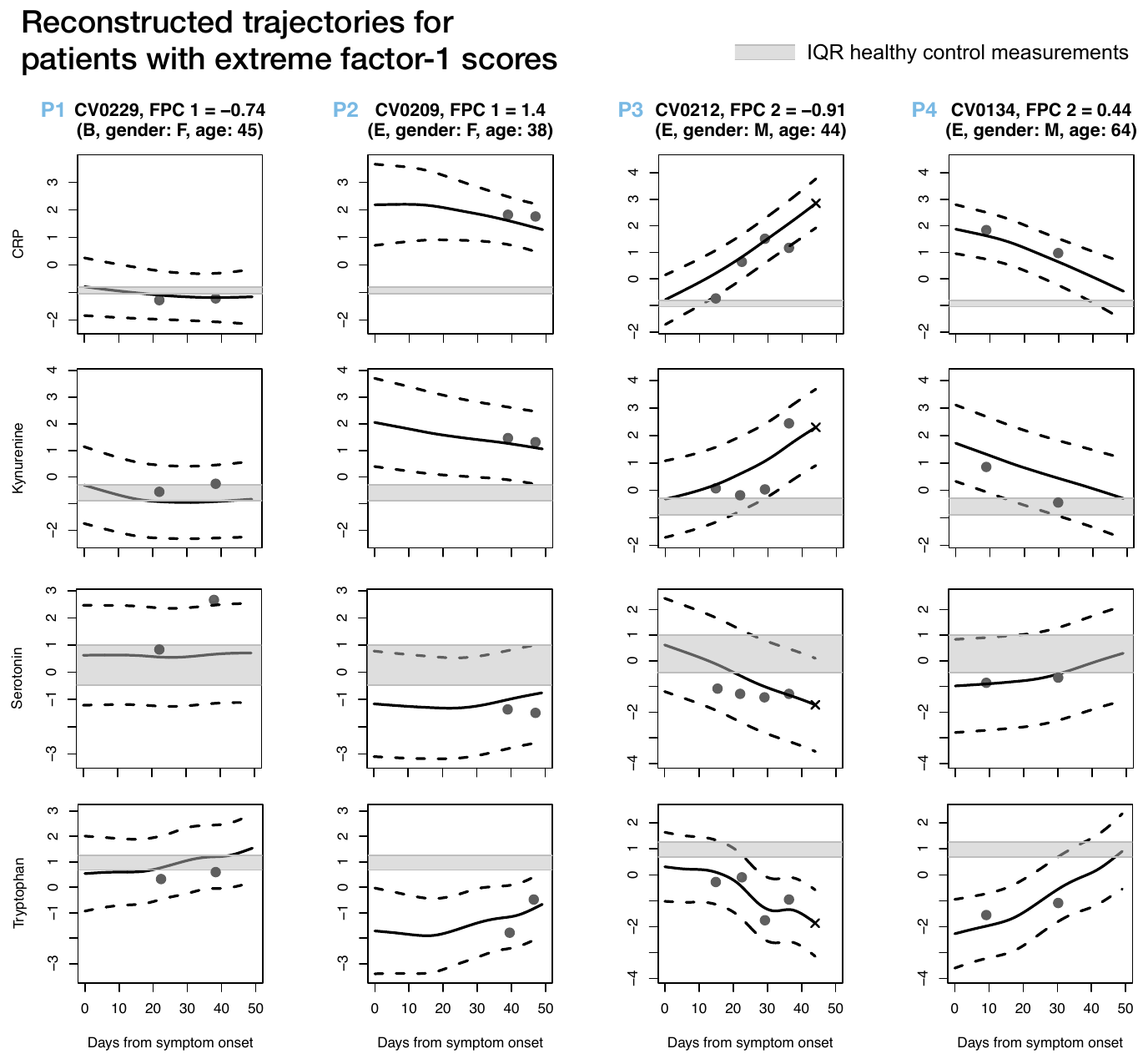}
\caption{\footnotesize
Reconstructed trajectories with corresponding $95\%$ pointwise prediction bands for selected markers representative of the first factor signature. Each row corresponds to one marker (CRP, kynurenine, serotonin, tryptophan), while columns show the four patients (P1–P4) with the most extreme scores for the first or second eigenfunctions (see Figure~\ref{fig6}). From left to right, the columns represent patients with low and high severity scores, followed by patients with low and high recovery scores. Patient P3, with worse recovery scores, died before the end of the 7-week window (black cross). The horizontal grey band corresponds to the interquartile range of the markers in healthy control measurements. }
\label{fig7}

\footnotesize Alt text: Grid of line plots with four rows, for CRP, kynurenine, serotonin and tryptophan, and four columns, for patients P1 to P4. Each plot shows the observed values of the patient, the reconstructed trajectory with its 95\% prediction band and a horizontal grey band for the interquartile range in healthy controls; a black cross marks the death of patient P3.
\end{figure}

The second factor is dominated by metabolites enriched for amino-acid-related
pathways, including branched-chain amino acids such as leucine, isoleucine and
valine, which load with the same sign, consistent with previously reported depletion of circulating amino acids in severe COVID-19 and related inflammatory states \citep{atila2021serum, masoodi2022disturbed}. GSEA also demonstrated 
enrichment for \textit{Biosynthesis of amino acids} (FDR
$=1.10\times10^{-3}$), \textit{Aminoacyl-tRNA biosynthesis} (FDR
$=0.042$) and \textit{2-Oxocarboxylic acid
metabolism} (FDR $=0.062$). Complementary hypergeometric over-representation analysis (based on markers with PPI $>0.5$ contributing to the factor signature) confirmed
enrichment of the same pathways (FDR 
between $4.80\times10^{-5}$ and $0.027$).

The eigenfunctions are estimated nonparametrically,
on a penalised spline basis, so that simple shapes, such as those found below for the first
factor, are a finding about the data rather than a constraint of the model.
Figure~\ref{fig6}(b) indicates that, for the first factor, the first two eigenfunctions explain more than
99\% of the variability, while for the second factor, four eigenfunctions are required to capture the same amount of variation. In the following, we focus on interpretation of the first factor, whose temporal structure and clinical associations admit a direct biological interpretation. The eigenfunctions associated with the second factor also suggest structured temporal heterogeneity, although their interpretation is less immediate and would benefit from further targeted biological investigation. 

The first eigenfunction of the first factor may be understood as a proxy for disease severity. Indeed, it  is positive over the entire time
window, implying that 
subjects with positive scores have an overall upward shift
of the factor trajectory $h^{(1)}$, corresponding to sustained elevation
of inflammatory activity (given the positive inflammation marker loadings). 
As shown in Figure~\ref{fig6}(c), the corresponding scores display a clear gradient across clinical severity
classes and this association is significant (anova $p < 0.001$), which further supports the interpretation of this component as a severity axis. The four
patients highlighted in Figure~\ref{fig6}(c) have two to four samples, against a median of one.
As expected under partial pooling, the scores of a patient with few samples are shrunk towards
zero, which prevents a single sample from producing an extreme score, and the absolute scores
accordingly increase with the number of samples (rank correlations between 0.19 and 0.31 over the
first two components of the two factors). Since the number of samples also differs between
severity classes, the mildest class having the most, we checked that the association with
severity does not arise from this. The rank correlation of this component with severity is
essentially unchanged after adjusting for the number of samples (0.83, against 0.84 without
adjustment). 
The second eigenfunction changes sign over time, so that subjects with large positive scores have
factor trajectories that return more rapidly towards baseline. 
This second component explains 6.7\% of the variation,
and it rests to a large extent on the five patients who died within the window, whose
follow-up was cut short by their death. They include P3, the patient with the lowest score,
and excluding them reduces the variance of the scores on this component by 41\%. In a refit
without them, the component explains 1\% of the variation of the factor, whereas the first
component is essentially unchanged (Supplementary Material~S7). We therefore regard the
evidence for differences in recovery as weak.

The model enables reconstruction of individual marker trajectories through
the combination of estimated loadings and factor-level dynamics.
Figure~\ref{fig7} shows these trajectories for a
subset of representative markers in patients with extreme scores for
factor $q=1$.
For CRP and kynurenine, patients with high severity scores display
persistently elevated concentrations relative to controls.
In contrast, tryptophan and serotonin are reduced in severe cases,
consistent with the inferred loading signs and known inflammatory
metabolic shifts.
 Importantly, these marker-level reconstructions are mirrored at the latent level: the factor trajectories for the same individuals (Figure~\ref{fig6}(c)) show persistently elevated activity for high-severity patients. The columns for the
recovery scores are shown for completeness, and for the reasons given above we draw no
conclusion about recovery from them. 

We also fitted MEFISTO-base to these data. It recovers
the same two axes, but separates them less cleanly, spreading the inflammatory markers and the
amino acids across both of its leading factors, and its patient-level factor values are less
strongly associated with the severity classes than ours. Supplementary Material~S8 reports the
comparison.

\subsection{Transcriptomic analysis}\label{sec_rnaseq}

We now turn to whole-blood transcriptomic data from the same
cohort, which let us examine the behaviour of the model at a scale close to that of the
simulations. After retaining genes detected in at least 20\% of samples, the panel comprises
13\,177 genes measured in 164 patients over 287 samples, with one to four samples per patient
(median two). We set $Q_{\mathrm{max}} = 10$ and $L_{\mathrm{max}} = 5$.

All ten factors are retained (Table~\ref{tab_rnaseq}). The factor most associated with severity
class (rank correlation $0.65$, Kruskal--Wallis $p = 4\times10^{-16}$, and $0.54$ with peak CRP)
contrasts neutrophils with lymphocytes. All twelve neutrophil marker genes we examined load
positively on it and all fourteen T- and B-cell markers negatively, so it captures the balance
between the two cell types, whose shift towards neutrophils is a recognised marker of severe
COVID-19 \citep{li2020predictive}. A change in cell composition of this kind moves most of the
transcriptome, so the factor selects 87\% of the genes, and since the hallmark collection holds no
cell-type sets it returns no enrichment. Of the 88 patients of the biomarker panel analysis of
Section~\ref{sec_metabo}, 83 are also in the transcriptomic data, and over these patients this
factor and the severity scores of the panel analysis are strongly correlated ($\rho = 0.73$), so
that two analyses of different molecular measurements order the patients in
much the same way. The agreement goes mostly through severity class, within which the correlation
is 0.21 (Supplementary Material~S9).

A second factor carries the type-I and type-II interferon response, with all ten canonical
interferon-stimulated genes among its 200 largest loadings (STAT2, RSAD2, IFIH1, IRF7; false
discovery rate $2.5\times10^{-9}$ for both hallmark sets), and a third an erythroid signature
(heme metabolism, $5.0\times10^{-9}$). Two others isolate technical variation, one loading on 5S
ribosomal RNA and the other on ribosomal pseudogenes and other short transcripts, both associated
with the extraction batch ($p = 6\times10^{-18}$ and $3\times10^{-9}$). Two further factors are
enriched for MYC targets and oxidative phosphorylation, broad sets that we do not interpret
further, and the remaining three have cell-cycle sets (factors 5 and 8) or no enrichment (factor 4). Supplementary Material~S9 also
reports a sensitivity analysis excluding the patients who died within the window, and
shows the loadings, eigenfunctions and scores of the neutrophil and
lymphocyte, interferon and erythroid factors.

\begin{table}[t!]
\centering
\caption{\footnotesize The ten factors of the transcriptomic analysis. Sev.\ $\rho$
is the rank correlation of the patient-level factor value, the mean of the estimated trajectory at
that patient's own sampling times, with the severity class, each factor being oriented so that
this value increases with severity. The hallmark column gives the most significant set by gene set
enrichment analysis of the ranked loadings, with its false discovery rate. The last three columns
describe the temporal decomposition, namely the share of the variation of the factor explained by
its leading component, the number of times that component changes sign between days 5 and 45, and
the number of components needed for 99\% of the variation. Shares are normalised within the
$L_{\mathrm{max}} = 5$ components fitted. A dagger marks a factor flagged as technical, by an
extreme transcript length of its strongest loadings or by association with the extraction batch.
Factors are labelled where their leading genes or their enrichment point to a specific cell
type or process.
Key genes for each factor are given in Supplementary Material~S9.}
\label{tab_rnaseq}
\footnotesize

\begin{tabular}{clrp{3.1cm}rrr}
\toprule
 & & Sev. & & \multicolumn{2}{c}{Leading comp.} & Comp. \\
\cmidrule(lr){5-6}
 & Label & $\rho$ & Leading hallmark set (FDR) & PVE (\%) & sign ch. & for 99\% \\
\midrule
1 & Neutrophil/lymphocyte & 0.65 & allograft rejection ($4.1\times10^{-2}$) & 61 & 0 & 4 \\
2 & --- & 0.41 & MYC targets V1 ($5.7\times10^{-7}$) & 53 & 3 & 3 \\
3 & 5S rRNA, pseudogenes$^{\dagger}$ & 0.08 & none & 65 & 0 & 4 \\
4 & --- & 0.02 & none & 53 & 1 & 4 \\
5 & --- & 0.29 & G2M checkpoint ($3.2\times10^{-4}$) & 55 & 0 & 4 \\
6 & Interferon & 0.16 & interferon alpha response ($2.5\times10^{-9}$) & 80 & 3 & 4 \\
7 & --- & 0.05 & MYC targets V1 ($7.9\times10^{-7}$) & 70 & 1 & 5 \\
8 & --- & 0.20 & E2F targets ($6.2\times10^{-3}$) & 58 & 2 & 4 \\
9 & Short transcripts, pseudogenes$^{\dagger}$ & 0.06 & oxidative phosphorylation ($5.9\times10^{-4}$) & 64 & 0 & 4 \\
10 & Erythroid & 0.08 & heme metabolism ($5.0\times10^{-9}$) & 50 & 1 & 4 \\
\bottomrule
\end{tabular}

\end{table}

The model describes the temporal structure of these factors at two levels. The estimated mean
functions carry the course of the response shared across the cohort. Since
patients were sampled at different times after symptom onset, an analysis that ignored time would
confound the stage of the response at which a patient was sampled with how that patient differs
from others, and separating the shared course from the departures from it avoids this. Averaged over the 50 genes
with the largest loadings, they fall by 1.7 standard deviations over the seven weeks for the
interferon factor and rise by 1.6 for the erythroid factor, which agrees with the decline of the
interferon signature and the late erythroid signal described in longitudinal blood transcriptomic
data \citep{bernardes2020longitudinal}. The functional principal components then show when
patients depart from this shared course, and the three factors differ in this respect. For the
neutrophil and lymphocyte factor, the leading component, which explains 61\% of the variation of
the factor, keeps one sign between days 5 and 45, so that patients differ by a shift in the
balance between the two cell types that is sustained over the seven weeks. Its scores separate
the severity classes (rank correlation $0.58$), and more clearly among the 71 patients whose
samples span at least two weeks, whose scores rest on more data ($0.74$). As for the inflammatory
factor of the panel analysis, the component associated with severity is therefore a sustained
shift rather than a difference in the time course. For the erythroid factor, the leading component
places nine tenths of its weight after day 10, so that patients differ mainly in how far erythroid
expression rises from the second week onwards, and this is not associated with severity. For the
interferon factor, three quarters of the weight of the leading component lies before day 10, and
most of it in the first five days, in which only 22 patients were sampled. Its
scores are weakly associated with severity (rank correlation $-0.27$), and, since the interferon
genes load negatively on the factor, they place the more severe patients above the shared course
between about days 6 and 22 and below it between days 22 and 36. The association does not
strengthen among the 92 patients with two or more samples ($-0.20$), so we checked it on the
expression scale, among patients whose own samples span day 10 and a later day, so that no
trajectory is extrapolated. There, the more severe patients have higher interferon expression
at day 10 and lose more of it over the following weeks, the rank correlation between the change and
severity class lying between $-0.46$ and $-0.60$ for end points between days 25 and 35 (23 to 40
patients; Supplementary Material~S9). When the severity class is
regressed on the leading scores of all ten factors together, in a proportional-odds model, the
scores of the neutrophil and lymphocyte factor, of the interferon factor and of factors 5 and 8,
whose leading hallmark sets relate to the cell cycle, are each associated with severity, so that
these factors carry distinct information about it (Supplementary Material~S9).

Without supervision, the model thus recovers from 13\,177 genes a shift from lymphocytes towards
neutrophils that follows severity and persists over the seven weeks, the shared resolution of the
interferon response, whose level at day 10 is higher in the more severe patients on the
expression scale, and a rise of erythroid
expression whose extent varies between patients independently of severity. It also keeps
technical variation apart from these signals. Unlike in the panel analysis, selection is not
sparse, each factor retaining between 20\% and 87\% of the genes, and no factor is pruned, two
features to which we return in Section~\ref{sec_dis}.

\section{Discussion}\label{sec_dis}

We have proposed a Bayesian functional factor model for high-dimensional longitudinal data that jointly represents dependence across variables and temporal variability within latent processes.
By combining latent factor modelling and FPCA within a coherent hierarchical framework, the approach yields an interpretable decomposition of multivariate trajectories into latent 
processes and their subject-level temporal heterogeneity.

 Notably, our numerical experiments indicated that annealing is not merely a computational refinement but a 
core component of the inference scheme that leads to more stable factor selection, improved reconstruction accuracy and systematically higher ELBO values compared with vanilla variational inference. Comparisons with MEFISTO \citep{velten2022identifying} further highlighted methodological and practical distinctions from Gaussian process–based factor models, in simulation and on the COVID-19 data (Supplementary Material~S8). Beyond its lack of an explicit FPCA-based decomposition of subject-level temporal variation, MEFISTO suffered from  substantial scalability limitations, with runtimes and memory requirements increasing prohibitively at sample sizes typical of modern longitudinal molecular studies.

Application to the COVID-19 study 
uncovered coordinated inflammatory programs, including cytokine and kynurenine-related pathways consistent with immune activation, and further identified a new axis of heterogeneity in disease severity, with weaker evidence of differences in recovery, quantified through subject-level FPCA scores. These scores provide low-dimensional scalar summaries of longitudinal behaviour that can be readily incorporated into downstream clinical or biological analyses. In
whole-blood transcriptomic data from the same cohort, the model recovered a contrast between
neutrophils and lymphocytes that follows severity and the resolution of the interferon response
over the seven weeks, and kept technical variation apart from both.
More broadly, the methodology is well suited to longitudinal studies involving large panels of biomarkers where the objective is to characterise how groups of biomarkers evolve jointly over time and how individuals differ in these dynamic patterns. 

The transcriptomic analysis of Section~\ref{sec_rnaseq}
shows how the sparsity mechanism behaves at this scale. The spike-and-slab prior selects between a
fifth and nearly nine tenths of the genes on each factor, far from the sparse loadings recovered on
the 57-marker panel, which we attribute to the coordinated shifts in cell composition that make
large numbers of genes covary in whole blood. No factor is pruned either, since the posterior
factor inclusion probability of Section~\ref{sec_Q} saturates once a few thousand genes load on a
factor. At this scale $Q_{\mathrm{max}}$ is therefore best read as a lower bound on the number of
factors the data support, and its choice is guided by the interpretability of the factors
recovered.

The assumption that the latent processes are
uncorrelated is made in the prior, so the estimated scores of different factors can still be
correlated where the data support it. The assumption costs little when
the correlated processes are linear combinations of uncorrelated ones, which the model represents
through different, and in general less sparse, loadings. The prior cannot represent a process that follows
another with a delay, because a delay makes the dependence between two variables asymmetric in
time, which no sum of factors can reproduce, although the estimated scores can carry such a
dependence, and in the simulations of Section~\ref{sec_misspec} correlated latent processes with
different temporal shapes left the estimates intact. Differences between subjects in the timing of a shared
response are a separate matter, since the model does not align the trajectories of different
subjects in time. It represents such differences as differences in amplitude, through
additional components or factors, which may then be read as distinct modes of variation.
Anchoring time at symptom onset limits such differences in the COVID-19 application without
excluding them. Aligning the trajectories in time within the model, known as registration
\citep{marron15}, is a natural extension.

Several extensions are of interest. Covariates could be incorporated directly into the factor score distributions to formally link latent trajectories to clinical outcomes. Likewise, a joint model for the trajectories and the time to
death would account for the informative truncation of the trajectories of patients who die
during follow-up, although the present data would give it little to work with, since 5 of the 88
patients of the panel analysis and 15 of the 164 patients of the transcriptomic analysis died
within the window. We examine the consequences of this truncation in simulation (Section~\ref{sec_misspec}) and
in the application (Supplementary Material~S7 and~S9). Biologically informed structured priors may also further improve interpretability in specific applications. From a methodological perspective, theoretical investigation of posterior behaviour in high-dimensional functional settings remains an important direction for future work.

 \section*{Software availability}

\noindent The R package \texttt{bayesSYNC} is available at \url{https://github.com/hruffieux/bayesSYNC}. The code reproducing the simulations and the analyses of this paper, and the code generating the simulated data that emulate the COVID-19 data, are available at \url{https://github.com/hruffieux/bayesSYNC_paper_code} and archived at \url{https://doi.org/10.5281/zenodo.23676333} \citep{jaoua2026code}.

 \section*{Data availability}
\noindent 
The cytokine, complement, flow cytometry and clinical data of the NIHR CITIID COVID-19 cohort, and whole-blood RNA-seq data for a subset of 183 samples, are publicly available at \url{https://www.covid19cellatlas.org/patient/citiid/} \citep{bergamaschi2021longitudinal}. The metabolomic and glycoprotein measurements, and the full longitudinal RNA-seq data analysed here, are not publicly available; simulated data emulating both analyses, with the code to generate them, are provided in the same repository.

 \section*{Ethics approval}
\noindent Ethical approval for the CITIID-NIHR BioResource COVID-19 study was obtained from the East of England-Cambridge Central Research Ethics Committee (``NIHR BioResource'', REC ref.\ 17/EE/0025, and ``Genetic variation AND Altered Leucocyte Function in health and disease - GANDALF'', REC ref.\ 08/H0308/176), and all participants provided written informed consent \citep{bergamaschi2021longitudinal}.

 \section*{Supplementary Material}
 The Supplementary Material, after the references, contains the full derivation of the annealed variational inference algorithm and additional results from the simulation studies and the COVID-19 application.

 \section*{Funding}
This work was supported by the Lopez–Loreta Foundation (D.T. and H.R.).

 \section*{Acknowledgements}

\noindent We thank Xiaoyue Xi and Aimee Hanson for their guidance on aspects of the COVID-19 RNA-seq dataset. We thank Anthony Davison for the helpful discussion on the methodology. 
We thank NIHR BioResource volunteers for their participation, and gratefully acknowledge NIHR BioResource centres, NHS Trusts and staff for their contribution. We thank the National Institute for Health and Care Research, NHS Blood and Transplant, and Health Data Research UK as part of the Digital Innovation Hub Programme. The views expressed are those of the author(s) and not necessarily those of the NHS, the NIHR or the Department of Health and Social Care. For the purpose of open access, the author(s) has applied a Creative Commons attribution (CC BY) licence to any Author Accepted Manuscript version arising.

 \section*{Conflict of interest}

The authors declare that they have no conflict of interest.

\bibliographystyle{plainnat}
\bibliography{references}

@article{fruhwirth2023generalized,
  title={{Generalized cumulative shrinkage process priors with applications to sparse Bayesian factor analysis}},
  author={Fr{\"u}hwirth-Schnatter, S.},
  journal={Philosophical Transactions of the Royal Society A},
  volume={381},
  number={2247},
  pages={20220148},
  year={2023},
  publisher={The Royal Society}
}

@article{inamo2025comparative,
  title={{Comparative single-cell and spatial profiling of anti-SSA-positive and anti-centromere-positive Sj{\"o}gren’s disease reveals common and distinct immune activation and fibroblast-mediated inflammation}},
  author={Inamo, J. and Takeshita, M. and Suzuki, K. and others},
  journal={{Nature Communications}},
  volume={16},
  number={1},
  pages={8299},
  year={2025},
  publisher={Nature Publishing Group UK London},
  doi={10.1038/s41467-025-63935-9}
}

@article{arutyunyan2023spatial,
  title={Spatial multiomics map of trophoblast development in early pregnancy},
  author={Arutyunyan, A. and Roberts, K. and Troulé, K. and others},
  journal={Nature},
  volume={616},
  number={7955},
  pages={143--151},
  year={2023},
  publisher={Nature Publishing Group UK London},
  doi={10.1038/s41586-023-05869-0}
}

@article{montagna2012bayesian,
  title={Bayesian latent factor regression for functional and longitudinal data},
  author={Montagna, S. and Tokdar, S. T. and Neelon, B. and Dunson, D. B.},
  journal={Biometrics},
  volume={68},
  pages={1064--1073},
  year={2012},
  publisher={{Oxford University Press}}
}

@article{legramanti2020bayesian,
  title={Bayesian cumulative shrinkage for infinite factorizations},
  author={Legramanti, S. and Durante, D. and Dunson, D. B.},
  journal={Biometrika},
  volume={107},
  number={3},
  pages={745--752},
  year={2020},
  publisher={Oxford University Press}
}

@article{carvalho10,
	author = {C. M. Carvalho and N. G. Polson and J. G. Scott},
	title = {The horseshoe estimator for sparse signals},
	journal = {Biometrika},
	volume = {97},
	number = {2},
	pages = {465--480},
	year = {2010}
}

@article{carvalho2008high,
  title={High-dimensional sparse factor modeling: applications in gene expression genomics},
  author={Carvalho, C. M. and Chang, J. and Lucas, J. E. and Nevins, J. R. and Wang, Q. and West, M.},
  journal={Journal of the American Statistical Association},
  volume={103},
  number={484},
  pages={1438--1456},
  year={2008},
  publisher={Taylor \& Francis}
}

@article{bernardo2003bayesian,
  title={Bayesian factor regression models in the "large p, small n" paradigm},
  author={Bernardo, J. M. and Bayarri, M. J. and Berger, J. O. and others},
  journal={{Bayesian Statistics}},
  volume={7},
  pages={733--742},
  year={2003}
}

@inproceedings{anderson1956statistical,
  title={Statistical inference in factor analysis},
  author={Anderson, T. W. and Rubin, H.},
  booktitle={{Proceedings of the third Berkeley symposium on mathematical statistics and probability}},
  volume={5},
  pages={111--150},
  year={1956}
}

@article{bhattacharya2011sparse,
  title={{Sparse Bayesian infinite factor models}},
  author={Bhattacharya, A. and Dunson, D. B.},
  journal={Biometrika},
  volume={98},
  pages={291--306},
  year={2011},
  publisher={Oxford University Press}
}

@article{velten2022identifying,
  title={{Identifying temporal and spatial patterns of variation from multimodal data using MEFISTO}},
  author={Velten, B. and Braunger, J. M. and Argelaguet, R. and others},
  journal={{Nature Methods}},
  volume={19},
  number={2},
  pages={179--186},
  year={2022},
  publisher={{Nature Publishing Group US New York}}
}

@article{kowal2023semiparametric,
  title={{Semiparametric functional factor models with Bayesian rank selection}},
  author={Kowal, D. R. and Canale, A.},
  journal={Bayesian Analysis},
  volume={18},
  pages={1161--1189},
  year={2023},
  publisher={International Society for Bayesian Analysis}
}

@article{lucas2020longitudinal,
  title={{Longitudinal analyses reveal immunological misfiring in severe COVID-19}},
  author={Lucas, C. and Wong, P. and Klein, J. and others},
  journal={Nature},
  volume={584},
  pages={463--469},
  year={2020},
  publisher={{Nature Publishing Group}},
  doi={10.1038/s41586-020-2588-y}
}

@article{bernardes2020longitudinal,
  title={{Longitudinal multi-omics analyses identify responses of megakaryocytes, erythroid cells, and plasmablasts as hallmarks of severe COVID-19}},
  author={Bernardes, J. P. and Mishra, N. and Tran, F. and others},
  journal={Immunity},
  volume={53},
  number={6},
  pages={1296--1314},
  year={2020},
  publisher={Elsevier},
  doi={10.1016/j.immuni.2020.11.017}
}

@article{li2020predictive,
  title={{Predictive values of neutrophil-to-lymphocyte ratio on disease severity and mortality in COVID-19 patients: a systematic review and meta-analysis}},
  author={Li, X. and Liu, C. and Mao, Z. and others},
  journal={Critical Care},
  volume={24},
  pages={647},
  year={2020},
  publisher={BioMed Central}
}

@article{bergamaschi2021longitudinal,
  title={{Longitudinal analysis reveals that delayed bystander CD8+ T cell activation and early immune pathology distinguish severe COVID-19 from mild disease}},
  author={Bergamaschi, L. and Mescia, F. and Turner, L. and others},
  journal={Immunity},
  volume={54},
  pages={1257--1275},
  year={2021},
  publisher={Elsevier},
  doi={10.1016/j.immuni.2021.05.010}
}

@article{ruffieux2023patient,
  title={A patient-centric modeling framework captures recovery from SARS-CoV-2 infection},
  author={Ruffieux, H. and Hanson, A. L. and Lodge, S. and others},
  journal={Nature Immunology},
  volume={24},
  number={2},
  pages={349--358},
  year={2023},
  publisher={Nature Publishing Group US New York},
  doi={10.1038/s41590-022-01380-2}
}

@article{katahira08,
	author = {K. Katahira and K. Watanabe and M. Okada},
	title = {Deterministic annealing variant of variational {B}ayes method},
	journal = {Journal of Physics: Conference Series},
	volume = {95},
	pages = {012015},
	year = {2008}
}

@article{hosszejni26,
	author = {D. Hosszejni and S. Fr\"{u}hwirth-Schnatter},
	title = {Cover it up! {B}ipartite graphs uncover identifiability in sparse factor analysis},
	journal = {Journal of Multivariate Analysis},
	volume = {211},
	pages = {105536},
	year = {2026}
}

@article{marron15,
	author = {J. S. Marron and J. O. Ramsay and L. M. Sangalli and A. Srivastava},
	title = {Functional data analysis of amplitude and phase variation},
	journal = {Statistical Science},
	volume = {30},
	number = {4},
	pages = {468--484},
	year = {2015}
}

@article{rohe23,
	author = {K. Rohe and M. Zeng},
	title = {Vintage factor analysis with {V}arimax performs statistical inference},
	journal = {Journal of the Royal Statistical Society Series B: Statistical Methodology},
	volume = {85},
	number = {4},
	pages = {1037--1060},
	year = {2023}
}

@article{ruppert2002selecting,
  title={Selecting the number of knots for penalized splines},
  author={Ruppert, D.},
  journal={{Journal of Computational and Graphical Statistics}},
  volume={11},
  pages={735--757},
  year={2002},
  publisher={Taylor \& Francis}
}

@article{masoodi2022disturbed,
  title={Disturbed lipid and amino acid metabolisms in COVID-19 patients},
  author={Masoodi, M. and Peschka, M. and Schmiedel, S. and others},
  journal={{Journal of Molecular Medicine}},
  volume={100},
  number={4},
  pages={555--568},
  year={2022},
  publisher={Springer},
  doi={10.1007/s00109-022-02177-4}
}

@article{korotkevich2016fast,
  title={Fast gene set enrichment analysis},
  author={Korotkevich, G. and Sukhov, V. and Budin, N. and Shpak, B. and Artyomov, M. N. and Sergushichev, A.},
  journal={biorxiv},
  pages={060012},
  year={2016},
  publisher={{Cold Spring Harbor Laboratory}}
}

@article{atila2021serum,
  title={{The serum amino acid profile in COVID-19}},
  author={Atila, A. and Alay, H. and Yaman, M. E. and others},
  journal={{Amino Acids}},
  volume={53},
  number={10},
  pages={1569--1588},
  year={2021},
  publisher={Springer},
  doi={10.1007/s00726-021-03081-w}
}

@article{nolan23,
	title = {Bayesian functional principal components analysis via variational message passing with multilevel extensions},
	author = {T. H. Nolan and J. Goldsmith and D. Ruppert},
	journal = {Bayesian Analysis},
  volume={1},
  number={1},
  pages={1--27},
  year={2023},
  publisher={{International Society for Bayesian Analysis}}
}

@article{osullivan86,
	author = {F. O'Sullivan},
	title = {A statistical perspective on ill-posed inverse problems},
	journal = {Statistical Science},
	pages = {502--527},
	volume = {1},
	year = {1986}
}

@book{bishop06,
	author = {C. M. Bishop},
	title = {{Pattern Recognition and Machine Learning}},
	publisher = {Springer},
	year = {2006},
	address = {New York}
}

@article{menictas13,
	author = {M. Menictas and M. P. Wand},
	title = {Variational inference for marginal longitudinal semiparametric regression},
	journal = {Stat},
	pages = {61--71},
	volume = {2},
	year = {2013}
}

@article{ruffieux2021epispot,
  title={{EPISPOT: an epigenome-driven approach for detecting and interpreting hotspots in molecular QTL studies}},
  author={Ruffieux, H. and Fairfax, B. P. and Nassiri, I. and others},
  journal={{The American Journal of Human Genetics}},
  volume={108},
  number={6},
  pages={983--1000},
  year={2021},
  publisher={Elsevier}
}

@article{rose1990deterministic,
  title={A deterministic annealing approach to clustering},
  author={Rose, K. and Gurewitz, E. and Fox, G.},
  journal={{Pattern Recognition Letters}},
  volume={11},
  pages={589--594},
  year={1990},
  publisher={Elsevier}
}

@article{ueda1998deterministic,
  title={{Deterministic annealing EM algorithm}},
  author={Ueda, N. and Nakano, R.},
  journal={{Neural Networks}},
  volume={11},
  pages={271--282},
  year={1998},
  publisher={Elsevier}
}

@article{ruffieux2020global,
  title={A global-local approach for detecting hotspots in multiple-response regression},
  author={Ruffieux, H. and Davison, A. C. and Hager, J. and others},
  journal={{The Annals of Applied Statistics}},
  volume={14},
  number={2},
  pages={905},
  year={2020}
}

@article{nolan2025efficient,
  title={{Efficient Bayesian functional principal component analysis of irregularly-observed multivariate curves}},
  author={Nolan, T. H. and Richardson, S. and Ruffieux, H.},
  journal={Computational Statistics \& Data Analysis},
  volume={203},
  pages={108094},
  year={2025},
  publisher={Elsevier}
}

@book{opper2001advanced,
  title={{Advanced mean field methods: Theory and practice}},
  author={Opper, M. and Saad, D.},
  year={2001},
  publisher={{MIT Press}},
  address={{Cambridge, Massachusetts}}
}

@article{scott2010bayes,
  title={{Bayes and empirical-Bayes multiplicity adjustment in the variable-selection problem}},
  author={Scott, J. G. and Berger, J. O.},
  journal={{Annals of Statistics}},
  volume={38},
  pages={2587--2619},
  year={2010},
  publisher={{Institute of Mathematical Statistics}}
}

@article{barbieri2004optimal,
  title={Optimal predictive model selection},
  author={Barbieri, M. M. and Berger, J. O.},
  journal={{Annals of Statistics}},
  pages={870--897},
  volume={32},
  year={2004},
  publisher={{JSTOR}}
}

@book{ramsay05,
	author = {J. O. Ramsay and B. W. Silverman},
	title = {{Functional Data Analysis}},
	publisher = {Springer},
	year = {2005},
	address = {New York}
}

@article{suarez17,
	author = {A. J. Suarez and S. Ghosal},
	title = {Bayesian estimation of principal components for functional data},
	journal = {Bayesian Analysis},
	pages = {311--333},
	volume = {12},
	year = {2017}
}

@article{brodin2021immune,
  title={{Immune determinants of COVID-19 disease presentation and severity}},
  author={Brodin, P.},
  journal={{Nature Medicine}},
  volume={27},
  pages={28--33},
  year={2021},
  publisher={{Nature Publishing Group}}
}

@article{huang20216,
  title={{6-month consequences of COVID-19 in patients discharged from hospital: a cohort study}},
  author={Huang, C. and Huang, L. and Wang, Y. and others},
  journal={{The Lancet}},
  volume={397},
  pages={220--232},
  year={2021},
  publisher={Elsevier}
}

@article{wand08,
	author = {M. P. Wand and J. T. Ormerod},
	title = {On semiparametric regression with {O'Sullivan} penalized splines},
	journal = {{Australian} \& {New Zealand} Journal of Statistics},
	pages = {179--198},
	volume = {50},
	year = {2008}
}

@article{wang16,
	author = {J. L. Wang and J. M. Chiou and H. G. M\"{u}ller},
	title = {Functional data analysis},
	journal = {Annual Review of Statistics and Its Applications},
	pages = {257--295},
	volume = {3},
	year = {2016}
}

@article{yao05,
	author = {F. Yao and H. G. M\"{u}ller and J. L. Wang},
	title = {Functional data analysis for sparse longitudinal data},
	journal = {Journal of the {American} Statistical Association},
	pages = {577--590},
	volume = {100},
	year = {2005}
}

@misc{jaoua2026code,
	title = {Code for: {A} scalable {B}ayesian functional factor model for high-dimensional longitudinal molecular data},
	author = {S. Jaoua and D. Temko and H. Ruffieux},
	note = {Version 1.0.1, \url{https://github.com/hruffieux/bayesSYNC_paper_code}},
	year = {2026},
	doi = {10.5281/zenodo.23676333},
	url = {https://doi.org/10.5281/zenodo.23676333},
}

\clearpage
\begin{center}
{\LARGE Supplementary Material}
\end{center}
\setcounter{section}{0}
\setcounter{figure}{0}
\setcounter{table}{0}
\setcounter{equation}{0}
\renewcommand{\thesection}{S\arabic{section}}
\renewcommand{\thesubsection}{S\arabic{section}.\arabic{subsection}}
\renewcommand{\thefigure}{S\arabic{figure}}
\renewcommand{\thetable}{S\arabic{table}}
\renewcommand{\theequation}{S\arabic{equation}}
\renewcommand{\theHsection}{S\arabic{section}}
\renewcommand{\theHsubsection}{S\arabic{section}.\arabic{subsection}}
\renewcommand{\theHfigure}{S\arabic{figure}}
\renewcommand{\theHtable}{S\arabic{table}}
\renewcommand{\theHequation}{S\arabic{equation}}

\section{Annealed variational inference updates}\label{sm_sec_updates}

In this appendix, we derive the annealed mean-field variational updates under the factorisation
\begin{eqnarray*}
q(\bm\Theta)&=&\prod_{j=1}^p q(\boldsymbol{\nu}_{\mu}^{(j)})
\prod_{q=1}^Q \prod_{\ell=1}^L q(\boldsymbol{\nu}_{\psi, \ell}^{(q)})
\prod_{i=1}^N \prod_{q=1}^Q\prod_{\ell=1}^L q(\zeta_{i\ell}^{(q)}) 
\prod_{j=1}^p q(\sigma_{\varepsilon,j}^{2}) q(a_{\varepsilon, j})
\prod_{j=1}^p q(\sigma_{\mu, j}^{2}) q(a_{\mu, j})\\
&&\times
\prod_{q=1}^Q \prod_{\ell=1}^L q(\sigma_{\psi, q\ell}^{2}) q(a_{\psi, q\ell})
\prod_{j,q} q(b_{jq},\gamma_{jq}),
\end{eqnarray*}
using the notation of the main text. Let $T> 1$ be the current temperature from the annealing schedule and let $q_T(\bm \Theta)$ be the \emph{heated}  variational approximation to the true posterior distribution, $\pi(\bm \Theta \mid \bm y)$.  We maximise the lower bound on the marginal log-likelihood,
$$\mathcal{L}_T(q) = \int q_T(\bm \Theta) \log \pi(\bm \Theta, \bm y) \mathrm{d}\bm \Theta -   T \int q_T(\bm \Theta) \log q_T(\bm \Theta) \mathrm{d}\bm \Theta.$$
The heated variational distributions for factor $q_T(\bm\Theta_m)$, $m = 1, \ldots, M$, are given by
\begin{equation*}\label{sm_upd}\log q_T(\bm\Theta_m) = T^{-1}\mathrm{E}_{-q_m} \{\log \pi(\bm y, \bm\Theta_m)\} + \mathrm{const.},\qquad\qquad m=1, \ldots, M,\end{equation*}
where $\mathrm{E}_{-q_m}(\cdot)$ is  the expectation with respect to the distributions $q_T(\bm\Theta_{m'})$, for all the variables $\bm\Theta_{m'}$ ($m'\neq m$), and $\mathrm{const.}$ is constant with respect to $\bm\Theta_m$. 
For ease of reading, we hereafter drop the subscript $T$ in $q_T(\cdot)$, and write $c = T^{-1}$. 


\underline{Update for $q(\boldsymbol{\nu}_{\mu}^{(j)})$}

Let $\bm C_i \equiv [ \textbf{1}_{n_i} , \bm t_i ,\bm  z_1(\bm t_i) , \ldots, \bm z_{K'}( \bm t_i)] $, the $n_i \times K$ O'Sullivan spline design matrix evaluated on the observation grid for subject $i$, with $K' = K-2$ and let the corresponding spline coefficients $\bm\nu_\mu^{(j)}= (
        \beta_{\mu 0}^{(j)},   
        \beta_{\mu 1}^{(j)}, 
        u_{\mu 1}^{(j)}, 
        \ldots, 
        u_{\mu K'}^{(j)})^\top\in \mathbb{R}^{K}$ and similarly for $\bm\nu_{\psi,\ell}^{(q)}\in \mathbb{R}^{K}$.

The likelihood contribution for variable $j = 1, \ldots, p$ and subject $i = 1, \ldots, N$ is
\[
\bm{y}_i^{(j)} \mid \boldsymbol{\nu}_{\mu}^{(j)}, \{\boldsymbol{\nu}_{\psi, \ell}^{(q)}\}, \{\boldsymbol{\zeta}_{i}^{(q)}\}, \sigma_{\varepsilon,j}^{2}
\sim \mathcal{N}\!\left(
\bm{C}_i \boldsymbol{\nu}_{\mu}^{(j)}
+ \bm{C}_i \sum_{q=1}^Q b_{jq} \bm{V}_{\psi}^{(q)} \boldsymbol{\zeta}_{i}^{(q)},
\,
\sigma_{\varepsilon,j}^{2} \bm I_{n_i}
\right),
\]
where  $\bm{V}_{\psi}^{(q)} = [\boldsymbol{\nu}_{\psi_1}^{(q)},\ldots,\boldsymbol{\nu}_{\psi, \ell}^{(q)}]$ and $\boldsymbol{\zeta}_{i}^{(q)} =  (\zeta_{i1}^{(q)},\ldots,\zeta_{iL}^{(q)})^\top$.

The prior is
\[
\boldsymbol{\nu}_{\mu}^{(j)} \mid \sigma_\beta^2, \sigma_{\mu,j}^{2} \sim 
\mathcal{N}\!\left(\bm{0}, \boldsymbol{\Sigma}_{\mu}^{(j)}\right),
\]
where $\bm \Sigma_\mu^{(j)} = \text{blockdiag}(\sigma_\beta^2 \bm I_2, \sigma_{\mu,j}^{2}\bm I_{K'}).$

Collecting quadratic and linear terms in $\boldsymbol{\nu}_{\mu}^{(j)}$, the optimal variational factor is Gaussian,
\[
q(\boldsymbol{\nu}_{\mu}^{(j)}) = 
\mathcal{N}\!\left(\boldsymbol{\mu}_{\mu}^{*(j)}, \boldsymbol{\Sigma}_{\mu}^{*(j)}\right),
\]
with
\begin{align*}
\boldsymbol{\Sigma}_{\mu}^{*(j)} &=
c^{-1}\left(
\mathbb{E}_q\left[\left(\boldsymbol{\Sigma}_{\mu}^{(j)}\right)^{-1}\right]
+ \mathbb{E}_q\!\left[\frac{1}{\sigma_{\varepsilon, j}^{2}}\right]
\sum_{i=1}^N \bm{C}_i^\top \bm{C}_i
\right)^{-1}, \\
\boldsymbol{\mu}_{\mu}^{*(j)} &= c\,
\boldsymbol{\Sigma}_{\mu}^{*(j)}
\mathbb{E}_q\!\left[\frac{1}{\sigma_{\varepsilon,j}^{2}}\right]
\sum_{i=1}^N \bm{C}_i^\top
\left(
\bm{y}_i^{(j)}
- \bm{C}_i
\sum_{q=1}^Q \mathbb{E}_q[b_{jq}]
\mathbb{E}_q[\bm{V}_{\psi}^{(q)}]
\mathbb{E}_q[\boldsymbol{\zeta}_{i}^{(q)}]
\right).
\end{align*}

\underline{Update for $q(\boldsymbol{\nu}_{\psi, \ell}^{(q)})$}

From the model, we have
\[
\boldsymbol{\nu}_{\psi, \ell}^{(q)} \mid \sigma_\beta^2, \sigma_{\psi, q\ell}^{2}
\sim \mathcal{N}\!\left(\bm{0}, \boldsymbol{\Sigma}_{\psi, \ell}^{(q)}\right).
\]
where $\boldsymbol{\Sigma}_{\psi, \ell}^{(q)}= \text{blockdiag}(\sigma_\beta^2 \bm I_2, \sigma_{\psi,q\ell}^{2}\bm I_{K'}).$

The log joint likelihood is quadratic in $\boldsymbol{\nu}_{\psi, \ell}^{(q)}$, yielding a Gaussian variational factor
\[
q(\boldsymbol{\nu}_{\psi, \ell}^{(q)})=
\mathcal{N}\!\left(\boldsymbol{\mu}_{\psi, \ell}^{*(q)}, \boldsymbol{\Sigma}_{\psi, \ell}^{*(q)}\right),
\]
where
\begin{align*}
\boldsymbol{\Sigma}_{\psi, \ell}^{*(q)} &= c^{-1}
\Bigg(
\mathbb{E}_q\left[\left(\boldsymbol{\Sigma}_{\psi, \ell}^{(q)}\right)^{-1}\right]
+ \sum_{i=1}^N \mathbb{E}_q[\zeta_{i\ell}^{(q)2}]\,
\bm{C}_i^\top \bm{C}_i
\sum_{j=1}^p 
\mathbb{E}_q[b_{jq}^2]
\mathbb{E}_q\!\left[\frac{1}{\sigma_{\varepsilon, j}^{2}}\right]
\Bigg)^{-1}, \\
\boldsymbol{\mu}_{\psi, \ell}^{*(q)} &=c\,
\boldsymbol{\Sigma}_{\psi, \ell}^{*(q)}
\sum_{i=1}^N \sum_{j=1}^p
\mathbb{E}_q[b_{jq}]
\mathbb{E}_q\!\left[\frac{1}{\sigma_{\varepsilon,j}^{2}}\right]
\mathbb{E}_q[\zeta_{i\ell}^{(q)}]
\bm{C}_i^\top
\bm{r}_{ij}^{(q,\ell)},
\end{align*}
with residual
\[
\bm{r}_{ij}^{(q,\ell)} =
\bm{y}_i^{(j)}
- \bm{C}_i \mathbb{E}_q[\boldsymbol{\nu}_{\mu}^{(j)}]
- \bm{C}_i
\sum_{(q',\ell')\neq(q,\ell)}
\mathbb{E}_q[b_{jq'}]
\mathbb{E}_q[\zeta_{i'\ell'}^{(q')}]
\mathbb{E}_q[\boldsymbol{\nu}_{\psi_{\ell'}}^{(q')}].
\]

\underline{Update for $q(\boldsymbol{\zeta}_{i}^{(q)})$}

Conditionally on the other parameters, $\boldsymbol{\zeta}_{i}^{(q)}\in\mathbb{R}^L$ only affects subject $i$. The prior is
\[
\boldsymbol{\zeta}_{i}^{(q)} \sim \mathcal{N}(\bm{0},  \bm I_L).
\]

Defining the $L \times L$ matrix $\bm{H}_{i,\psi}^{(q)} =
\bm{V}_{\psi}^{(q)\top}\bm{C}_i^\top \bm{C}_i \bm{V}_{\psi}^{(q)}$,
the optimal factor is
\[
q(\boldsymbol{\zeta}_{i}^{(q)})=
\mathcal{N}\!\left(\boldsymbol{m}_{\zeta,{iq}}, \boldsymbol{S}_{\zeta,{iq}}\right),
\]
with
\begin{align*}
\boldsymbol{S}_{\zeta, {iq}} &=c^{-1}
\left(
\sum_{j=1}^p 
\mathbb{E}_q[b_{jq}^2]
\mathbb{E}_q\!\left[\frac{1}{\sigma_{\varepsilon,j}^{2}}\right]
\mathbb{E}_q[\bm{H}_{i,\psi}^{(q)}]
+ \bm I_L
\right)^{-1}, \\
\boldsymbol{m}_{\zeta,{iq}} &=c\,
\boldsymbol{S}_{\zeta,{iq}}
\sum_{j=1}^p
\mathbb{E}_q[b_{jq}]
\mathbb{E}_q\!\left[\frac{1}{\sigma_{\varepsilon,j}^{2}}\right]
\mathbb{E}_q[\bm{V}_{\psi}^{(q)}]^\top
\bm{C}_i^\top
\tilde{\bm{r}}_{ij}^{(q)},
\end{align*}
where
\[
\tilde{\bm{r}}_{ij}^{(q)} =
\bm{y}_i^{(j)}
- \bm{C}_i \mathbb{E}_q[\boldsymbol{\nu}_{\mu}^{(j)}]
- \bm{C}_i
\sum_{q'\neq q}
\mathbb{E}_q[b_{jq'}]
\mathbb{E}_q[\bm{V}_{\psi}^{(q')}]
\mathbb{E}_q[\boldsymbol{\zeta}_{i}^{(q')}].
\]

\underline{Update for $q(\sigma_{\varepsilon,j}^{2})$}

Assuming the hierarchical inverse-Gamma prior
\(
\sigma_{\varepsilon,j}^{2} \mid a_{\varepsilon,j} \sim \text{IG}(1/2, 1/a_{\varepsilon,j}),
\)
the variational posterior is
\[
q(\sigma_{\varepsilon,j}^{2})=\text{IG}(\kappa_{\varepsilon,j}, \lambda_{\varepsilon,j}),
\]
with
\begin{align*}
\kappa_{\varepsilon,j} &= c\,\left(
\frac{1}{2}\sum_{i=1}^N n_i + \frac{1}{2}\right) +c -1, \\
\lambda_{\varepsilon,j} &=c\,\left\{
\mathbb{E}_q\!\left[\frac{1}{a_{\varepsilon,j}}\right]
+ \frac{1}{2}\sum_{i=1}^N 
\mathbb{E}_q\!\left[
\left\|\bm{y}_i^{(j)}-\bm{C}_i\left(\boldsymbol{\nu}_{\mu}^{(j)}+\sum_{q=1}^Q b_{jq}\bm{V}_{\psi}^{(q)}\boldsymbol{\zeta}_{i}^{(q)}\right)\right\|_2^2
\right]\right\}.
\end{align*}

The quadratic expectation is computed using
\[
\mathbb{E}_q[\bm{x}^\top \bm{A} \bm{x}]
=
\operatorname{tr}\!\left(\bm{A}\,\mathbb{C}\text{ov}_q(\bm{x})\right)
+ \mathbb{E}_q[\bm{x}]^\top \bm{A}\,\mathbb{E}_q[\bm{x}].
\]

\underline{Updates for local scale parameters}

By conjugacy of the inverse-Gamma hierarchy, we have 
\begin{align*}
q(a_{\varepsilon,j}) &= \text{IG}\!\left(
2c-1,
c\,\mathbb{E}_q\!\left[\frac{1}{\sigma_{\varepsilon,j}^{2}}\right]
+ \frac{c}{A^2}
\right), \\
q(\sigma_{\mu,j}^{2}) &= \text{IG}\!\left(
\frac{K'}{2}+\frac{1}{2},
c\,\mathbb{E}_q\!\left[\frac{1}{a_{\mu,j}}\right]
+ \frac{c}{2}\mathbb{E}_q\!\left[
\boldsymbol{u}_{\mu}^{(j)\top}\boldsymbol{u}_{\mu}^{(j)}
\right]
\right), \\
q(a_{\mu,j}) &= \text{IG}\!\left(
2c-1,
c\,\mathbb{E}_q\!\left[\frac{1}{\sigma_{\mu,j}^{2}}\right]
+ \frac{c}{A^2}
\right), \\
q(\sigma_{\psi, q\ell}^{2}) &= \text{IG}\!\left(
\frac{K'}{2}+\frac{1}{2},
c\,\mathbb{E}_q\!\left[\frac{1}{a_{\psi, q\ell}}\right]
+ \frac{c}{2}\mathbb{E}_q\!\left[
\boldsymbol{u}_{\psi, \ell}^{(q)\top}
\boldsymbol{u}_{\psi, \ell}^{(q)}
\right]
\right), \\
q(a_{\psi, q\ell}) &= \text{IG}\!\left(
2c-1,
c\,\mathbb{E}_q\!\left[\frac{1}{\sigma_{\psi, q\ell}^{2}}\right]
+ \frac{c}{A^2}
\right),
\end{align*}
where $\bm u_{\mu}^{(j)} = \left(  u_{\mu 1}^{(j)}, 
        \ldots, 
        u_{\mu K'}^{(j)}\right)^{\top}$ and $\bm u_{\psi,\ell}^{(q)} = \left(  u_{\psi,\ell1}^{(q)}, 
        \ldots, 
        u_{\psi,\ell K'}^{(q)}\right)^{\top}$.
        
\underline{Update for $q(b_{jq}, \gamma_{jq})$}

Using the spike-and-slab prior
\[
b_{jq} \mid \gamma_{jq} \sim 
\gamma_{jq}\,\mathcal{N}(0,1)
+ (1-\gamma_{jq})\delta_0,
\quad
\gamma_{jq}\sim\text{Bernoulli}(\omega_q),
\]
the conditional log-density is quadratic in $b_{jq}$ when $\gamma_{jq}=1$.

Hence,
\[
q(b_{jq}\mid \gamma_{jq}=1)=
\mathcal{N}(\mu_{b,jq}, \sigma_{b,jq}^2),
\]
with
\begin{align*}
\sigma_{b,jq}^2 &=c^{-1}
\left(
1
+ \mathbb{E}_q\!\left[\frac{1}{\sigma_{\varepsilon,j}^{2}}\right]
\sum_{i=1}^N 
\mathbb{E}_q\!\left[
\boldsymbol{\zeta}_{i}^{(q)\top}
\bm{H}_{i,\psi}^{(q)}
\boldsymbol{\zeta}_{i}^{(q)}
\right]
\right)^{-1}, \\
\mu_{b,jq} &=c\,
\sigma_{b,jq}^2\,
\mathbb{E}_q\!\left[\frac{1}{\sigma_{\varepsilon,j}^{2}}\right]
\sum_{i=1}^N
\mathbb{E}_q[\boldsymbol{\zeta}_{i}^{(q)}]^\top
\mathbb{E}_q[\bm{V}_{\psi}^{(q)}]^\top
\bm{C}_i^\top
\bm{e}_{ij}^{(q)},
\end{align*}
where
\[
\bm{e}_{ij}^{(q)}=
\bm{y}_i^{(j)}
- \bm{C}_i \mathbb{E}_q[\boldsymbol{\nu}_{\mu}^{(j)}]
- \bm{C}_i
\sum_{q'\neq q}
\mathbb{E}_q[b_{jq'}]
\mathbb{E}_q[\bm{V}_{\psi}^{(q')}]
\mathbb{E}_q[\boldsymbol{\zeta}_{iq'}].
\]


The optimal variational factor for the inclusion indicator is Bernoulli,
\[
q(\gamma_{jq}) = \text{Bernoulli}(\gamma_{jq}^*),
\]
with
\[
\gamma_{jq}^*
= \left[
1 + \exp\!\left(
- c\,\mathbb{E}_q[\log \omega_q - \log(1-\omega_q)]
-\frac{1}{2}\left(
 \log \sigma_{b,jq}^2
+ \frac{\mu_{b,jq}^2}{\sigma_{b,jq}^2}
\right)
\right)
\right]^{-1}.
\]
Note that \(\mathbb{E}_q[\gamma_{jq}] = \gamma_{jq}^*\) and
\[
\mathbb{E}_q[b_{jq}] = \gamma_{jq}^* \mu_{b,jq},
\qquad
\mathbb{E}_q[b_{jq}^2] = \gamma_{jq}^*\left(\mu_{b,jq}^2 + \sigma_{b,jq}^2\right).
\]

\underline{Update for $q(\omega_q)$}

Assume the Beta prior \(\omega_q \sim \text{Beta}(c_0,d_0)\) and recall that
\[
\pi(\gamma_{jq}\mid \omega_q)=
\omega_q^{\gamma_{jq}}(1-\omega_q)^{1-\gamma_{jq}}.
\]
We have 
\[
q(\omega_q) = \text{Beta}(c_q^*, d_q^*),
\]
with
\[
c_q^* = c\left(c_0 + \sum_{j=1}^p \gamma_{jq}^*\right)-c+1,
\qquad
d_q^* = c\left(d_0 + p - \sum_{j=1}^p \gamma_{jq}^*\right)-c+1.
\]
The required expectations are
\[
\mathbb{E}_q[\log \omega_q] = \psi(c_q^*) - \psi(c_q^* + d_q^*),
\qquad
\mathbb{E}_q[\log (1-\omega_q)] = \psi(d_q^*) - \psi(c_q^* + d_q^*),
\]
where \(\psi(\cdot)\) denotes the digamma function.

\section{Computation of the evidence lower bound}\label{sm_sec_elbo}

The evidence lower bound (ELBO) is
\[
\mathcal{L}(q)
=
\mathbb{E}_q[\log \pi(\bm y,\bm\Theta)]
-
\mathbb{E}_q[\log q(\bm\Theta)],
\]
where $\bm\Theta$ collects all  parameters.

Under the mean-field factorisation specified above, this decomposes additively as
\begin{eqnarray*}
\mathcal{L}(q)
&=&
\mathcal{L}_{\text{lik}}
+ \sum_{j=1}^p \mathcal{L}_{\nu_\mu^{(j)}}
+ \sum_{q=1}^Q \sum_{\ell=1}^L \mathcal{L}_{\nu_{\psi,\ell}^{(q)}}
+ \sum_{i=1}^N \sum_{q=1}^Q \mathcal{L}_{\zeta_i^{(q)}}
+ \sum_{j=1}^p \Big(
\mathcal{L}_{\sigma_{\varepsilon,j}^2}
+ \mathcal{L}_{a_{\varepsilon,j}}
+ \mathcal{L}_{\sigma_{\mu,j}^2}
+ \mathcal{L}_{a_{\mu,j}}
\Big)
\\
&&
+ \sum_{q=1}^Q \sum_{\ell=1}^L
\Big(
\mathcal{L}_{\sigma_{\psi,q\ell}^2}
+ \mathcal{L}_{a_{\psi,q\ell}}
\Big)
+ \sum_{j=1}^p \sum_{q=1}^Q
\mathcal{L}_{b_{jq},\gamma_{jq}}
+ \sum_{q=1}^Q \mathcal{L}_{\omega_q},
\end{eqnarray*}
where each term is of the form
\(
\mathcal{L}_{x}
=
\mathbb{E}_q[\log \pi(x \mid \cdot)]
-
\mathbb{E}_q[\log q(x)].
\)

\underline{Likelihood contribution}

From the Gaussian observation model,
\[
\bm{y}_i^{(j)} \mid \cdot \sim
\mathcal{N}\!\left(
\bm{C}_i \boldsymbol{\nu}_{\mu}^{(j)}
+ \bm{C}_i \sum_{q=1}^Q b_{jq}\bm{V}_{\psi}^{(q)}\boldsymbol{\zeta}_{i}^{(q)},
\,
\sigma_{\varepsilon,j}^{2}\bm I_{n_i}
\right),
\]
we obtain
\begin{align*}
\mathcal{L}_{\text{lik}}
&=
\sum_{i=1}^N \sum_{j=1}^p
\mathbb{E}_q
\big[
\log \pi(\bm y_i^{(j)} \mid \cdot)
\big] \\
&=
-\frac{1}{2}
\sum_{i=1}^N \sum_{j=1}^p
\Bigg\{
n_i \log(2\pi)
+ n_i\,\mathbb{E}_q[\log \sigma_{\varepsilon,j}^{2}]
+ \mathbb{E}_q\!\left[\frac{1}{\sigma_{\varepsilon,j}^{2}}\right]
\mathbb{E}_q\!\left[
\left\|
\bm{y}_i^{(j)}
-
\bm{C}_i\boldsymbol{\nu}_{\mu}^{(j)}
-
\bm{C}_i \sum_{q=1}^Q
b_{jq}\bm{V}_{\psi}^{(q)}\boldsymbol{\zeta}_{i}^{(q)}
\right\|_2^2
\right]
\Bigg\}.
\end{align*}

The quadratic expectation is evaluated using
$
\mathbb{E}_q[\bm x^\top \bm A \bm x]
=
\operatorname{tr}\!\big(
\bm A\,\mathbb{C}\mathrm{ov}_q(\bm x)
\big)
+
\mathbb{E}_q[\bm x]^\top
\bm A\,
\mathbb{E}_q[\bm x].
$

\underline{Contribution of $q(\boldsymbol{\nu}_{\mu}^{(j)})$}

Since both the prior and variational factor are Gaussian, we have
\[
\mathcal{L}_{\nu_\mu^{(j)}}
=
\mathbb{E}_q[\log \pi(\boldsymbol{\nu}_{\mu}^{(j)})]
-
\mathbb{E}_q[\log q(\boldsymbol{\nu}_{\mu}^{(j)})].
\]

We obtain
\begin{align*}
\mathcal{L}_{\nu_\mu^{(j)}}
&=
-\frac{1}{2}
\Big(
\log \det \boldsymbol{\Sigma}_{\mu}^{(j)}
+
\operatorname{tr}\!\big(
\mathbb{E}_q[(\boldsymbol{\Sigma}_{\mu}^{(j)})^{-1}]
\boldsymbol{\Sigma}_{\mu}^{*(j)}
\big)
+
\boldsymbol{\mu}_{\mu}^{*(j)\top}
\mathbb{E}_q[(\boldsymbol{\Sigma}_{\mu}^{(j)})^{-1}]
\boldsymbol{\mu}_{\mu}^{*(j)}
\Big)
+\frac{1}{2}\log \det \boldsymbol{\Sigma}_{\mu}^{*(j)}
+\frac{K}{2}.
\end{align*} 

\underline{Contribution of $q(\boldsymbol{\nu}_{\psi,\ell}^{(q)})$}

We have
\begin{align*}
\mathcal{L}_{\nu_{\psi,\ell}^{(q)}}
&=
-\frac{1}{2}
\Big(
\log \det \boldsymbol{\Sigma}_{\psi,\ell}^{(q)}
+
\operatorname{tr}\!\big(
\mathbb{E}_q[(\boldsymbol{\Sigma}_{\psi,\ell}^{(q)})^{-1}]
\boldsymbol{\Sigma}_{\psi,\ell}^{*(q)}
\big)
+
\boldsymbol{\mu}_{\psi,\ell}^{*(q)\top}
\mathbb{E}_q[(\boldsymbol{\Sigma}_{\psi,\ell}^{(q)})^{-1}]
\boldsymbol{\mu}_{\psi,\ell}^{*(q)}
\Big)
+\frac{1}{2}\log \det \boldsymbol{\Sigma}_{\psi,\ell}^{*(q)}
+\frac{K}{2}.
\end{align*}

\underline{Contribution of $q(\boldsymbol{\zeta}_{i}^{(q)})$}

With standard Gaussian prior $\mathcal{N}(\mathbf{0},\mathbf{I}_L)$, we have

\begin{align*}
\mathcal{L}_{\zeta_i^{(q)}}
&=
-\frac{1}{2}
\left(
\operatorname{tr}(\boldsymbol{S}_{\zeta,iq})
+
\boldsymbol{m}_{\zeta,iq}^\top
\boldsymbol{m}_{\zeta,iq}
\right)
+\frac{1}{2}
\log \det \boldsymbol{S}_{\zeta,iq}
+\frac{L}{2}.
\end{align*}

\underline{Contribution of $q(\sigma_{\varepsilon,j}^{2})$}

We have
\begin{align*}
\mathcal{L}_{\sigma_{\varepsilon,j}^2}
&=
-\frac{1}{2}
\mathbb{E}_q[\log(a_{\varepsilon,j})]
-
\log \Gamma(1/2)
-
\mathbb{E}_q\!\left[\frac{1}{\sigma_{\varepsilon,j}^{2}}\right]
\mathbb{E}_q\!\left[\frac{1}{a_{\varepsilon,j}}\right]
-
\kappa_{\varepsilon,j}\log \lambda_{\varepsilon,j}
+
\log \Gamma(\kappa_{\varepsilon,j})
\\
&\quad
+
(\kappa_{\varepsilon,j}-\frac{1}{2})\mathbb{E}_q[\log \sigma_{\varepsilon,j}^2]
+
\lambda_{\varepsilon,j}
\mathbb{E}_q\!\left[\frac{1}{\sigma_{\varepsilon,j}^2}\right].
\end{align*}

\underline{Contribution of $q(a_{\varepsilon,j})$}

We have
\begin{align*}
\mathcal{L}_{a_{\varepsilon,j}}
&=
-\frac{1}{2}\log(A^2)
-
\log \Gamma(1/2)
+\frac{1}{2}\mathbb{E}_q[\log a_{\varepsilon,j}]
-
\frac{1}{A^2}
\mathbb{E}_q\!\left[\frac{1}{a_{\varepsilon,j}}\right]
-
\log \lambda_{a_{\varepsilon,j}}
+
\lambda_{a_{\varepsilon,j}}
\mathbb{E}_q\!\left[\frac{1}{a_{\varepsilon,j}}\right].
\end{align*}

\underline{Contribution of $q(\sigma_{\mu,j}^{2})$}

We have
\begin{align*}
\mathcal{L}_{\sigma_{\mu,j}^2}
&=
-\frac{1}{2}
\mathbb{E}_q[\log(a_{\mu,j})]
-
\log \Gamma(1/2)
-\frac{1}{2}\mathbb{E}_q[\log \sigma_{\mu,j}^2]
-
\mathbb{E}_q\!\left[\frac{1}{\sigma_{\mu,j}^{2}}\right]
\mathbb{E}_q\!\left[\frac{1}{a_{\mu,j}}\right]
\\
&\quad
-
\kappa_{\mu,j}\log \lambda_{\mu,j}
+
\log \Gamma(\kappa_{\mu,j})
+
\kappa_{\mu,j}\mathbb{E}_q[\log \sigma_{\mu,j}^2]
+
\lambda_{\mu,j}
\mathbb{E}_q\!\left[\frac{1}{\sigma_{\mu,j}^2}\right].
\end{align*}

\underline{Contribution of $q(a_{\mu,j})$}

We have
\begin{align*}
\mathcal{L}_{a_{\mu,j}}
&=
-\frac{1}{2}\log(A^2)
-
\log \Gamma(1/2)
+\frac{1}{2}\mathbb{E}_q[\log a_{\mu,j}]
-
\frac{1}{A^2}
\mathbb{E}_q\!\left[\frac{1}{a_{\mu,j}}\right]
-
\log \lambda_{a_{\mu,j}}
+
\lambda_{a_{\mu,j}}
\mathbb{E}_q\!\left[\frac{1}{a_{\mu,j}}\right].
\end{align*}

\underline{Contribution of $q(\sigma_{\psi,q\ell}^{2})$}

We have
\begin{align*}
\mathcal{L}_{\sigma_{\psi,q\ell}^2}
&=
-\frac{1}{2}
\mathbb{E}_q[\log(a_{\psi,q\ell})]
-
\log \Gamma(1/2)
-\frac{1}{2}\mathbb{E}_q[\log \sigma_{\psi,q\ell}^2]
-
\mathbb{E}_q\!\left[\frac{1}{\sigma_{\psi,q\ell}^{2}}\right]
\mathbb{E}_q\!\left[\frac{1}{a_{\psi,q\ell}}\right]
\\
&\quad
-
\kappa_{\psi,q\ell}\log \lambda_{\psi,q\ell}
+
\log \Gamma(\kappa_{\psi,q\ell})
+
\kappa_{\psi,q\ell}\mathbb{E}_q[\log \sigma_{\psi,q\ell}^2]
+
\lambda_{\psi,q\ell}
\mathbb{E}_q\!\left[\frac{1}{\sigma_{\psi,q\ell}^2}\right].
\end{align*}

\newpage
\underline{Contribution of $q(a_{\psi,q\ell})$}

We have
\begin{align*}
\mathcal{L}_{a_{\psi,q\ell}}
&=
-\frac{1}{2}\log(A^2)
-
\log \Gamma(1/2)
+\frac{1}{2}\mathbb{E}_q[\log a_{\psi,q\ell}]
-
\frac{1}{A^2}
\mathbb{E}_q\!\left[\frac{1}{a_{\psi,q\ell}}\right]
-
\log \lambda_{a_{\psi,q\ell}}
+
\lambda_{a_{\psi,q\ell}}
\mathbb{E}_q\!\left[\frac{1}{a_{\psi,q\ell}}\right].
\end{align*}

\underline{Contribution of $q(b_{jq},\gamma_{jq})$}

Collecting the terms derived above, we have
\begin{align*}
\mathcal{L}_{q(b_{jq},\gamma_{jq})}
&=\mathbb{E}_q[\log \pi(b_{jq}\mid \gamma_{jq})]
+
\mathbb{E}_q[\log \pi(\gamma_{jq}\mid \omega_q)]
-
\mathbb{E}_q[\log q(b_{jq},\gamma_{jq})] \\
&=\frac{1}{2}\gamma_{jq}^*
\mathbb{E}_q[\log \sigma_{b,jq}^2]
+
\frac{1}{2}\gamma_{jq}^*
-\frac{1}{2}\gamma_{jq}^* \, ( \mu_{b,jq}^2 + \sigma_{b,jq}^{2})
+ \gamma_{jq}^* \, \mathbb{E}_q[\log \omega_q]
+ (1-\gamma_{jq}^*) \, \mathbb{E}_q[\log(1-\omega_q)] \\
&\quad
- \gamma_{jq}^* \log \gamma_{jq}^*
- (1-\gamma_{jq}^*) \log (1-\gamma_{jq}^*).
\end{align*}

\underline{Contribution of $q(\omega_q)$}

With the Beta prior $\omega_q \sim \mathrm{Beta}(c_0,d_0)$ and
\(
q(\omega_q)=\mathrm{Beta}(c_q^*,d_q^*)
\),
we have
\begin{align*}
\mathcal{L}_{\omega_q}
&=
\mathbb{E}_q[\log \pi(\omega_q)]
-
\mathbb{E}_q[\log q(\omega_q)] \\
&=
(c_0-1)\mathbb{E}_q[\log \omega_q]
+
(d_0-1)\mathbb{E}_q[\log(1-\omega_q)]
-
\log B(c_0,d_0)
\\
&\quad
-
\Big[
(c_q^*-1)\mathbb{E}_q[\log \omega_q]
+
(d_q^*-1)\mathbb{E}_q[\log(1-\omega_q)]
-
\log B(c_q^*,d_q^*)
\Big],
\end{align*}
where
\[
\mathbb{E}_q[\log \omega_q]
=
\psi(c_q^*)-\psi(c_q^*+d_q^*),
\qquad
\mathbb{E}_q[\log(1-\omega_q)]
=
\psi(d_q^*)-\psi(c_q^*+d_q^*).
\]

Summing all components together, the ELBO used for optimisation is evaluated at each iteration using the current variational moments
(e.g. $\mathbb{E}_q[\cdot]$, $\mathbb{C}\mathrm{ov}_q(\cdot)$) and serves as the objective for monitoring convergence of the annealed variational inference algorithm.

\newpage
\section{Simulated and estimated loadings in the large-scale example}\label{sm_sec_loadings}

\begin{figure}[h!]
\centering 
\includegraphics[width=0.7\linewidth]{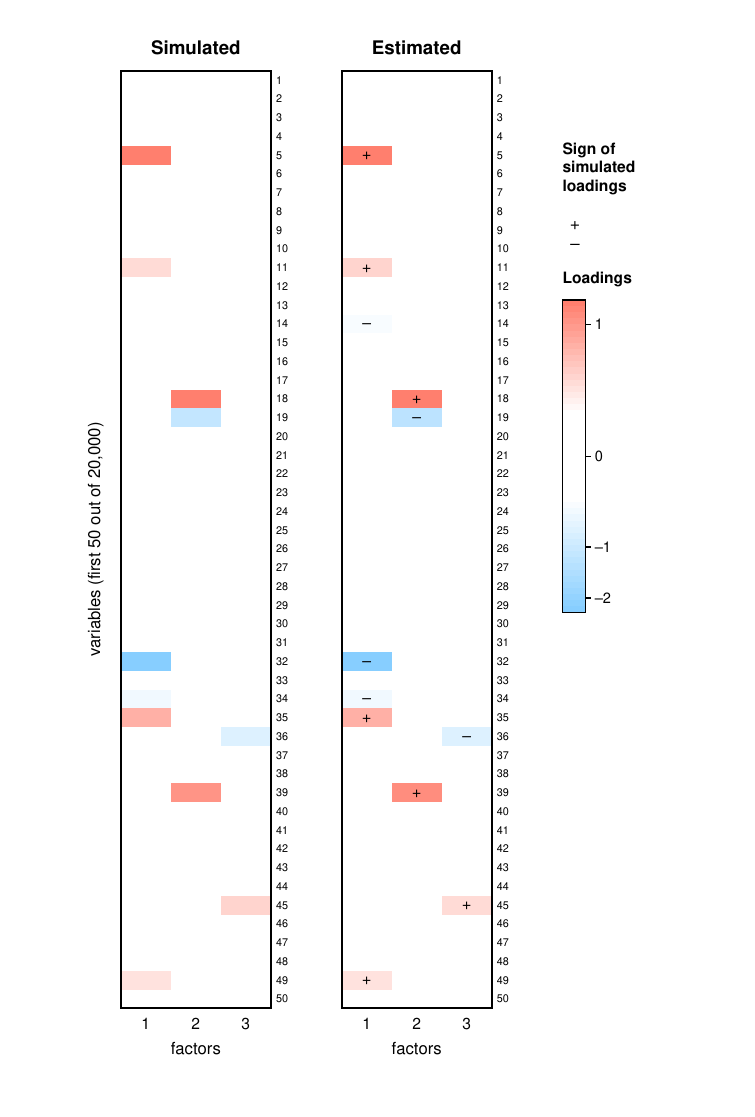}
\caption{\footnotesize Simulated and estimated loadings of the three factors for the
first 50 of the $20\,000$ variables in the single replicate of Section~\ref{sec_sim_single}
($N = 100$, $Q = 3$, $L = 2$), after alignment of the estimated factors with the simulated ones.
The two panels share one colour scale, and $+$ and $-$ mark on the estimated panel the signs of
the nonzero simulated loadings, as in Figure~\ref{fig2}(b).}\label{fig:supp_loadings}
\end{figure}

For the single replicate of
Section~\ref{sec_sim_single}, Figure~\ref{fig:supp_loadings} shows the simulated loadings
next to the estimated ones for the first 50 of the $20\,000$ variables, so that the
magnitudes of the simulated effects can be compared with the estimates as well as their
signs.

\newpage
\section{Estimated factor trajectories in the MEFISTO comparison}
\label{sec:supp_factor_traj}

\begin{figure}[!ht]
\centering
\includegraphics[width=0.84\linewidth]{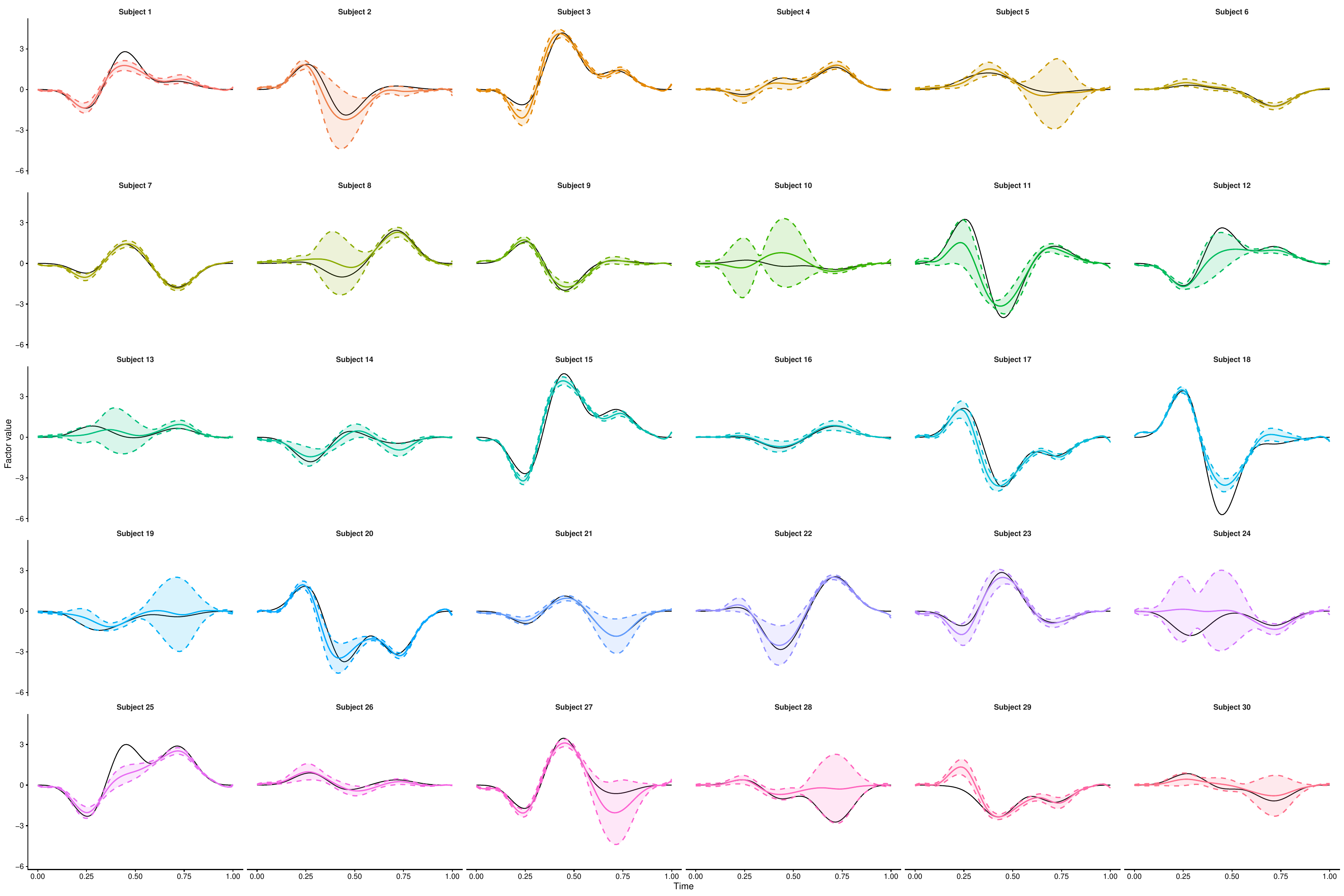}
\caption{\footnotesize
Reconstructed subject-specific trajectories for latent factor $q=1$ obtained with bayesSYNC for one data replicate of the simulation scenario with $N=30$ subjects, with $95\%$ credible bands (colors). True simulated latent trajectories are shown in black. Each panel corresponds to one subject.
\vspace{0.2cm}
}
\label{fig:supp_factor_q1_sync}
\end{figure}

\begin{figure}[!ht]
\centering
\includegraphics[width=0.84\linewidth]{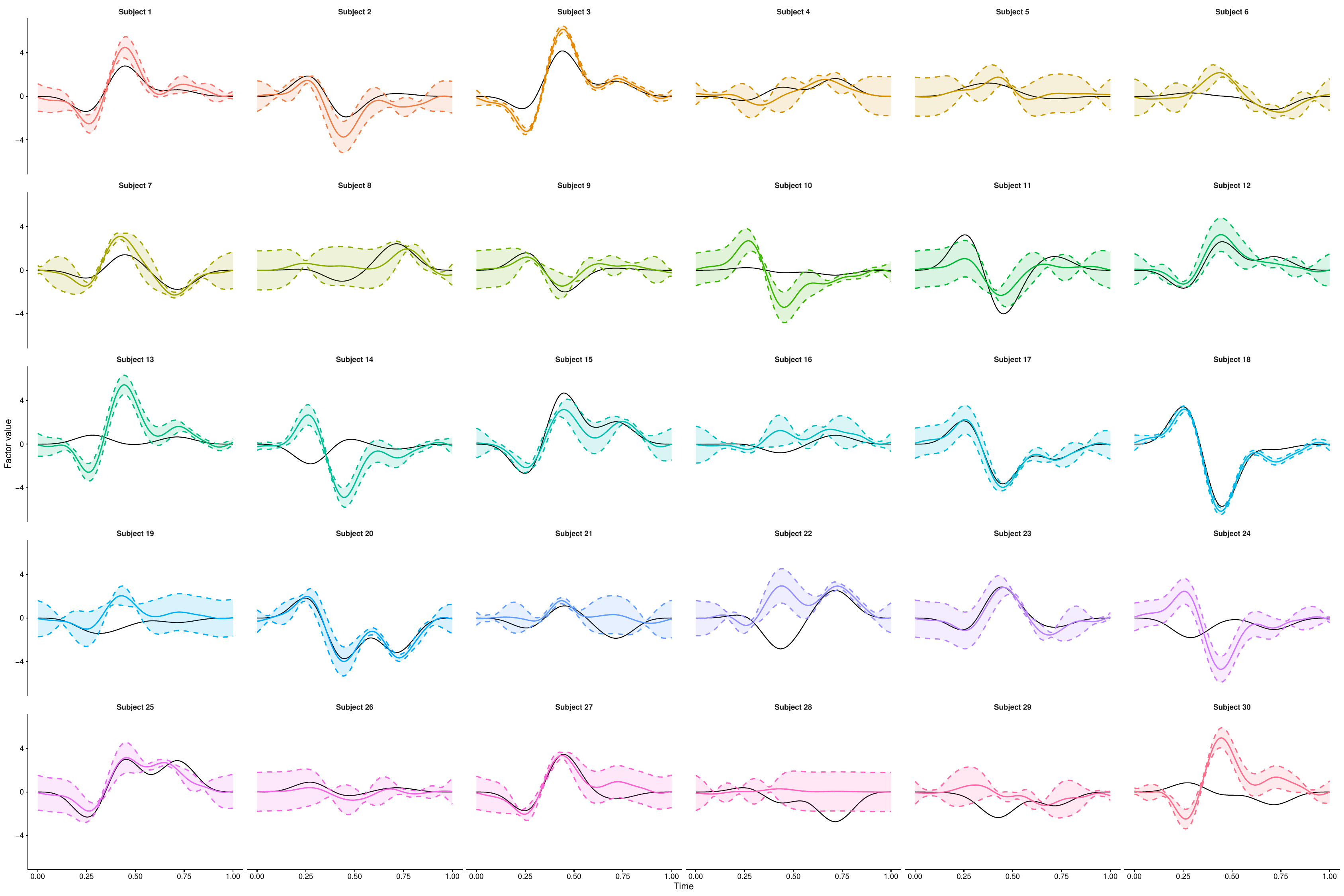}
\caption{\footnotesize
Reconstructed subject-specific trajectories for latent factor $q=1$ obtained with MEFISTO; description as per Figure~\ref{fig:supp_factor_q1_sync}. 
}
\label{fig:supp_factor_q1_mefisto}
\end{figure}

\begin{figure}[!ht]
\centering
\includegraphics[width=0.84\linewidth]{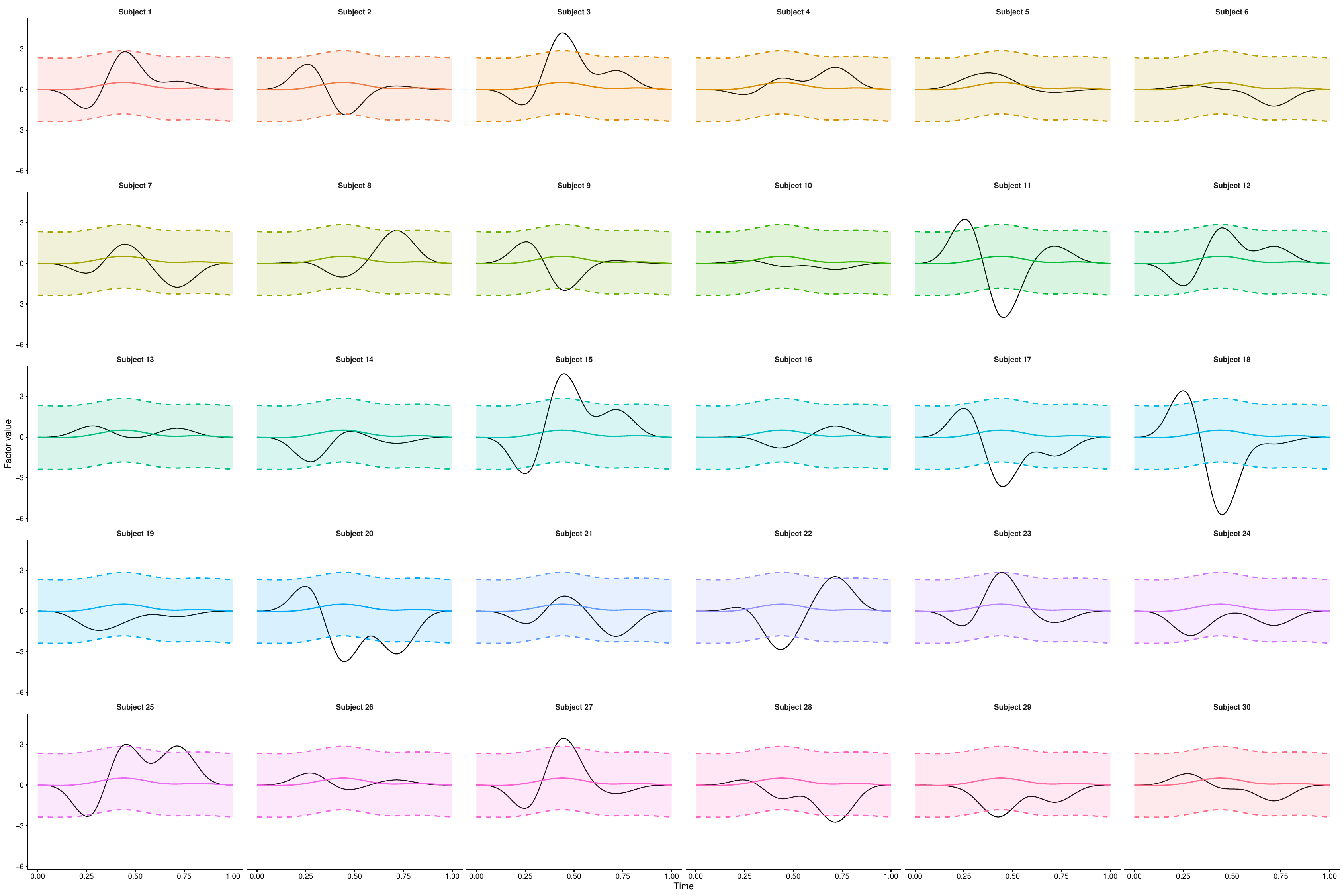}
\caption{\footnotesize
Reconstructed trajectories for latent factor $q=1$ obtained with MEFISTO-base (the inferred trajectories are the same for all subjects); description as per Figure~\ref{fig:supp_factor_q1_sync}. 
\vspace{0.5cm}
}
\label{fig:supp_factor_q1_mefisto_base}
\end{figure}

\begin{figure}[!ht]
\centering
\includegraphics[width=0.84\linewidth]{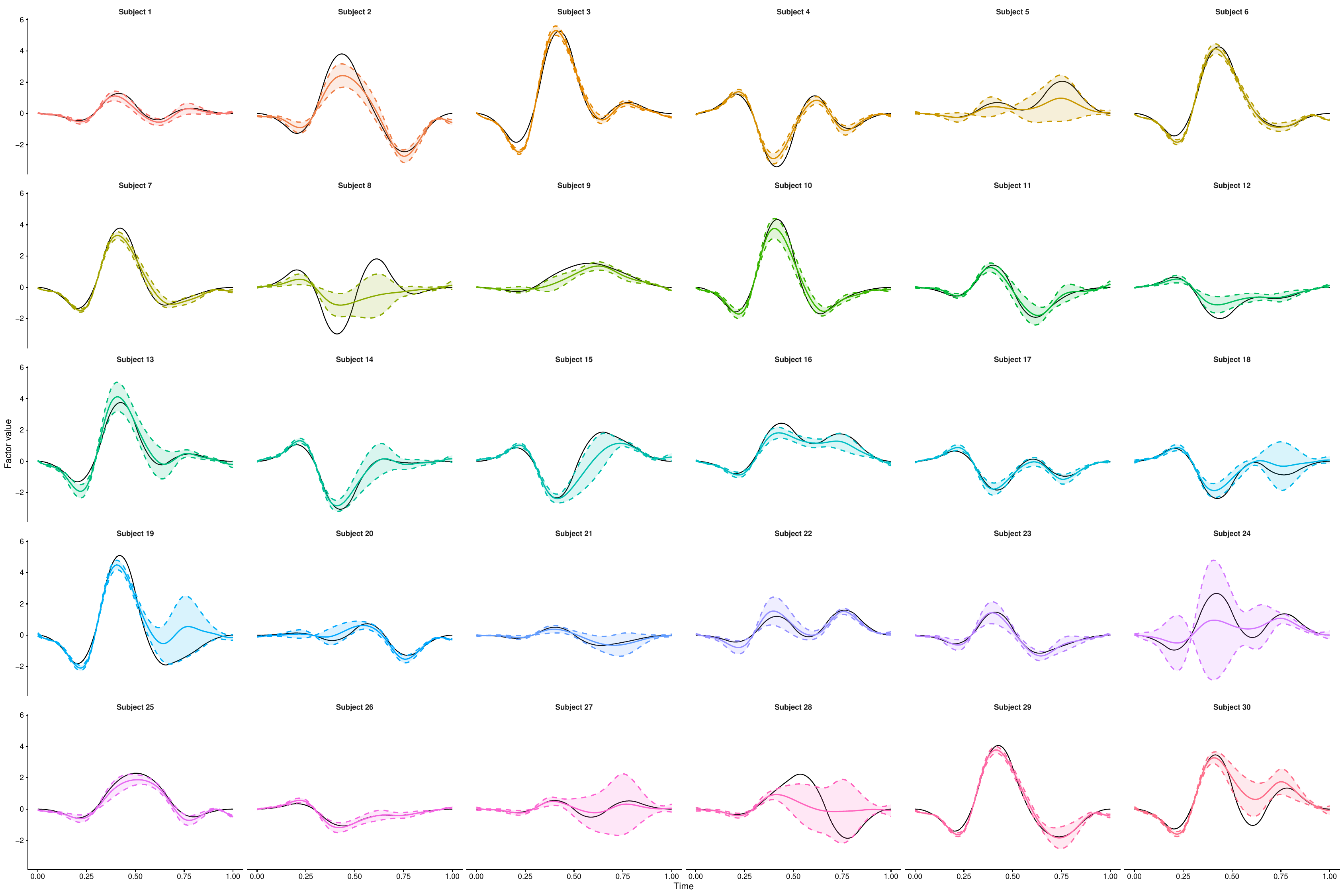}
\caption{\footnotesize
Reconstructed subject-specific trajectories for latent factor $q=2$ obtained with bayesSYNC; description as per Figure~\ref{fig:supp_factor_q1_sync}. 
}
\label{fig:supp_factor_q2_sync}
\end{figure}

\begin{figure}[!h]
\centering
\includegraphics[width=0.84\linewidth]{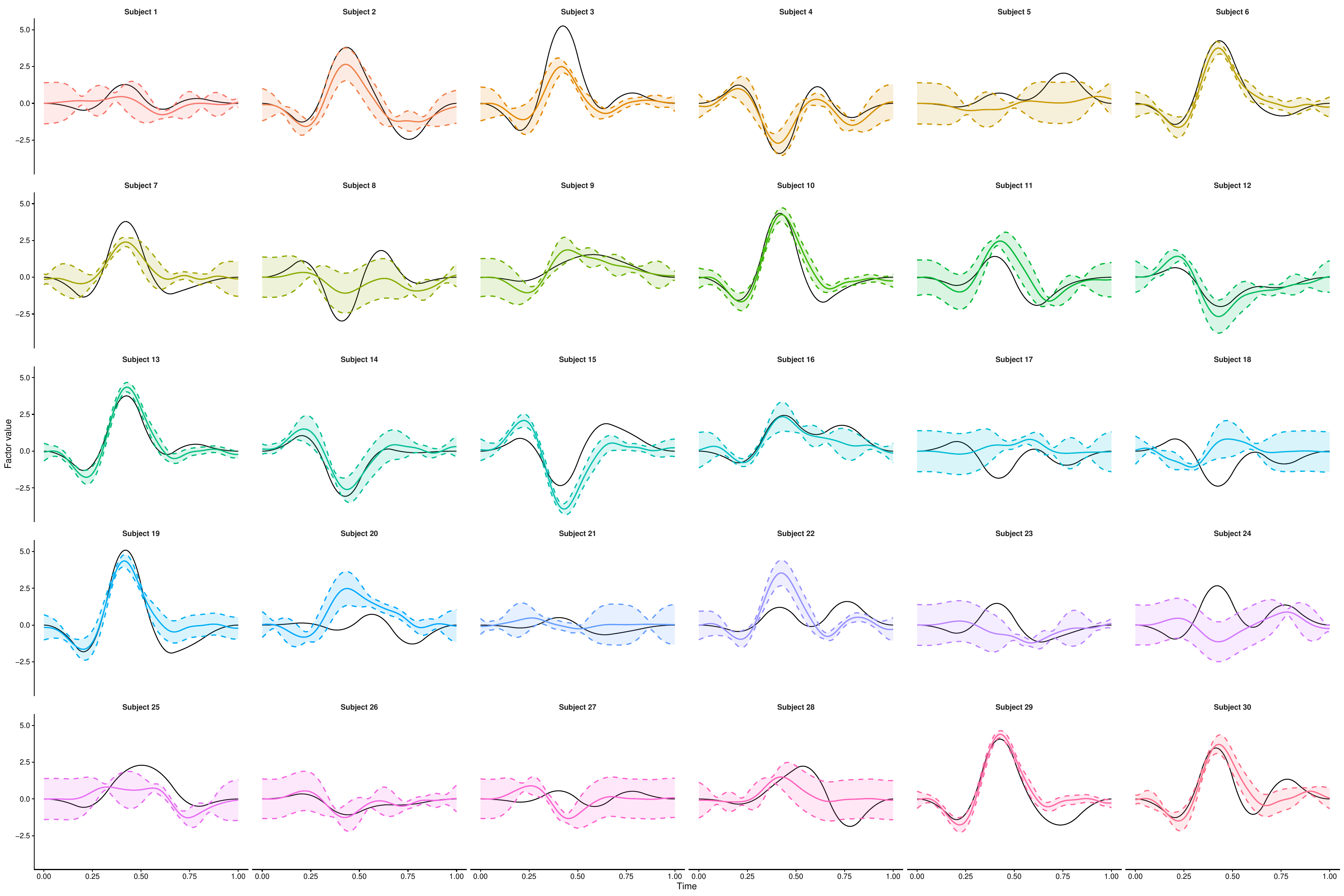}
\caption{\footnotesize
Reconstructed subject-specific trajectories for latent factor $q=2$ obtained with MEFISTO; description as per Figure~\ref{fig:supp_factor_q1_sync}. 
\vspace{0.5cm}
}
\label{fig:supp_factor_q2_mefisto}
\end{figure}

\begin{figure}[!h]
\centering
\includegraphics[width=0.84\linewidth]{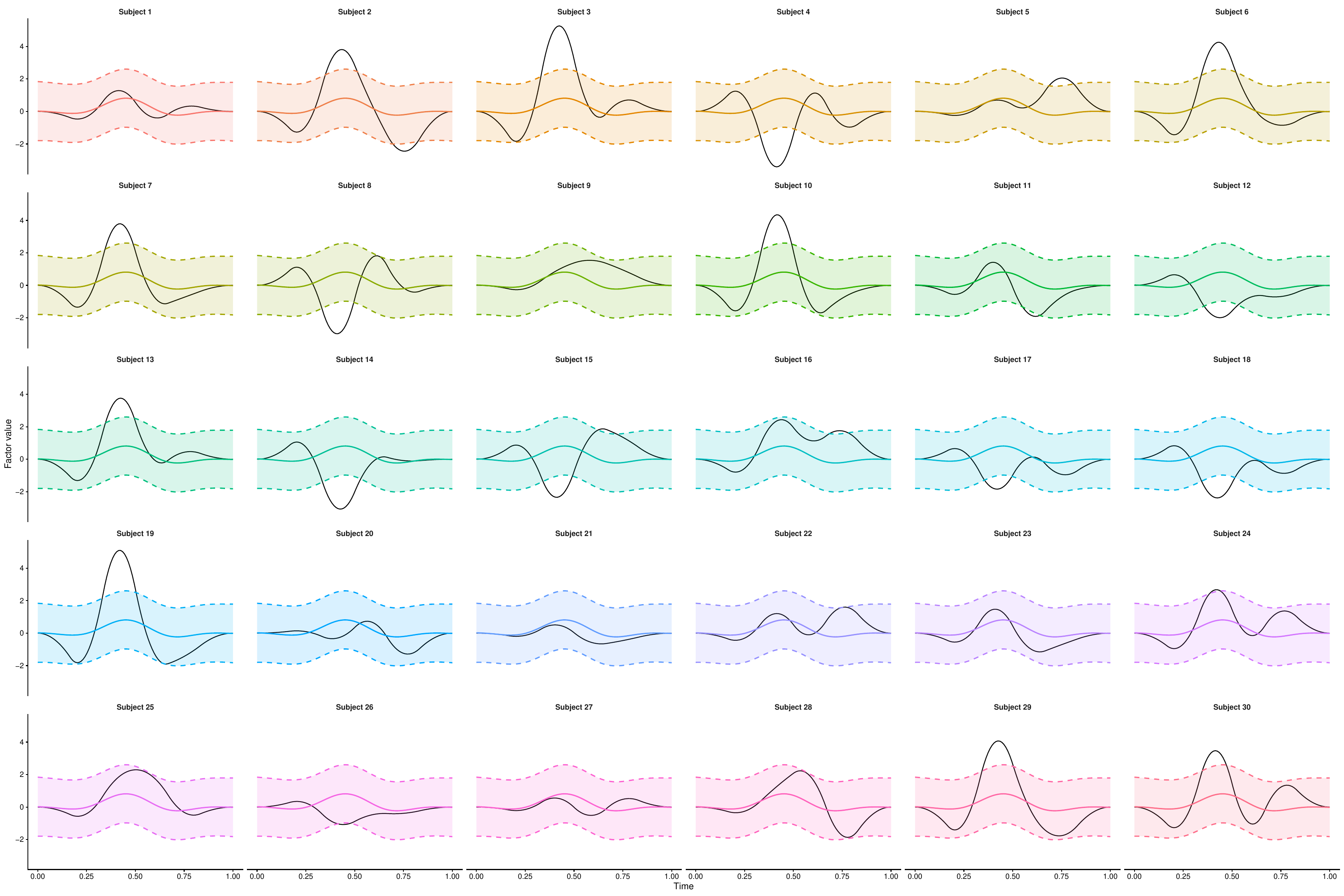}
\caption{\footnotesize
Reconstructed trajectories for latent factor $q=2$ obtained with MEFISTO-base (the inferred trajectories are the same for all subjects); description as per Figure~\ref{fig:supp_factor_q1_sync}. 
}
\label{fig:supp_factor_q2_mefisto_base}
\end{figure}

We provide additional visualisation of the reconstructed subject-specific latent factor trajectories for the simulation scenario with $N=30$ subjects described in Section~\ref{sec_sim_mefisto}.
Figures~\ref{fig:supp_factor_q1_sync}--\ref{fig:supp_factor_q2_mefisto_base} display the estimated trajectories for the two latent factors ($q=1,2$) across all subjects for one data replicate, as obtained using bayesSYNC, MEFISTO and MEFISTO-base.

Overall, bayesSYNC and MEFISTO achieve a decent recovery of the simulated trajectories across subjects. Our approach accurately captures 
subject-specific variability with narrow credible bands, while 
MEFISTO suffers from 
occasional failure to recover the correct temporal structure of the
latent factors. MEFISTO-base estimates, by design, a single set of latent factor trajectories shared across all subjects. It therefore cannot represent subject-level variation in the latent space. As a result, the posterior credible bands become wider in order to absorb the unexplained between-subject heterogeneity. 
\bigskip

\newpage

\section{Additional runtime profiling visualisation in the MEFISTO comparison}
\label{sec:supp_runtime}

We provide an additional log-scale visualisation of the runtime for our method (bayesSYNC) and the two Gaussian process method settings (MEFISTO and MEFISTO-base) across the simulation scenarios described in  Section~\ref{sec_sim_mefisto}. Both bayesSYNC and MEFISTO-base run within a few minutes, while MEFISTO (which models subject-specific latent trajectories, as bayesSYNC) requires hours to complete. 

\begin{figure}[h!]
\centering
\includegraphics[scale=0.65]{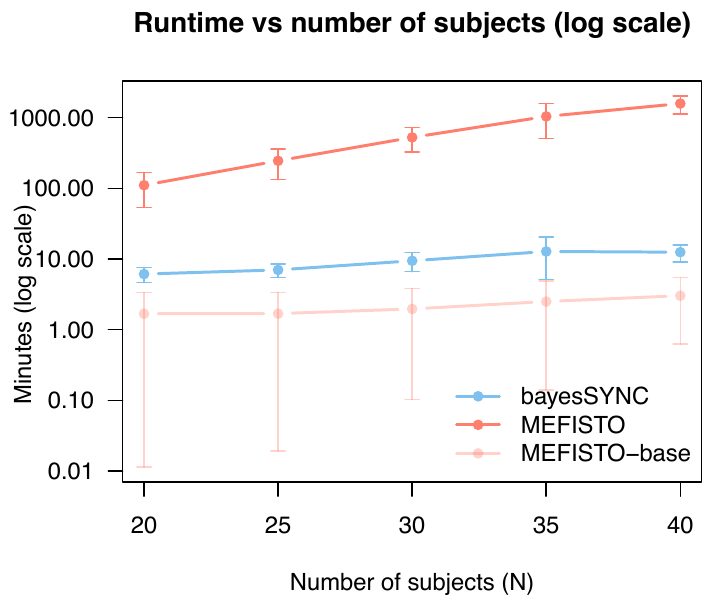}
\caption{\footnotesize Computational performance of our method (bayesSYNC) and the Gaussian process-based method with or without subject-specific latent factor trajectories (MEFISTO and MEFISTO-base). 
Runtime is displayed in minutes on a log scale, as a function of sample size $N$, averaged over replicates on an Intel Xeon CPU, 2.60 GHz machine. Error bars denote $\pm$ one standard error, based on 25 replicates ($14$ replicates only for MEFISTO in the $N=40$-scenario). 
All methods are applied with $Q_{\max}=5$; serial runtimes are reported for comparability.
}\label{fig:supp_runtime}
\end{figure}

\newpage
\section{Residual correlation in the COVID-19 application}\label{sm_sec_resid_covid}

\begin{figure}[h!]
\centering
\includegraphics[width=0.5\linewidth]{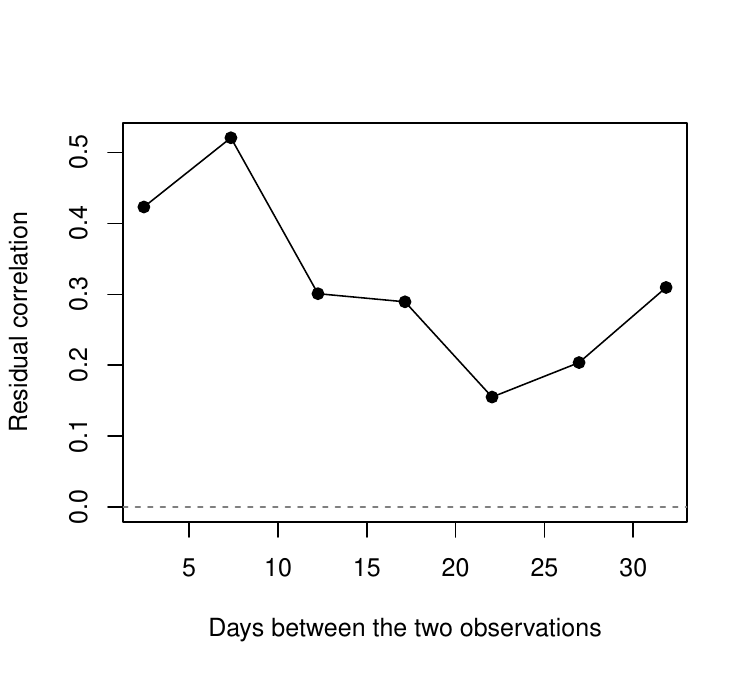}
\caption{\footnotesize Correlation between the residuals of the same marker at two
samples of the same patient, against the number of days between the samples, in the COVID-19
application ($N = 88$ patients, 57 markers). Residuals are pooled over markers and the bins
are about five days wide. Under errors uncorrelated along $t$, the correlation would be zero
at every lag.}\label{fig:supp_correlogram}
\end{figure}

Figure~\ref{fig:supp_correlogram} shows, for the
application of Section~\ref{sec_application}, the correlation between the residuals of the
same marker at two sampling times of the same patient, against the number of days between
them. It is the statistic reported for the simulations in Section~\ref{sec_misspec}. Of
the 88 patients, 39 have more than one sample, and no two samples of a patient are more than
34 days apart. The correlation is positive at every lag, between 0.16 and 0.52, so the
variable-specific deviations are correlated along $t$. Residuals of different markers taken
at the same visit are, however, nearly uncorrelated (mean correlation 0.03), which indicates
that the correlation is specific to each marker, a patient's level of a given marker staying
above or below what the factors predict. Deviations of this kind enter only the covariance
of a marker with itself and not the covariance between different markers
(Section~\ref{sec_latent}), and in the simulations of Section~\ref{sec_misspec}
correlated errors left the recovery of the loadings intact. Their main cost there was an
inflated number of factors, nearly all replicates retaining all five candidate factors. Here two of the five
are retained, although, since the true number of factors is unknown, this does not exclude
that part of the retained structure reflects these deviations.

\newpage
\section{Sensitivity of the COVID-19 application to the patients who died}\label{sm_sec_deaths}

\begin{figure}[h!]
\centering
\includegraphics[width=0.85\linewidth]{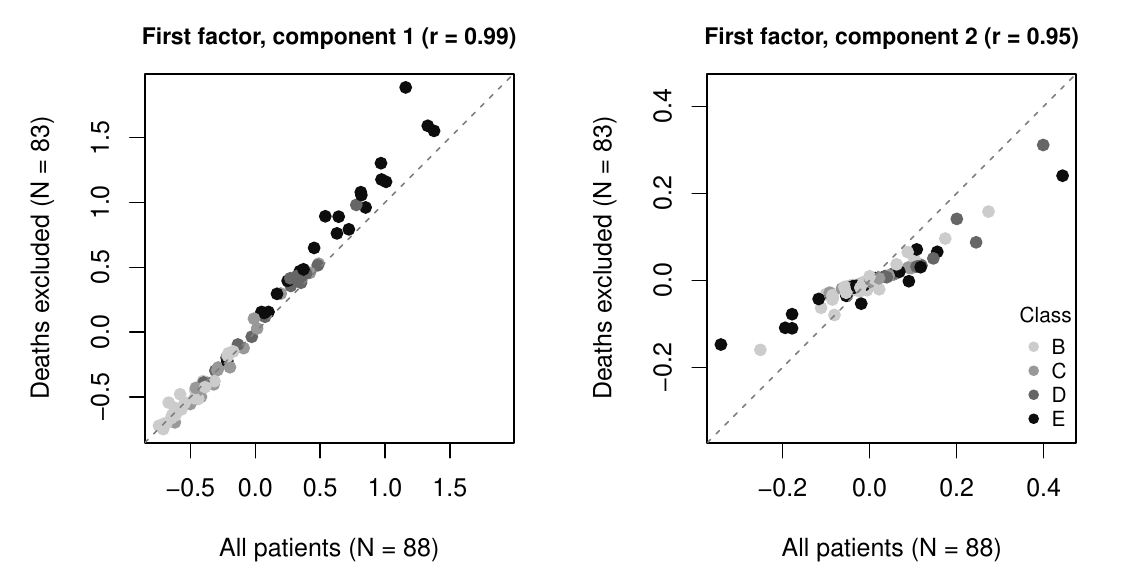}
\caption{\footnotesize Scores of the first factor in the fit with all 88 patients
(horizontal axis) and in the fit excluding the five patients who died within the window
(vertical axis), for the 83 patients common to both, shaded by severity class. Left, first
component; right, second component. The dashed line is the identity.}
\label{fig:supp_deaths_scores}
\end{figure}

Five of the 88 patients analysed in
Section~\ref{sec_application} died within the seven-week window, so their trajectories are
truncated by death. We refitted the model without them, on the remaining 83 patients, with
the same settings. The same two factors are retained. Table~\ref{tab:supp_deaths} compares the
two fits factor by factor, after matching the factors on their loadings, and
Figure~\ref{fig:supp_deaths_scores} compares the scores of the first factor for the 83
patients common to both fits. The loadings of both factors and the first component of the
first factor are essentially unchanged. The second component of the first factor, which
Section~\ref{sec_application} relates to recovery, keeps the same ordering of patients
but explains a much smaller share of the variation once the five patients are excluded, and
the temporal components of the second factor change more than its loadings.

\begin{table}[h!]
\centering
\caption{\footnotesize Agreement between the fits with ($N = 88$) and without
($N = 83$) the five patients who died within the window. The loadings are compared by their
absolute correlation over the 57 markers, the first two eigenfunctions by their absolute cosine
similarity and the scores by their absolute correlation over the 83 patients common to both
fits. The last two columns give the share of the variation of the factor explained by its
first and second components in each fit. Absolute values are given because the sign of a
factor or of a component is not determined by the data.}
\label{tab:supp_deaths}
\footnotesize
\begin{tabular}{lccccccc}
\toprule
 & Loadings & \multicolumn{2}{c}{Eigenfunctions} & \multicolumn{2}{c}{Scores} & \multicolumn{2}{c}{Variance explained (\%)} \\
\cmidrule(lr){3-4} \cmidrule(lr){5-6} \cmidrule(lr){7-8}
Factor & & comp.\ 1 & comp.\ 2 & comp.\ 1 & comp.\ 2 & $N = 88$ & $N = 83$ \\
\midrule
First (inflammation) & 0.993 & 0.998 & 0.996 & 0.992 & 0.950 & 93.3 / 6.7 & 99.0 / 1.0 \\
Second (amino acids) & 0.996 & 0.891 & 0.758 & 0.832 & 0.627 & 48.0 / 27.4 & 62.0 / 26.5 \\
\bottomrule
\end{tabular}

\end{table}

\newpage
\section{Comparison with MEFISTO on the COVID-19 data}\label{sm_sec_mefisto_covid}

\begin{figure}[h!]
\centering
\includegraphics[width=0.6\linewidth]{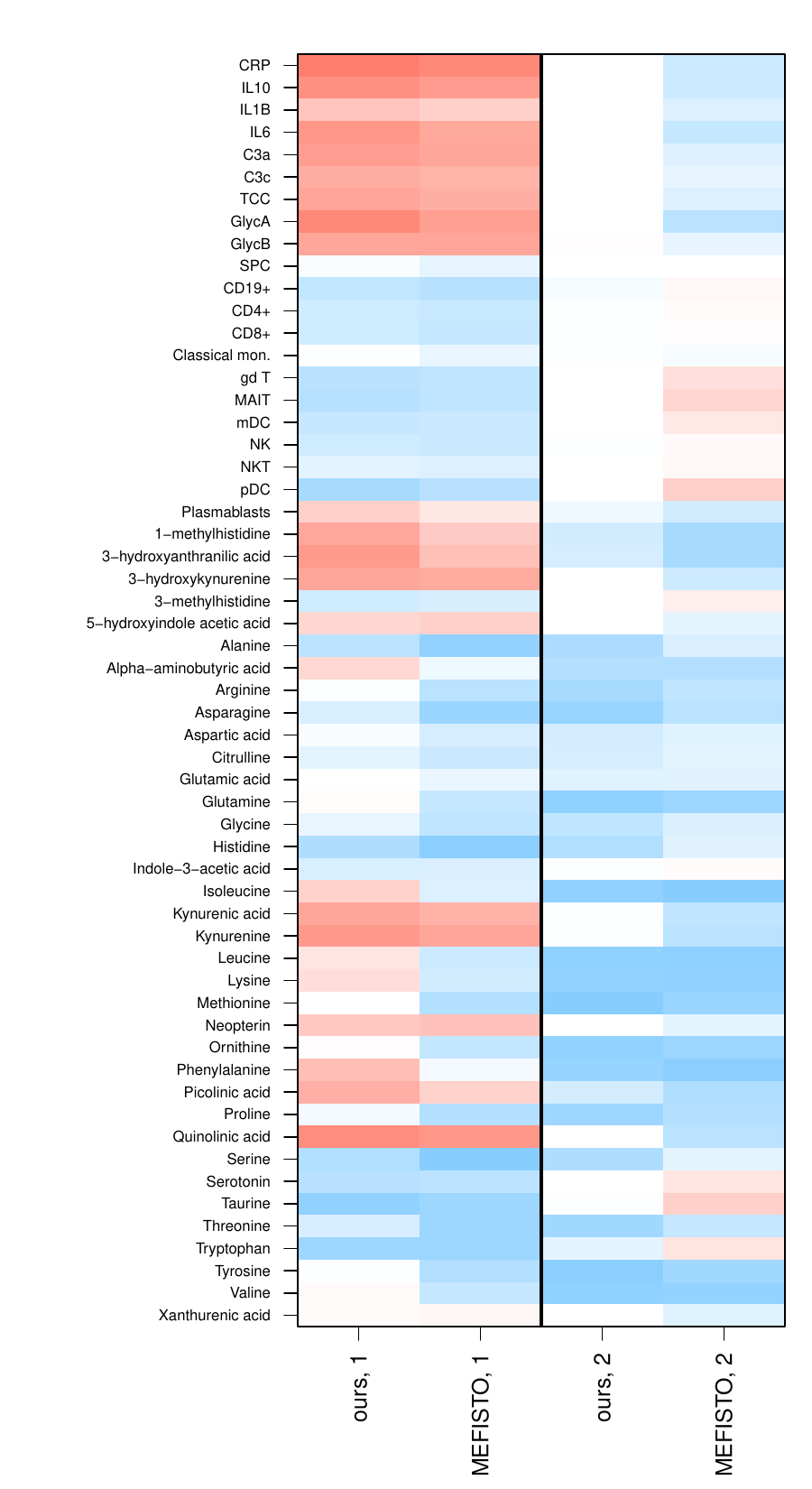}
\caption{\footnotesize Loadings of the two factors retained by our model and of the
matching MEFISTO-base factors, on the COVID-19 data. Each column is scaled to a largest
absolute value of one, since the two methods scale factors differently, and oriented as in
Figure~\ref{fig6}.}\label{fig:supp_mefisto_loadings}
\end{figure}

We fitted MEFISTO-base, which estimates one latent
trajectory per factor shared by all patients, to the data of
Section~\ref{sec_application}, with the same number of candidate factors, $Q_{\max} = 5$.
The version of MEFISTO with subject-specific trajectories was not fitted, since in
Section~\ref{sec_sim_mefisto} it did not complete within 36 hours for 11 of 25 replicates
already at $N = 40$, whereas the application has $N = 88$. MEFISTO-base retains all five
factors, which explain between 4\% and 23\% of the variance. Its two leading factors
correspond to ours, with loading correlations of 0.91 for the inflammatory factor and 0.71 for
the amino-acid factor (Figure~\ref{fig:supp_mefisto_loadings}), but they are less clearly
separated and less sparse. The first also gives weight to several amino acids, and the second
to the inflammatory markers and to metabolites of the kynurenine pathway; 49 of the 57 markers
load on the second with more than 10\% of its largest loading, against 26 for our second
factor. The estimated smoothness of the MEFISTO factors in time is close to zero, so their
shared trajectories carry almost no temporal structure. Averaged over each patient's samples,
the values of MEFISTO's first factor are associated with severity (Kruskal--Wallis
$p = 2 \times 10^{-9}$), as are our first-component scores ($p = 2 \times 10^{-13}$), and the
two order the patients similarly (rank correlation 0.85). MEFISTO-base has no representation
of how patients differ in their trajectories, so the functional principal components and the
patient-level scores of Figure~\ref{fig6}(c) have no counterpart in its output.

\newpage
\section{The transcriptomic factors}\label{sm_sec_rnaseq}

Table~\ref{tab:supp_rnaseq} gives every column of the
summary reported for the transcriptomic analysis in Section~\ref{sec_rnaseq}, for all ten
factors, and the last paragraph of this section gives the over-representation results that the
main text omits. The
patient-level factor value used for the severity and CRP columns is the mean of the estimated
factor trajectory at that patient's own sampling times, so it uses no extrapolated part of the
trajectory and does not depend on the rotation within a factor made by the post-processing. The
length-bias index is the within-sample rank correlation between each gene's deviation from its
mean and its log transcript length, so a sample with degraded RNA, which is depleted in long
transcripts, takes a negative value. We do not use it to flag a factor as technical, because
shifts in cell composition move it as well, which may explain why the neutrophil and lymphocyte
factor has the largest correlation with it ($-0.46$). The neutrophil marker genes examined for that
factor are FCGR3B, CXCR2, CSF3R, S100A8, S100A12, MMP8, MMP9, ARG1, CEACAM8, LCN2, FPR1 and MME,
and the T- and B-cell markers CD3E, CD3D, CD247, CD2, IL7R, CCR7, LEF1, TCF7, MS4A1, CD79A, CD19,
BCL11B, CD5 and CD6.

As a sensitivity analysis, we refitted the model
without the 15 patients who died within the window, with the same settings. The interferon,
erythroid and 5S rRNA factors are recovered, with overlaps of 0.91, 0.89 and 0.92 between the sets
of 200 genes with the largest loadings in the two fits and loading correlations of 0.98, 0.95 and
0.97. The factor most associated with severity again contrasts neutrophils with lymphocytes, with
ten of the twelve neutrophil markers loading with one sign and all fourteen T- and B-cell markers
with the other, although its 200 largest loadings change, as can be expected for a factor that
selects most of the genes. The fit without the patients who died also contains a proliferation
factor, with nine of ten cell-cycle markers among its 200 largest loadings (E2F targets
$5.5\times10^{-9}$), and an inflammatory factor (TNF-$\alpha$ signalling via NF-$\kappa$B,
$4.7\times10^{-7}$), neither of which has a counterpart in the main fit. These two factors therefore
depend on the exclusion of the patients who died, and we do not interpret them.

Several factors are associated with severity one at
a time (Table~\ref{tab:supp_rnaseq}). To see whether they carry distinct information, we regressed
the severity class on the leading scores of all ten factors together, in a proportional-odds
model, since the class is ordinal. The leading scores of different factors are nearly
uncorrelated, with absolute correlations of at most 0.27. When the other factors are allowed for,
the scores of four factors remain associated with severity at the Bonferroni-corrected level of
0.005, namely the neutrophil and lymphocyte factor (odds ratio 4.7 per standard deviation of the
score, 95\% interval 3.0 to 7.1), the interferon factor (1.9, 1.3 to 2.6), and factors 5 (1.9, 1.4
to 2.7) and 8 (2.0, 1.3 to 2.9), whose leading hallmark sets are the G2M checkpoint and the E2F
targets. Together, the nine other factors add to the neutrophil and lymphocyte factor
(likelihood-ratio test, $p = 2\times10^{-9}$). Factors 2 and 9 reach $p = 0.02$ without passing
the correction; factor 9 is flagged as technical, and its association may reflect a difference in
RNA quality between the severity classes. The scores are estimates, shrunk more for patients with
fewer samples, and their uncertainty is not carried into the model, so these odds ratios describe
associations rather than the ability of the factors to predict severity.

\begin{figure}[h!]
\centering
\includegraphics[width=\linewidth]{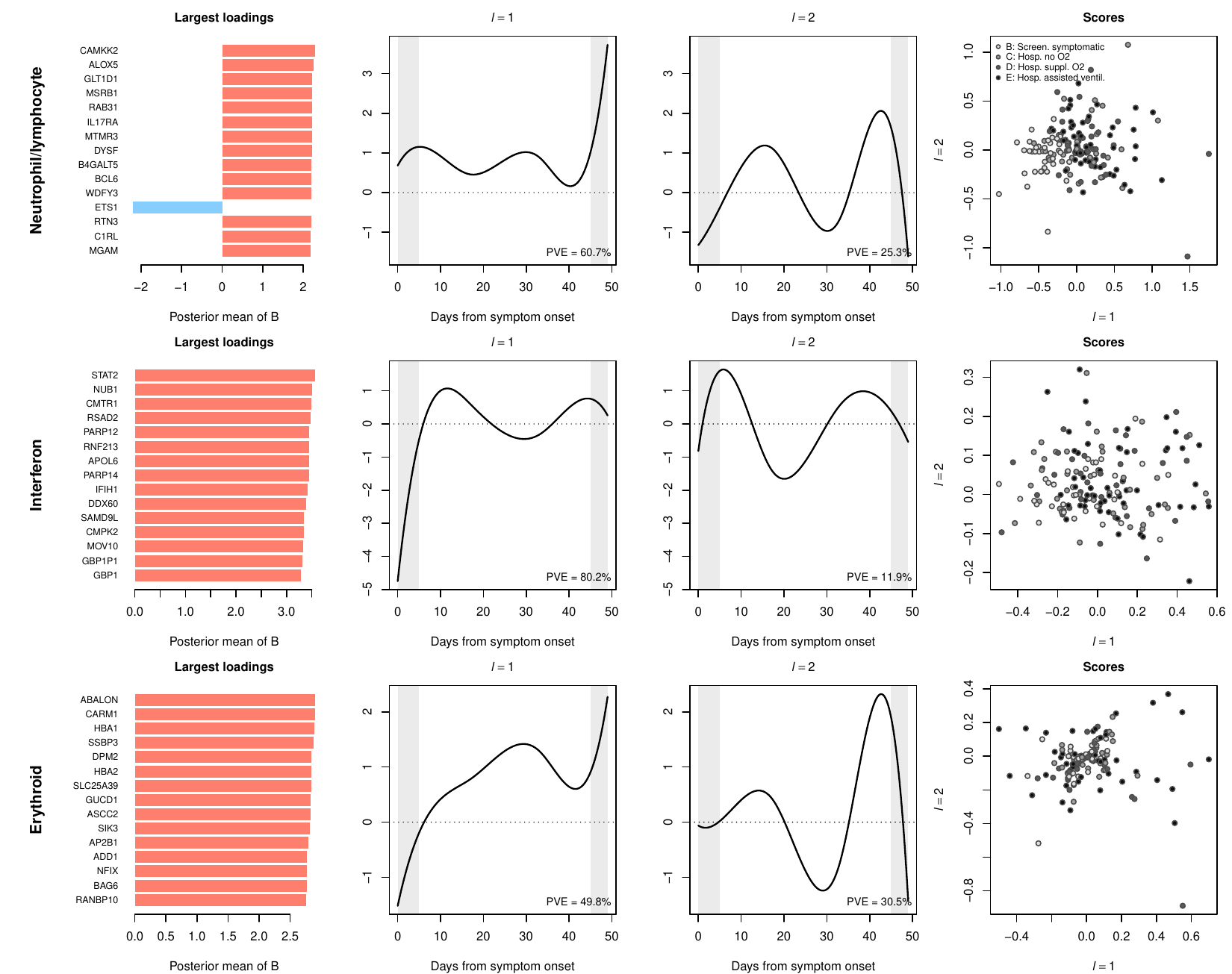}
\caption{\footnotesize The neutrophil and lymphocyte,
interferon and erythroid factors of the transcriptomic analysis (Section~\ref{sec_rnaseq}). Left, the 15 largest loadings of each
factor. Middle, the first two eigenfunctions, with the share of the variation of the factor that
each explains. Days before day 5 and after day 45, in which 22 and 10 of the 164 patients have a
sample, are shaded, since the eigenfunctions rest on few observations there. Right, the scores on
the two components, shaded by severity class. Each factor is oriented so that most of its 50
largest loadings are positive, so that a higher trajectory means higher expression of these
genes, and each eigenfunction so that its mean over days 5 to 45 is positive, its scores changing
sign with it.}
\label{fig:supp_rnaseq_factors}
\end{figure}

\begin{table}[h!]
\centering
\caption{\footnotesize The ten factors of the transcriptomic analysis
(Section~\ref{sec_rnaseq}), $N = 164$ patients, 287 samples and 13{,}177 genes. Length pctile
is the percentile, among the analysed genes, of the median transcript length of the 100 largest
loadings. Severity $p$ is a Kruskal--Wallis test across the four classes and batch $p$ the same
test across the extraction batch of the patient's first sample. CRP $\rho$ is the rank correlation with the patient's peak hsCRP. A dagger marks a factor flagged
as technical, by a length percentile below 10 or above 90 or by a batch $p$-value below
$10^{-6}$.}
\label{tab:supp_rnaseq}
\scriptsize

\begin{tabular}{clrrrrrrc}
\toprule
 & Label & Selected & Length & Severity & Severity & CRP & Batch & Length \\
 & & (\%) & pctile & $\rho$ & $p$ & $\rho$ & $p$ & bias $\rho$ \\
\midrule
1 & Neutrophil/lymphocyte & 87 & 66.8 & 0.65 & $3.9\times10^{-16}$ & 0.54 & $6.6\times10^{-4}$ & $-0.46$ \\
2 & --- & 76 & 46.6 & 0.41 & $3.2\times10^{-8}$ & 0.34 & $1.0\times10^{-5}$ & $-0.07$ \\
3 & 5S rRNA, pseudogenes$^{\dagger}$ & 20 & 4.6 & 0.08 & 0.39 & 0.02 & $6.0\times10^{-18}$ & $-0.33$ \\
4 & --- & 44 & 64.2 & 0.02 & $3.9\times10^{-2}$ & $-0.01$ & 0.19 & $-0.23$ \\
5 & --- & 60 & 83.5 & 0.29 & $3.3\times10^{-3}$ & 0.25 & $9.1\times10^{-2}$ & $-0.35$ \\
6 & Interferon & 42 & 52.0 & 0.16 & $1.5\times10^{-2}$ & 0.32 & 0.70 & 0.01 \\
7 & --- & 57 & 60.0 & 0.05 & 0.12 & 0.18 & 0.27 & 0.35 \\
8 & --- & 40 & 60.9 & 0.20 & $1.5\times10^{-3}$ & 0.16 & 0.31 & 0.04 \\
9 & Short transcripts, pseudogenes$^{\dagger}$ & 49 & 4.5 & 0.06 & $1.3\times10^{-5}$ & 0.09 & $3.3\times10^{-9}$ & $-0.39$ \\
10 & Erythroid & 36 & 48.3 & 0.08 & 0.46 & $-0.01$ & $5.6\times10^{-2}$ & $-0.20$ \\
\bottomrule
\end{tabular}

\end{table}

\begin{table}[h!]
\centering
\caption{\footnotesize Key genes of the ten transcriptomic factors, the four largest
loadings of each sign, positive before the semicolon. Each factor is oriented so that its
patient-level value increases with the severity class, so for a factor associated with severity
the genes before the semicolon are the ones higher in the more severe patients. For a factor with
no such association the orientation is arbitrary.}
\label{tab:supp_rnaseq_genes}
\footnotesize

\begin{tabular}{clp{8.6cm}}
\toprule
 & Label & Key genes, the four largest loadings of each sign \\
\midrule
1 & Neutrophil/lymphocyte & CAMKK2, ALOX5, GLT1D1, MSRB1; ETS1, DYRK2, KLF12, TTC3 \\
2 & --- & SRP9, PSMC6, CBX3, USP1; SCAF1, CHERP, SF3A1, MEF2D \\
3 & 5S rRNA, pseudogenes & RPS2P5, RPL18AP3, RPS2P55, RPL23AP2; RNA5S6, RNA5S12, NBPF19, RNA5S4 \\
4 & --- & AP1G2, NEAT1, SRRM2, PPP6R2; SLBP, YWHAZ, SPCS1, MT\_CO2 \\
5 & --- & PIH1D1, TUBG1, H2BC15, DNAH10; KDM5A, STK4, NUFIP2, RALGAPB \\
6 & Interferon & STAT2, NUB1, CMTR1, RSAD2; RFLNB, ZNF398, SKI, NUDT3 \\
7 & --- & EIF3A, SON, ENO1, ACLY; MT\_TA, MT\_TQ, MT\_TP, MT\_TS1 \\
8 & --- & IGLJ2, IGLV1\_40, H3C8, IGLV2\_8; CD86, CD1D, MS4A7, MPEG1 \\
9 & Short transcripts, pseudogenes & PPIAP29, PPIAP31, TMSB4XP8, RNU6\_15P; NPIPB8, BAGE2, CFAP70, POTEI \\
10 & Erythroid & ABALON, CARM1, HBA1, SSBP3; SPDYE16, GOLGA8M, ABCC2, GOLGA6L4 \\
\bottomrule
\end{tabular}

\end{table}

\begin{figure}[h!]
\centering
\includegraphics[width=0.9\linewidth]{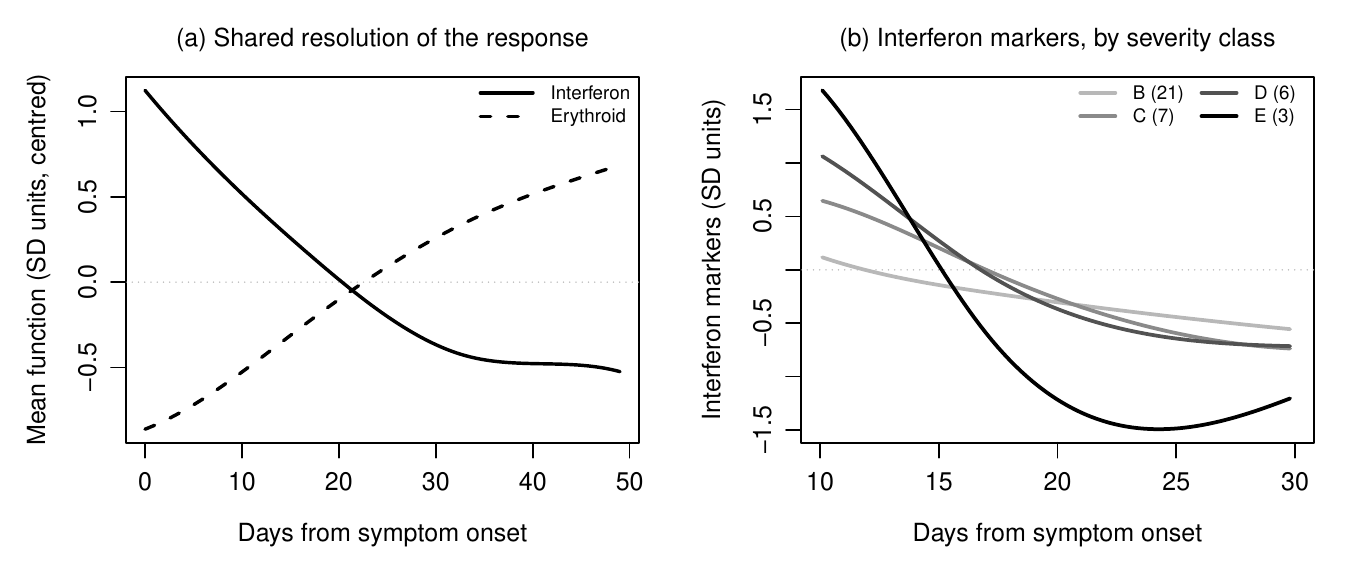}
\caption{\footnotesize The temporal structure of the interferon and erythroid factors of
the transcriptomic analysis, on the scale of the standardised expression data. (a) The estimated
mean functions, averaged over the 50 genes with the largest loadings on each factor and centred on
their mean over the seven weeks. The top-50 loadings of each factor share a sign, so the average
summarises how those genes move. (b) Expression of the ten canonical interferon-stimulated genes,
averaged within severity class, over days 10 to 30 and over the 37 patients whose own sampling
spans that interval, so that no patient trajectory is extrapolated. Classes B to E contain 21, 7,
6 and 3 of these patients, so the curve of class E rests on three patients.}
\label{fig:supp_rnaseq_temporal}
\end{figure}

Of the 88 patients of the panel analysis, 83 are also
in the transcriptomic analysis, with the same severity class in both (B 29, C 17, D 13, E 24).
Over these patients, the first-component scores of the inflammatory factor of the panel analysis
and the patient-level value of the neutrophil and lymphocyte factor have rank correlation 0.73
($p = 5\times10^{-15}$). Both are strongly associated with severity class in these patients
($\rho = 0.85$ and $0.78$), and given the class their partial rank correlation is 0.21
($p = 0.055$), or 0.17 ($p = 0.12$) when the mean sampling day in each analysis is also taken
into account. Within the four classes it is 0.11, 0.00, 0.59 and 0.33, with 29, 17, 13 and 24
patients. The agreement between the two analyses therefore goes mostly through severity class.
Part of the weakness of the association within classes may come from the samples of the two
analyses being taken on different days, a median of 7 days apart, and from the shrinkage of the
values of patients with one or two samples.

The association with severity of the change in
expression over time depends on the interval used, which is restricted to the patients whose own
sampling spans it. Over days 10 to 35, 10 to 30 and 10 to 25 (23, 37 and 40 patients), the rank
correlation between severity class and the change in the expression of the ten interferon marker
genes is $-0.60$, $-0.46$ and $-0.51$ ($p = 0.003$, $0.004$ and $0.001$). Over days 14 to 28 and 7
to 21 (53 and 28 patients) it is $-0.23$ and $-0.21$ ($p = 0.09$ and $0.27$). For the erythroid
marker genes the same five intervals give $+0.52$, $+0.31$, $+0.25$, $+0.34$ and $0.00$, which we do
not regard as a consistent association. The leading functional principal component of the
interferon factor, which places 76\% of its weight before day 10, is associated with severity class
over all 164 patients ($\rho = -0.27$, $p = 5\times10^{-4}$), more strongly among the 72 patients
with a single sample ($-0.32$) than among the 92 with two or more ($-0.20$). Since the score of a
patient sampled once rests on a single time point, this association says more about the level of
the response around the time of sampling than about its course, which is why we examine the
course on the expression scale.

Hallmark sets by over-representation among the 200 genes
with the largest loadings, against the analysed genes as background, complement the gene set
enrichment results of Section~\ref{sec_rnaseq}. For the interferon, erythroid and 5S rRNA
factors the two agree. Factor~6 gives the interferon gamma and interferon alpha responses
($2.3\times10^{-79}$ and $9.4\times10^{-79}$), factor~10 heme metabolism ($1.1\times10^{-26}$),
and factor~3 nothing, as by gene set enrichment. For the neutrophil and lymphocyte factor,
factor~1, both give only weak immune sets, complement ($3.7\times10^{-3}$) by over-representation
and allograft rejection ($4.1\times10^{-2}$) by gene set enrichment, as expected for a collection
without cell-type sets. Among the other factors, factor~2 gives MYC targets ($1.3\times10^{-5}$)
and oxidative phosphorylation ($2.9\times10^{-5}$), as by gene set enrichment, and factors~5
and~8 return nothing by over-representation.
Because each factor selects between 20\% and 87\% of the genes, over-representation on the
selected sets themselves has little power, which is why we rank by loading magnitude throughout.

\newpage
\section{Analysis of simulated data emulating the biomarker panel}\label{sm_sec_sim_panel}

Part of the COVID-19 data cannot be made public, so the
code accompanying the paper includes a script that generates simulated data emulating them, in
the format read by the analysis scripts. The data are generated from our model, with the
population-level parameters estimated on the real biomarker panel (Section~\ref{sec_metabo}),
namely the mean function, the loadings on the two retained factors and the residual standard
deviation of each of the 57 markers, the eigenfunctions of the two factors, and, for each
component, the mean and the within-class standard deviation of the scores in each severity
class. The 88 patients, their severity classes, their numbers of samples and the distribution of
the sampling days within each class follow the real design.

\begin{figure}[h!]
\centering
\includegraphics[width=0.78\linewidth]{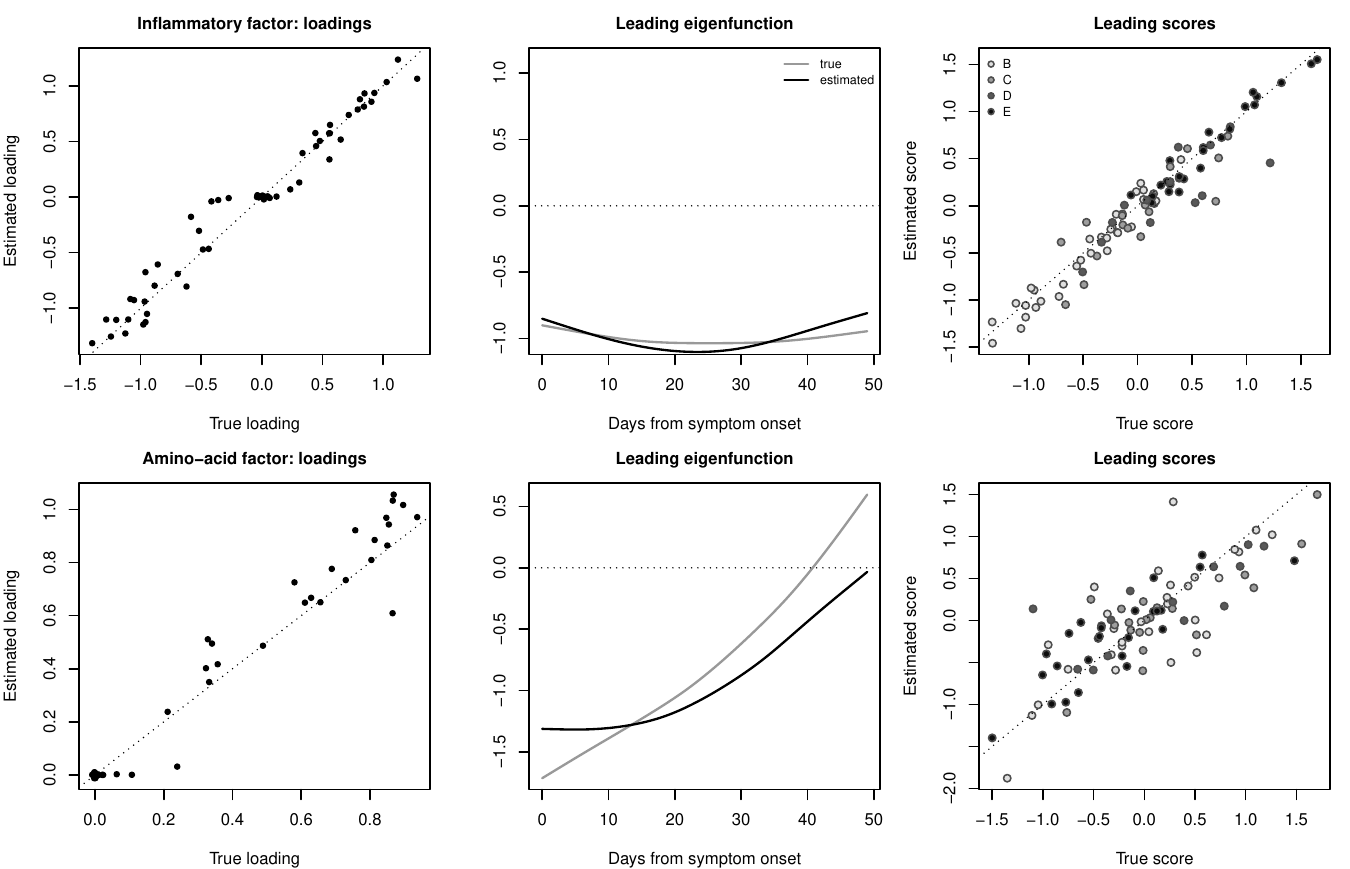}
\caption{\footnotesize Analysis of a simulated data set emulating the biomarker panel
(Supplementary Material~\ref{sm_sec_sim_panel}). Top, the simulated factor emulating the
inflammatory factor of Figure~\ref{fig6}; bottom, the one emulating the amino-acid factor.
Left, estimated against simulated loadings. Middle, the simulated (grey) and estimated (black)
leading eigenfunctions. Right, estimated against simulated leading scores, shaded by severity
class. The model identifies loadings, eigenfunctions and scores up to sign, so the signs of the
estimates are set to agree with the simulated values. Dotted lines mark the identity in the left
and right panels and zero in the middle ones.}
\label{fig:supp_sim_panel}
\end{figure}

The model retains two of the five candidate factors. Each
retained factor matches one simulated factor (Figure~\ref{fig:supp_sim_panel}), with a correlation
of 0.98 between the estimated and simulated loadings for both. We compare eigenfunctions by the
cosine of the angle between them in $L^2$. For the factor emulating the
inflammatory factor of Figure~\ref{fig6}, this cosine is 1.00 for the leading eigenfunction and
the correlation between the estimated and simulated leading scores is 0.96; the second component
is recovered as well, with a cosine of 1.00 and a score correlation of 0.67. For the factor
emulating the amino-acid factor, the leading component is recovered less closely, with a cosine
of 0.96 and a score correlation of 0.81, and its second component, which carries about a third
of the simulated variation of this factor, is not recovered (cosine 0.19, score correlation
0.23). The leading scores of the inflammatory factor are associated with the severity class as
the simulated scores are, with rank correlations of 0.70 and 0.73, and those of the amino-acid
factor only weakly, with $-0.14$ and $-0.22$. These results illustrate the use of the code and are
not expected to reproduce the values obtained on the real data.

\end{document}